\documentclass{jpp}
\usepackage{graphicx}
\usepackage{natbib}
\usepackage{amsmath}
\usepackage{amssymb}
\usepackage{color}
\usepackage{bm}
\usepackage{authblk}
\usepackage{multicol}

\ifCUPmtlplainloaded \else
  \checkfont{eurm10} 
  \iffontfound
    \IfFileExists{upmath.sty}
      {\typeout{^^JFound AMS Euler Roman fonts on the system,
                   using the 'upmath' package.^^J}%
       \usepackage{upmath}}
      {\typeout{^^JFound AMS Euler Roman fonts on the system, but you
                   dont seem to have the}%
       \typeout{'upmath' package installed. JPP.cls can take advantage
                 of these fonts, if you use 'upmath' package.^^J}%
      }
  \else
  \fi
\fi

\ifCUPmtlplainloaded \else
  \checkfont{msam10}
  \iffontfound
    \IfFileExists{amssymb.sty}
      {\typeout{^^JFound AMS Symbol fonts on the system, using the
                'amssymb' package.^^J}%
       \usepackage{amssymb}%
       \let\le=\leqslant  
       \let\ge=\geqslant  
      }{}
  \fi
\fi

\ifCUPmtlplainloaded \else
  \checkfont{msam10}
  \iffontfound
    \IfFileExists{amssymb.sty}
      {\typeout{^^JFound AMS Symbol fonts on the system, using the
                'amssymb' package.^^J}%
       \usepackage{amssymb}%
       \let\le=\leqslant  
       \let\ge=\geqslant  
      }{}
  \fi
\fi

\ifCUPmtlplainloaded \else
\IfFileExists{amsbsy.sty}
             {\typeout{^^JFound the 'amsbsy' package on the system, using it.^^J}%
     \usepackage{amsbsy}}
    {}
\fi

\newsavebox{\astrutbox}
\sbox{\astrutbox}{\rule[-5pt]{0pt}{20pt}}

\newcommand{\Alfven}{Alfv\'{e}n }
\newcommand{\Alfvenic}{Alfv\'{e}nic }

\newcommand{\V}[1]{\mathbf{#1}}

\newcommand{\figref}[1]{Fig.~\ref{#1}}   
\newcommand{\tabref}[1]{Table~\ref{#1}}   
\newcommand{\secref}[1]{\S\ref{#1}}

\title[Fundamental Parameters of Turbulence]{The Fundamental Parameters of Astrophysical Plasma Turbulence and its Dissipation: Nonrelativistic Limit }

\author{Gregory~G.~Howes}

\affiliation{Department of Physics and Astronomy,
University of Iowa, Iowa City IA 54224, USA}

\date{?; revised ?; accepted ?. - To be entered by editorial office}

\begin{document}
\maketitle

\begin{abstract}
A specific set of dimensionless plasma and turbulence parameters is
introduced to characterize the nature of turbulence and its
dissipation in weakly collisional space and astrophysical plasmas.
Key considerations are discussed for the development of predictive
models of the turbulent plasma heating that characterize the
partitioning of dissipated turbulent energy between the ion and
electron species and between the perpendicular and parallel degrees of
freedom for each species. Identifying the kinetic physical mechanisms
that govern the damping of the turbulent fluctuations is a critical first
step in constructing such turbulent heating models.  A set of ten
general plasma and turbulence parameters are defined, and reasonable
approximations along with the exploitation of existing scaling theories
for magnetohydrodynamic turbulence are used to reduce this general set
of ten parameters to just three parameters in the isotropic
temperature case. A critical step forward in this study is to
identify the dependence of all of the proposed kinetic mechanisms for
turbulent damping in terms of the same set of fundamental plasma and
turbulence parameters.  Analytical estimations of the scaling of each
damping mechanism on these fundamental parameters are presented, and
this information is synthesized to produce the first phase diagram for
the turbulent damping mechanisms as a function of driving scale and
ion plasma beta.
\end{abstract}


\section{Introduction}

Turbulence is a fundamental, yet incompletely understood, process in
space and astrophysical plasmas that mediates the transfer of the
energy of chaotic plasma flows and electromagnetic fields into the
energy of the plasma particles, either as heat of the plasma species
or as acceleration of a small population of particles
\citep{Howes:2017c}. The 2013 NRC Heliophysics Decadal Survey
\citep{NRCspace:2013} identifies plasma turbulence as a ubiquitous
phenomenon occurring both within the heliosphere and throughout the
Universe.  Predicting the heating or acceleration of the different
plasma species by turbulence, based on the observable turbulence and
plasma parameters at large scales, constitutes one of the grand
challenge problems in heliophysics and astrophysics.

Turbulent plasma heating plays a key role in governing the flow of
energy in the heliosphere, impacting the mesoscale and macroscale
evolution of key heliospheric environments comprising the coupled
solar-terrestrial system and connecting the solar corona, solar wind,
and Earth’s magnetosphere. The accuracy of ongoing efforts
\citep{Mikic:1999, Lionello:2009, vanderHolst:2014, Adhikari:2020}
to model globally the
flow of energy from the Sun through the interplanetary medium to the
Earth, to the other planets, and on to the boundary of the heliosphere with
the surrounding interstellar medium \citep{Opher:2020} relies on the
availability of prescriptive models of the turbulent
plasma heating \citep{Howes:2010d, Chandran:2011, Rowan:2017}.

Such turbulent heating prescriptions are also critical for the
interpretation of remote astronomical observations of emissions from
black hole accretion disks, such as the ground-breaking observations
by the Event Horizon Telescope of the supermassive black holes at the
center of M87 \citep{EHT:2019} and at Sagittarius A$^*$ in the Milky
Way \citep{EHT:2022}. In these cases, alternative turbulent heating
prescriptions \citep{Howes:2010d, Rowan:2017} have been shown to yield
drastically different predictions for the emitted radiation
\citep{Chael:2019}.

\subsection{How Does Turbulence Mediate Energy Transport and Plasma Heating?}

\begin{figure}
  \centering
\resizebox{4.5in}{!}{\includegraphics{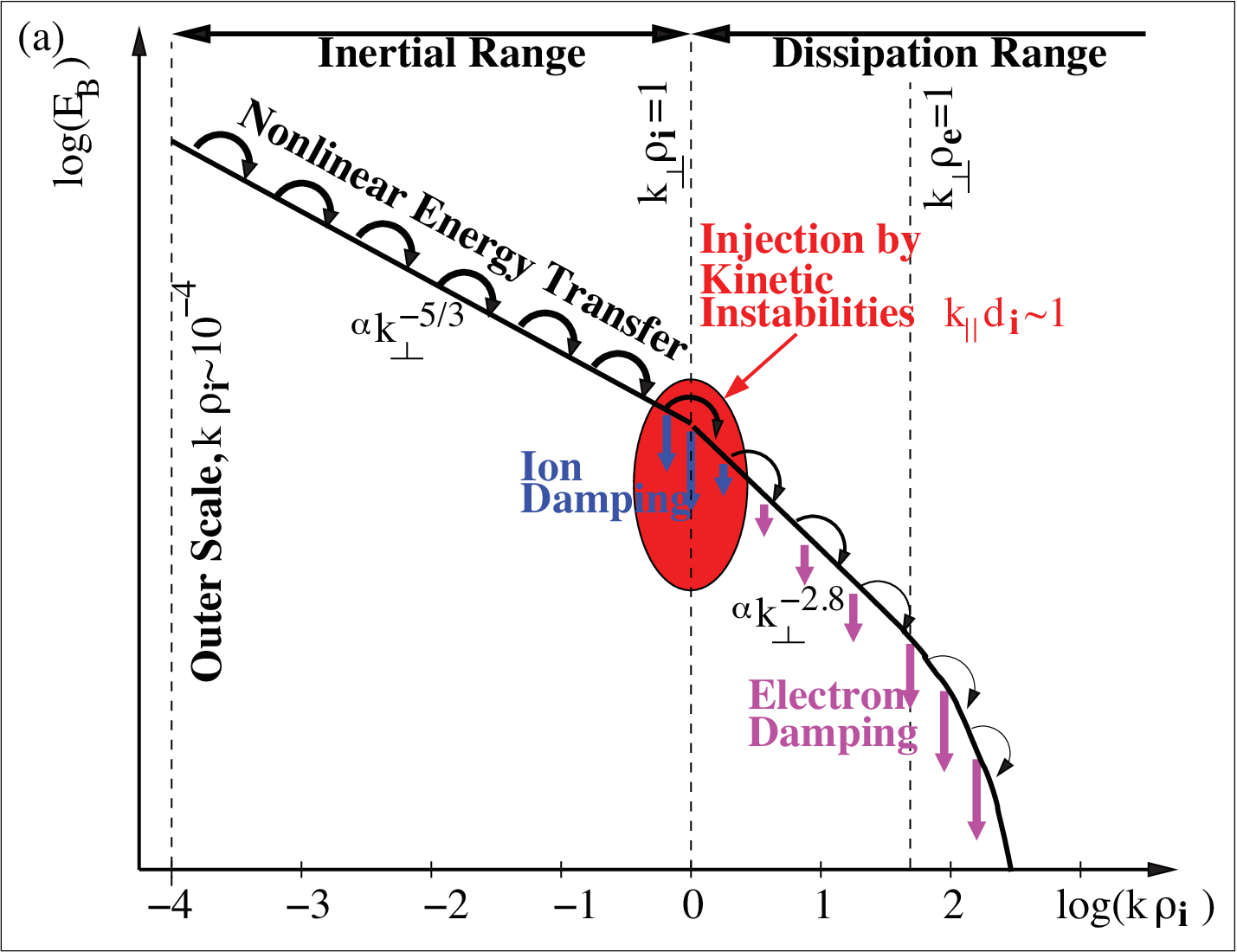}}\\
\resizebox{4.5in}{!}{\includegraphics{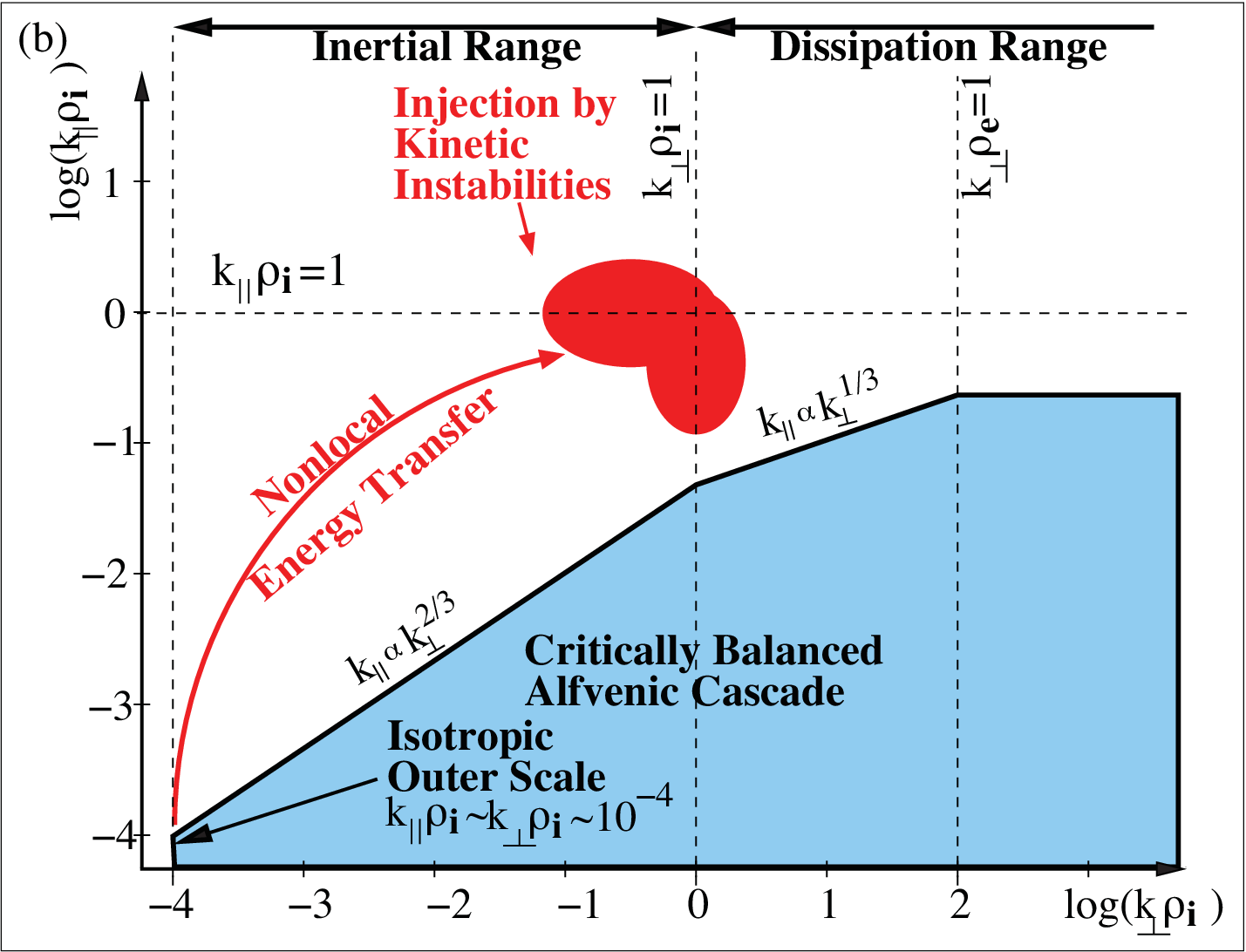}}
\caption{ (a) Schematic diagram of
  the turbulent magnetic-energy spectrum in the solar wind, depicting
  the local transfer of energy from large scales (small wavenumbers
  $k$) to small scales (large $k$) through a turbulent cascade (black
  arrows). Instabilities may alter this energy flow by nonlocally
  transporting energy directly to small, kinetic scales (red
  arrow). (b) In wavevector space, the local transfer of the
  Alfv\'enic turbulent cascade follows the anisotropic, critically
  balanced path, here plotted quantitatively for the case of the \emph{GS95} scaling theory. \label{fig:cascade}}
\end{figure}

The transport of energy mediated by plasma turbulence is depicted in
\figref{fig:cascade}, where (a) a plot of the turbulent magnetic
energy wavenumber spectrum shows that energy injected into the
turbulent cascade at the outer scale ($k_0 \rho_i =10^{-4}$) is
transferred nonlinearly without loss through the inertial range to the
transition into the dissipation range at ion scales ($k \rho_i \sim
1$).  At the ion scales and smaller ($k \rho_i \gtrsim 1$), poorly
understood mechanisms can remove energy from the turbulence,
transferring energy to the ions (blue arrows) or electrons (magenta
arrows).  In \figref{fig:cascade}(b), modern plasma turbulence theory
[\citep[][hereafter \emph{GS95}]{Goldreich:1995} or \citep[][hereafter
    \emph{B06}]{Boldyrev:2006}] predicts that the turbulent cascade
generates a scale-dependent anisotropy in wavevector space (blue
shaded region): isotropy at the driving scale ($k_{\perp 0} \rho_i=
k_{\parallel 0} \rho_i =10^{-4}$) develops into a significant
anisotropy $k_\perp \gg k_\parallel$ as the cascade transfers energy
to smaller scales\footnote{\emph{Parallel} and \emph{perpendicular}
here are defined relative to the local mean magnetic field
direction.}.  In addition, kinetic instabilities can generate unstable
turbulent fluctuations with wavevectors $k \rho_i \sim 1$ (red ovals
in both panels), which can mediate the nonlocal transfer of energy
from the large-scale motions directly to the ion kinetic scales. I
explicitly define ``turbulence'' here as the collection of physical
mechanisms involved in the conversion of the energy of large-scale
plasma flows and magnetic fields into plasma heat or the energy of
accelerated particles, including the nonlinear scale-to-scale energy
transfer of the turbulent cascade, the mechanisms damping the
turbulence at small scales, and any possible instabilities that may
generate turbulent fluctuations or govern nonlocal energy transfer.

\subsection{What Do We Want to Predict?}
A long-term goal of the heliophysics community is to characterize the
partitioning of turbulent energy in terms of observable plasma and
turbulence parameters, producing a predictive heating model that can
be used in numerical modeling of the global evolution of the
heliosphere, thereby advancing our capability for predictive modeling
of the coupled solar-terrestrial system.  Understanding how the
multiscale physics of plasma turbulence couples the large-scale
plasma conditions and evolution to the microphysical heating and
particle energization is essential to creating predictive heating
models.  Unfortunately, current numerical codes cannot simulate the
vast dynamic range of scales inherent to the multiscale nature of
heliospheric plasma turbulence while simultaneously capturing the
kinetic physics governing the conversion of turbulent energy to some
form of particle energy.  Developing turbulent heating models that
capture this coupling is one promising avenue for formulating
predictive models at a computationally feasible cost.  To develop such
models, it is essential to establish a thorough understanding of the
microphysical kinetic processes that govern the energization of
particles.

In a statistically steady-state turbulent cascade in a weakly
collisional astrophysical plasma, the energy injected into the
turbulence at the outer scale is transferred locally in scale down to
increasingly smaller scales until reaching the kinetic length scales
at which damping mechanisms can collisionlessly remove energy from the
turbulence and energize the different particle species, as depicted in
\figref{fig:cascade}(a).  This local cascade of energy is mediated by
nonlinear interactions among the turbulent fluctuations, such as
between counterpropagating \Alfven wave packets, often denoted
``\Alfven wave collisions''
\citep{Kraichnan:1965,Sridhar:1994,Goldreich:1995,Ng:1996,Galtier:2000,Howes:2012b,Howes:2013a,Nielson:2013a,Howes:2013b,Drake:2013,Howes:2016b,Verniero:2018a,Verniero:2018b,TenBarge:2021,Ripperda:2021}.
Nonlocal energy transfer mechanisms---such as kinetic instabilities
driven by the large-scale dynamics that may cause the plasma
conditions to deviate from local thermodynamic equilibrium---can also
directly generate turbulent fluctuations at the small kinetic length
scales, as illustrated in \figref{fig:cascade}(b).  Together, these
mechanisms lead to a turbulent energy density cascade rate $\varepsilon$
from the large scales down to the small scales of the turbulence.
This nonlinear transfer of the turbulent energy density is ultimately
damped at small scales by kinetic plasma physics mechanisms, converting the energy
density of the small-scale turbulent plasma flows and magnetic fields
into particle energy with a statistically steady-state total plasma
energy density heating rate $Q \sim \varepsilon$.

\begin{table}
  \begin{center}
  \begin{tabular}{|l|c|c|}
    \hline
    Quantity & Symbol & Bounded \\
    \hline
    Turbulent energy density cascade rate & $\varepsilon$& \\
    Plasma energy density heating rate& $Q$&  \\
    \hline
    Ion-to-Electron heating ratio &  $Q_i/Q_e$& $Q_i/Q$\\
    Anisotropic ion energy density heating ratio & $Q_{\perp,i}/Q_{\parallel,i}$ & $Q_{\perp,i}/Q_{i}$\\
    Anisotropic electron energy density heating ratio & $Q_{\perp,e}/Q_{\parallel,e}$ &$Q_{\perp,e}/Q_{e}$\\
    \hline
  \end{tabular}
  \end{center}
  \caption{The key input quantities (top section) and deliverables
    (bottom section) for predictive turbulent heating models. Bounded
    quantities are normalized such that their values vary only
    over the interval $[0, 1]$.
  \label{tab:quantities}}
  \end{table}

For a single-ion-species plasma, such as a fully ionized hydrogenic
plasma of protons and electrons, the primary deliverable of a
turbulent heating model is the determination of  the partitioning of the energy
by the kinetic turbulent damping mechanisms between the ions and
electrons, $Q_i/Q_e$.  Unlike in the case of strongly collisional
plasmas, in which the dissipated turbulent energy would be rapidly
equilibrated between ion and electron species by collisions, under the
weakly collisional plasma conditions typical of space and
astrophysical plasmas, the differential heating of the plasma species
can impact the thermodynamic evolution of the plasma on mesoscales and
macroscales.

Furthermore, energy transferred to a given plasma species may lead to
energization of the parallel degree of freedom or the perpendicular
degrees of freedom relative to the direction of the local magnetic
field.  Thus, secondary deliverables of a turbulent heating model are
measures of the anisotropic ion energy density heating rate
$Q_{\perp,i}/Q_{\parallel,i}$ and the anisotropic electron energy
density heating rate $Q_{\perp,e}/Q_{\parallel,e}$.  In a strongly
collisional plasma, the anisotropic heating for each species would be
rapidly isotropized by collisions, but weakly collisional plasmas
allow the parallel and perpendicular\footnote{Note that, under typical
astrophysical plasma conditions in which the turbulent fluctuation
frequencies are not much faster than the species cyclotron
frequencies, $\omega \lesssim \Omega_s$, the two perpendicular degrees
of freedom are rapidly equilibrated by the gyromotion of the particles
about the magnetic field, leading to gyrotropic plasma conditions with
$T_{\perp,1} \sim T_{\perp,2}$.}  degrees of freedom to be energized
independently \citep{Kawazura:2019}.

Different kinetic particle energization mechanisms may energize
different degrees of freedom---\emph{e.g.}, Landau damping energizes
particles parallel to the magnetic field, but cyclotron damping
energizes particles perpendicular to the magnetic field---so
determining $Q_{\perp,i}/Q_{\parallel,i}$ and
$Q_{\perp,e}/Q_{\parallel,e}$ first requires one to identify which
kinetic physical mechanisms govern the damping of the turbulent
fluctuations as a function of the plasma and turbulence parameters.
Therefore, identifying the contributing turbulent damping mechanisms
as a function of the plasma and turbulence parameters is an additional
deliverable of a predictive turbulent heating model.  Such information
about the dominant turbulent damping mechanisms, as a function of the
key dimensionless parameters describing the turbulent plasma, can be
concisely presented in a \emph{phase diagram} for the dissipation of
kinetic plasma turbulence, analogous to that developed for the
behavior of magnetic reconnection in astrophysical plasmas
\citep{Ji:2011}.  A first attempt at a phase diagram for the turbulent
damping mechanisms in weakly collisional space and astrophysical
plasmas is presented here in  \figref{fig:phase_k0bi}.  

The key input quantities for a turbulent heating model ($\varepsilon$,
$Q$) and the deliverables ($Q_i/Q_e$, $Q_{\perp,i}/Q_{\parallel,i}$,
$Q_{\perp,e}/Q_{\parallel,e}$) for a turbulent heating model are
summarized in \tabref{tab:quantities}.  Note that for ion cyclotron
damping, which energizes only the perpendicular degrees of freedom in
the ion velocity distribution function, the anisotropic ion heating
ratio $Q_{\perp,i}/Q_{\parallel,i}$ would be formally infinite.  To
avoid these inelegant infinities, we may choose to use alternative
\emph{bounded} versions of the heating ratios in the lower part of
\tabref{tab:quantities} that vary only over the range $[0, 1]$.  These
bounded quantities are related to the original quantities by $Q_i/Q=
(Q_i/Q_e)/[1+(Q_i/Q_e)]$ for the species partition and
$Q_{\perp,s}/Q_{s}=(Q_{\perp,s}/Q_{\parallel,s})/[1+(Q_{\perp,s}/Q_{\parallel,s})]$
for the anisotropic heating ratio for each species.

It is worthwhile noting that the determination of the partitioning of
the turbulent energy between ions and electrons $Q_i/Q_e$ can be
simplified by the properties of the turbulent cascade that are evident
in \figref{fig:cascade}(a).  The inertial range is defined as the
range of scales over which the turbulent energy density is transferred
to smaller scales without loss, leading to a constant energy density
cascade rate $\varepsilon$ throughout this range\footnote{This standard
definition of the inertial range from fluid turbulence theory neglects
the possibility of a nonlocal transfer of energy that may arise in
weakly collisional plasmas.}.  Therefore, the turbulent cascade rate
$\varepsilon$ at the onset of damping at ion kinetic scales at $k \rho_i
\sim 1$ is the same as that determined at the outer scale of the
turbulence at $k_0 \rho_i \ll 1$.  The collisionless wave-particle
interactions that serve to remove energy from the turbulent
fluctuations and transfer that energy to the \emph{ions} generally play a
significant role only within the range of length scales similar to the
ion kinetic length scales, $k \rho_i \sim 1$: (i) at larger scales $k
\rho_i \ll 1$, the collisionless damping rate with the ions $\gamma_i$
is typically negligible in comparison to the effective nonlinear frequency of the turbulent energy transfer
$\gamma_i \ll \omega_{nl}$; and (ii) at smaller scales $k \rho_i \gg
1$, the gyromotion of the ions averages out the effect of the
small-scale electromagnetic fluctuations, so the ions decouple from
the turbulent fields and do not exchange significant energy. Whatever
turbulent energy manages to cascade beyond the ion scales to smaller
scales at $k \rho_i \gg 1$ will ultimately be deposited with the
electrons.  Thus, the task of determining $Q_i/Q_e$ boils down to
determining what fraction of the turbulent energy is damped on the
ions as the cascade progresses through the ion kinetic scales at $k
\rho_i \sim 1$.


Ultimately, the kinetic mechanisms that serve to damp the turbulent
fluctuations at kinetic length scales depend on the nature of the
turbulent fluctuations at those scales. Therefore, a viable strategy
for developing a useful phase diagram for plasma turbulence is to use
turbulence scaling theories (\emph{e.g.}, \emph{GS95} or \emph{B06})
to predict the properties of the turbulent fluctuations upon reaching
the ion and electron kinetic scales from the plasma and turbulence
parameters at large scales.  The primary aim of this investigation is
to establish a theoretical framework upon which to base such
predictions of the dominant turbulent damping mechanisms as a function
of the plasma and turbulence parameters.

\subsection{What is the Approach for Developing a Predictive Turbulent Heating Model?}
The Buckingham Pi Theorem \citep{Buckingham:1914} uses dimensional
analysis \citep{Barenblatt:1996} to determine the minimum number of
dimensionless parameters upon which the physical behavior of a system
depends. This approach is the key principle underlying the development
of predictive models of turbulent plasma heating in terms of the
fundamental dimensionless parameters of kinetic plasma turbulence.
For sufficiently simple systems, in which there are not multiple
physical quantities of importance that have the same dimensions, one
can define a unique set of fundamental dimensionless parameters.  But
for more complicated systems, in which there may be more than one
relevant physical quantity with the same dimensions, such as multiple
characteristic length scales, the determination of the fundamental
dimensionless parameters is generally non-unique.  The case of kinetic
turbulence in a weakly collisional plasma certainly falls into this
more complicated category.  In that situation, it is essential to use
physical insight into the system's properties and behavior to define a
particular set of dimensionless parameters that is most useful in
characterizing the behavior of the system.

The goal of this paper is to define a particular set of fundamental
dimensionless parameters that can be used to characterize the
properties of kinetic plasma turbulence and its damping mechanisms.
Modern scaling theories for anisotropic plasma turbulence can then be
expressed in terms of these fundamental parameters to determine the
properties of the turbulent fluctuations at the kinetic length scales
at which kinetic physical mechanisms can act to remove energy from the
turbulent fluctuations and energize the plasma particles. How those
kinetic damping mechanisms depend on the plasma parameters and
properties of the turbulent fluctuations can then be used to predict
the dominant mechanisms for the damping of turbulence as a function of
the dimensionless parameters.  All of this information can then be
synthesized to generate improved turbulent heating prescriptions for
comparison to direct numerical simulations of turbulent dissipation
and for use in global modeling efforts for space and astrophysical
plasma systems, such as the heliosphere or supermassive black hole
accretion disks.

\section{The Fundamental Dimensionless Parameters of Kinetic Plasma Turbulence}

The first step in developing a predictive model for plasma turbulence
in weakly collisional space and astrophysical plasmas is to identify
the key dimensionless parameters upon which the turbulent energy
cascade and its dissipation depend. In this paper, the focus is to
characterize kinetic plasma turbulence in the non-relativistic limit
for sub-\Alfvenic turbulent motions, a case widely applicable to the
turbulent plasmas found ubiquitously in the heliosphere and throughout
the universe.  There are two categories of governing parameters:
\emph{Plasma Parameters} and \emph{Turbulence Parameters}.  Modern
scaling theories for anisotropic plasma turbulence can be employed to
reduce the number of parameters needed to characterize the turbulence,
as described in \secref{sec:reducing}.  The reduced set of
dimensionless parameters can then be used to develop useful phase
diagrams that identify the dominant physical mechanisms of turbulent
dissipation and highlight which of these key parameters has the strongest impact
on how particles are energized by the turbulence.

\begin{table}
  \begin{center}
  \begin{tabular}{|l|c|}
    \hline
    \multicolumn{2} {|c|} {Characteristic Plasma Quantities}\\
    \hline
    Parameter & Symbol and Definition \\
    \hline
    Parallel Species Thermal Velocity & $v_{t\parallel,s} = \sqrt{2 T_{\parallel, s}/m_s}$\\
    Perpendicular Species Thermal Velocity & $v_{t\perp,s} = \sqrt{2 T_{\perp, s}/m_s}$\\
    \Alfven Velocity & $v_A=B_0/\sqrt{4 \pi n_i m_i}$ \\
    Ion Acoustic Velocity&  $c_s = \sqrt{T_{\parallel,e}/m_i}$\\
    Species Cyclotron Frequency (Angular) & $\Omega_s = q_s B_0/m_s c$ \\
    Species Plasma Frequency (Angular) & $\omega_{ps}= \sqrt{4 \pi n_s q_s^2/m_s}$ \\
    Electron-electron Collision Frequency & $\nu_{ee} = 2^{3/2} \pi n_e e^4 \ln \Lambda/(m_e^{1/2} T_{\parallel,e}^{3/2})$ \\
    Ion  Larmor Radius & $\rho_i=v_{t\perp,i}/\Omega_i $\\
    Ion Sound Larmor Radius &  $\rho_s=c_s/\Omega_i $\\
    Electron  Larmor Radius & $\rho_e=v_{t\perp,e}/\Omega_e $\\
    Species Inertial Length & $d_s = c/\omega_{ps}$\\
    Species Debye Length &  $\lambda_{D,s} =\sqrt{T_s/(4 \pi n_s q_s^2)}$ \\
    Electron Mean Free Path &$\lambda_{mfp,e}= v_{t \parallel,e}/\nu_{ee}$\\
    Ion Mean Free Path &$\lambda_{mfp,i}= v_{t \parallel,i}/\nu_{ii}$\\
    Plasma Parameter & $ \Lambda = n_e \lambda_{D,e}^3$ \\
    Parallel Driving Scale & $L_{\parallel 0}$ \\
    Perpendicular Driving Scale & $L_{\perp 0}$ \\
    Parallel Driving Wavenumber & $k_{\parallel 0}=2 \pi/L_{\parallel 0}$ \\
    Perpendicular Driving Wavenumber & $k_{\perp 0}=2 \pi/L_{\perp 0}$ \\
    \hline
    \multicolumn{2} {|c|} {Useful Conversion Formulas}\\
    \hline
   \multicolumn{1} {|c|} { $d_i= c/\omega_{pi}=v_A/\Omega_i=\rho_i/\sqrt{\beta_{\parallel,i} A_i}$} &
     $\nu_{ei} \sim\nu_{ee} $     \\
   \multicolumn{1} {|c|}   {   $d_e = v_A \mu^{1/2} /\Omega_e = \rho_i/\sqrt{\beta_{\parallel, i} A_i \mu} = \rho_e \sqrt{\tau_\parallel \mu/(\beta_{\parallel,i}A_e)}$} &
    $\nu_{ii} \sim \mu^{-1/2} \tau_\parallel^{-3/2} \nu_{ee}  $\\
   \multicolumn{1} {|c|}  {    $\rho_s = \rho_i /\sqrt{2 \tau_\parallel A_i}$ }&
     $\nu_{ie} \sim \mu^{-1}\tau_\parallel^{-3/2}\nu_{ee} $ \\
    \multicolumn{1} {|c|} {  $\rho_e = \rho_i/\sqrt{\tau_\parallel A_e \mu/A_i}$   }
   & $\lambda_{mfp,i} = \lambda_{mfp,e} \tau_\parallel^2$\\
    \multicolumn{1} {|c|}   { $\beta_{\parallel,i} = v_{t\parallel,i}^2 /v_A^2$} & 
      $\beta_{\parallel,e} = \beta_{\parallel,i} /\tau_\parallel$    \\
    \hline
  \end{tabular}
  \end{center}
  \caption{Definitions of characteristic plasma quantities in cgs units and useful conversion relations, where $ \beta_{\parallel,i}$, $\tau_\parallel$, $A_i$, and  $\mu$ are defined in \tabref{tab:params}.
     \label{tab:defs} }
\end{table}


\subsection{Plasma Parameters}
The \emph{Plasma Parameters} play a critical role in determining the
nature of the turbulent fluctuations and their interaction with the
underlying particle velocity distributions. For simplicity of
notation, we assume here a fully ionized, hydrogenic plasma of protons
and electrons with bi-Maxwellian equilibrium velocity distributions
characterized by separate parallel and perpendicular
temperatures\footnote{Note that we absorb the Boltzmann constant into
the temperature throughout this paper, giving temperature in units of
energy.} for each species, $T_{\parallel,s}$ and $T_{\perp,s}$.  In
this idealized case, the equilibrium ion and electron number densities
are equal, $n_i=n_e$.

The first and most important plasma parameter is the parallel ion
plasma beta, $\beta_{\parallel,i} = 8 \pi n_i T_{\parallel i}/B^2$, which
characterizes the ratio of parallel thermal pressure to the magnetic
pressure in the plasma and dominantly controls the phase speeds of
different linear wave modes in a magnetized plasma.  Note that with
definitions for the parallel ion thermal velocity $v_{t\parallel,i}^2 = 2
T_{\parallel, i}/m_i$ and the \Alfven velocity $v_A^2 = B^2/(4 \pi n_i
m_i)$, the parallel ion plasma beta can be alternatively expressed as
$\beta_{\parallel,i} = v_{t\parallel,i}^2 /v_A^2$.  The second parameter is the
ion-to-electron parallel temperature ratio $\tau_\parallel=T_{\parallel
  ,i}/T_{\parallel, e}$. The third and forth parameters describe the ion
and electron temperature anisotropies,
$A_i=T_{\perp,i}/T_{\parallel,i}$ and
$A_e=T_{\perp,e}/T_{\parallel,e}$.  The fifth parameter characterizes
the collisionality of the plasma by comparing the mean free path scale
for electron-electron collisions relative to the parallel driving
scale $k_{\parallel 0} \lambda_{mfp,e}$.

Although for a hydrogenic plasma of protons and electrons the
ion-to-electron mass ratio $\mu=m_i/m_e$ has a fixed physical value of
1836, this ratio is included as an additional independent parameter in
the calculations below since a reduced mass ratio is often used in
numerical simulations of plasma turbulence for reasons of
computational efficiency.  Note also that, once $\mu$ and $\tau_\parallel$ are
specified, the single dimensionless parameter $k_{\parallel 0}
\lambda_{mfp,e}$ is sufficient to specify all possible charged particle
collision rates, since  $\nu_{ee}/(k_{\parallel 0} v_{t\parallel,e}) = 1/(k_{\parallel 0}
\lambda_{mfp,e})$ and  $\nu_{ei} \sim\nu_{ee} $,  $\nu_{ii} \sim \mu^{-1/2} \tau_\parallel^{-3/2} \nu_{ee}  $, and  $\nu_{ie} \sim \mu^{-1}\tau_\parallel^{-3/2}\nu_{ee} $.

Therefore, $(\beta_{\parallel,i},\tau_\parallel, A_i, A_e, k_{\parallel 0} \lambda_{mfp,e})$
are five key dimensionless plasma parameters that govern the
properties of plasma turbulence, along with the fixed mass ratio
$\mu$.  These Plasma Parameters are summarized in \tabref{tab:params}.

\begin{table}
  \begin{center}
  \begin{tabular}{|l|c|}
    \hline
    \multicolumn{2} {|l|} {Plasma Parameters}\\
    \hline
    Parameter & Symbol and Definition \\
    \hline
    Parallel Ion Plasma Beta & $\beta_{\parallel,i}=8 \pi n_i T_{\parallel, i}/B^2$ \\
    Ion-to-Electron Parallel Temperature Ratio &  $\tau_\parallel=T_{\parallel, i}/T_{\parallel ,e}$\\
    Ion Temperature Anisotropy & $A_i=T_{\perp,i}/T_{\parallel,i}$\\
    Electron Temperature Anisotropy & $A_e=T_{\perp,e}/T_{\parallel,e}$\\
    Plasma Collisionality & $k_{\parallel 0} \lambda_{mfp,e}$ \\
    Ion-to-Electron Mass Ratio & $\mu = m_i/m_e$\\
    \hline
    \multicolumn{2} {|l|} {Turbulence Parameters}\\
    \hline
    Parameter & Symbol and Definition \\
    \hline
    Perpendicular Driving Scale & $k_{\perp 0} \rho_i$\\
    Driving Scale Anisotropy & $k_{\parallel 0}/k_{\perp 0}$ \\
    Nonlinear Parameter & $\chi_0 = k_{\perp 0} \delta B_{\perp 0} \theta_0/(k_{\parallel 0} B_0)$\\
    Imbalance Parameter &  $Z_0^+/Z_0^-$ \\
    Driving Compressibility & $E_{comp}/E_{inc}$\\
    \hline
    \end{tabular}
  \end{center}
  \caption{For the general case of homogeneous turbulence in a fully
    ionized, hydrogenic plasma with bi-Maxwellian equilibrium velocity
    distributions, these are the fundamental dimensionless parameters
    of kinetic plasma turbulence, including Plasma Parameters and
    Turbulence Parameters. Note that the Boltzmann constant is absorbed to give
    temperature in units of energy. \label{tab:params} }
  \end{table}

Note that if the plasma contains multiple ion species, possibly each
with multiple charge states, the number of parameters increases
dramatically, requiring a minimum of five new dimensionless parameters
for each new species $s$: (i) mass ratio $m_s/m_i$, (ii) number density
ratio $n_s/n_e$, (iii) charge ratio $q_s/q_e$, (iv) parallel
temperature ratio $T_{\parallel, s}/T_{\parallel, i}$, and (v)
temperature anisotropy $T_{\perp,s}/T_{\parallel,s}$.  Partially
ionized plasmas will similarly require additional dimensionless
parameters to be characterized. Since the intention here is \emph{not}
to treat kinetic plasma turbulence in full generality, but rather to
describe how one can use the Pi Theorem and turbulence scaling
theories to minimize the number of relevant dimensionless parameters,
the remainder of this paper will focus on the case of a fully ionized,
hydrogenic plasma of protons and electrons with bi-Maxwellian
equilibrium velocity distributions. Refinements of turbulent heating
models to include minor ions, such as $\alpha$-particles that comprise
a small fraction of ions in space and astrophysical plasmas, can be
constructed following the same principles presented here.

\subsection{Turbulence Parameters}
The \emph{Turbulence Parameters} characterize the scale, amplitude,
and properties of the turbulent driving though five additional
parameters. For simplicity, the focus here is restricted to only the
case of turbulence in a spatially homogeneous system in which the
scale length of gradients in the equilibrium quantities are much
larger than the turbulent outer scale of the system.  Dimensionless
ratios of length scales are notationally simplified by characterizing
the length-scales of the turbulent fluctuations in terms of their
perpendicular and parallel wavenumbers. Below quantities at the outer
scale of the turbulence, commonly denoted the driving scale or energy
injection scale, are denoted by the subscript ``0''.

The driving of the turbulence can be characterized by three
dimensionless parameters. The first parameter is the perpendicular
wavenumber of the driving-scale fluctuations normalized by the ion
Larmor radius, $k_{\perp 0} \rho_i$.  The second parameter is the
wavevector anisotropy of the fluctuations at the driving scale,
$k_{\parallel 0} / k_{\perp 0}$. The third parameter is the
nonlinearity parameter that characterizes the amplitude of the
driving, $\chi_0 = (k_{\perp 0} \delta B_{\perp 0}
\theta_{0})/(k_{\parallel 0} B_0)$, where $\delta B_{\perp 0}$ is the
amplitude of the perpendicular magnetic field fluctuations, $B_0$ is
the equilibrium magnetic field magnitude, and $\theta_0$ is the angle
of dynamic alignment at the driving scale in radians
\citep{Boldyrev:2006}.  In the next section, we demonstrate how modern
scaling theories for anisotropic MHD turbulence can be used to combine
these three dimensionless parameters into a single parameter, the
\emph{isotropic driving wavenumber}, $k_0 \rho_i$. Note also that, in
principle, the dynamic alignment angle $\theta_0$ at driving scale is
an additional dimensionless parameter (measured in radians), but
dynamic alignment is typically understood to be a phenomenon that
develops self-consistently due to the turbulence as it cascades to
smaller scales, so we neglect this parameter at the outer scale,
generally taking $\theta_0 \sim 1$~rad.

The nature of the turbulent fluctuations at the driving scale also
plays an important role in determining the ultimate fate of the
turbulent energy.  The fourth turbulence parameter is the imbalance
parameter, measuring the ratio of the amplitudes in upward-to-downward
Elsasser fields, $Z_0^+/Z_0^-$, where the Elsasser fields at the outer
scale are defined by $\mathbf{Z}_0^\pm = \delta \V{U}_{\perp, 0} \pm
\delta \mathbf{B}_{\perp, 0}/\sqrt{4 \pi n_i m_i}$
\citep{Elsasser:1950}, and where $ \delta \V{U}_{\perp, 0}$ and
$\delta \mathbf{B}_{\perp, 0}$ are the perpendicular perturbed plasma
flow velocity and perpendicular perturbed magnetic field of the
turbulent fluctuations at the driving scale. Formally, $\delta
\V{U}_{\perp, 0}$ and $\delta \mathbf{B}_{\perp, 0}$ are vectors in
the plane perpendicular to the equilibrium magnetic field $\V{B}_0$,
but to calculate the ratio $Z_0^+/Z_0^-$, we simply take the amplitudes
of the Elsasser fields, $Z_0^\pm =|\mathbf{Z}_0^\pm|$.
Physically, \Alfven wave packets propagating up the mean
magnetic field have $Z^+=0$ and $Z^-\ne 0$, and those propagating down
the field have $Z^+\ne 0$ and $Z^-= 0$. The Elsasser fields
effectively characterize the wave energy flux in each direction along
the magnetic field\footnote{Note that the imbalance of
counterpropagating wave energy fluxes in plasma turbulence has
alternatively been characterized in the literature using the cross
helicity
\citep{Dobrowolny:1980b,Marsch:1990,Matthaeus:1982a,Perez:2009,Perez:2010b}
or the normalized Poynting flux \citep{Hnat:2003,DiMare:2024}.}  of
the typically dominant \Alfvenic fluctuations in heliospheric plasma
turbulence \citep{Tu:1995,Schekochihin:2009,Bruno:2013}.  Since the
nonlinear interactions that mediate the turbulent cascade in \Alfvenic
turbulence at MHD scales arise only between counterpropagating and perpendicularly
polarized \Alfven waves
\citep{Kraichnan:1965,Sridhar:1994,Goldreich:1995,Ng:1996,Galtier:2000,Howes:2013a,Howes:2015b},
the imbalance plays an important role in characterizing the
turbulence.  The importance of $Z_0^+/Z_0^-$ under sufficiently
imbalanced conditions, such as occurs in the inner heliosphere where
wave energy fluxes are dominantly anti-sunward, has been demonstrated
clearly with the recent discovery of a new phenomenon known as the
\emph{helicity barrier} in plasma turbulence
\citep{Meyrand:2021,Squire:2022,Squire:2023}.

A fifth turbulence parameter is the ratio of the energy in
compressible fluctuations to that in incompressible fluctuations,
$E_{comp}/E_{inc}$.  Since a magnetized plasma supports two distinct
compressible wave modes, the fast and slow magnetosonic waves
(differing by whether the thermal and magnetic pressure perturbations
associated with the wave act in conjunction or in opposition), one
could further characterize the compressible component of the turbulent
fluctuations by the additional ratio of the energy in the fast
magnetosonic fluctuations to the total energy of compressible
fluctuations $E_F/E_{comp}$, where the total compressible energy is
the sum of the contributions from fast-mode and slow-mode fluctuations,
$E_{comp}=E_F+E_s$. The ratio $E_F/E_{comp}$ is not included here as
one of the fundamental parameters based on a study of the inertial
range in solar wind turbulence at 1~AU that showed negligible energy
in fast-mode mode fluctuations, or $E_F/E_{comp} \ll 1$
\citep{Howes:2012a}.  In significant energy exists in fast-mode
fluctuations in other turbulent astrophysical plasmas, $E_F/E_{comp}$
may need to be included.

Therefore, $(k_{\perp 0} \rho_i,k_{\parallel
  0}/k_{\perp0},\chi_0,Z_0^+/Z_0^-,E_{comp}/E_{inc})$ are five key
dimensionless turbulence parameters that govern the properties of
plasma turbulence.  These Turbulence Parameters are summarized in
\tabref{tab:params}.

If the restriction to spatially homogeneous turbulent plasmas is relaxed,
the large-scale context of the turbulent plasma can also impact the
nonlinear dynamics of the turbulence and the evolution of the
turbulent fluctuations.  In the solar wind, the spherical expansion of
the solar wind governs the evolution of the plasma equilibrium
temperatures with heliocentric radius, possibly triggering kinetic
temperature anisotropy instabilities that can mediate nonlocal
transfer of energy to small scales, as depicted in
\figref{fig:cascade}(b). In rotating plasmas, such as found in the solar
convection zone, the turbulence may be driven with helical motions
that can initiate the growth of magnetic field energy through a dynamo
effect
\citep{Parker:1955,Glatzmaier:1984,Glatzmaier:1985,Parker:1993,Ossendrijver:2003}.
Differential rotation in astrophysical accretion disks, which can
trigger plasma turbulence through the magnetorotational instability
\citep{Balbus:1991,Hawley:1991,Balbus:1998} has been recently shown to
generate turbulence with a dominant fraction of its energy in
compressible fluctuations \citep{Kawazura:2022}.  Finally, kinetic
instabilities arising in the foot and ramp regions of collisionless shocks
\citep{Brown:2023} can generate turbulent fluctuations that are swept
into the downstream region.  In all of these astrophysically relevant
cases, additional dimensionless turbulence parameters will be required
to characterize properly the turbulent fluctuations and their
evolution.

\subsection{What Happened to the Reynolds Number?}
Readers familiar with studies of MHD turbulence may find what appear
to be two glaring omissions in the proposed list of Plasma and
Turbulence parameters in \tabref{tab:params} since it does not include
the Reynolds number, $\mbox{Re} = L U_0/\nu$, or the magnetic Reynolds
number, $\mbox{Re}_M = L U_0/\eta$.  The Reynolds number is a
dimensionless quantity that characterizes the ratio of the amplitude
of the convective term $\V{U} \cdot \nabla \V{U}$ to that of the
viscous diffusion term $\nu \nabla^2 \V{U}$ in the MHD momentum
equation; the magnetic Reynolds number characterizes the ratio of the
amplitude of the inductive term $\nabla \times (\V{U} \times \V{B})$
to that of the resistive diffusion term $\eta \nabla^2 \V{B}$ in the
MHD induction equation.

The kinematic viscosity $\nu$ and resistivity $\eta$ are transport
coefficients that are rigourously defined by an extension of the
Chapman-Enskog procedure for magnetohydrodynamic systems
\citep{Spitzer:1962,Grad:1963,Braginskii:1965} in the limit of small
but finite mean free path relative to the gradient scale,
$\lambda_{mfp}/L \ll 1$.  As illustrated by \figref{fig:k0rhoi_betai},
however, for many space and astrophysical plasma systems of interest,
the scales at which energy is removed from the turbulence (beginning
at the small-scale end of the inertial range) are weakly collisional,
with $\lambda_{mfp}/L \gtrsim 1$ or $\lambda_{mfp}/L \gg 1$. In these
complementary limits relevant to space and astrophysical plasmas, the
procedure to calculate the viscosity and resistivity ceases to be
valid.  In fact, the standard Laplacian viscosity and resistivity
terms are likely to be poor approximations to the physical mechanisms
that are believed to remove energy from weakly collisional plasma
turbulence (see \secref{sec:proposed} for a list of proposed damping
mechanisms).  Since it is not possible to define clearly the viscosity
or resistivity in a weakly collisional system, it is also not possible
to define the Reynolds number or magnetic Reynolds number in weakly
collisional, kinetic plasma turbulence.  Thus, we are forced to choose
alternative dimensionless numbers that characterize the plasma and
turbulence in weakly collisional systems, so one of the primary aims
of this paper is to propose the specific set of fundamental
dimensionless Plasma and Turbulence parameters that are suitable for
characterizing kinetic plasma turbulence, presented here in
\tabref{tab:params}.

\section{Scaling Theories for Anisotropic Plasma Turbulence}
\label{sec:mhdtheory}
The key to predicting the evolution of the turbulent fluctuations as
energy is cascaded to ever smaller scales, and therefore also to
developing a useful phase diagram for plasma turbulence, is to use
turbulence scaling theories to predict the properties of the
turbulence upon reaching the ion kinetic scales from the
characteristics of the turbulence at large scales.  In particular, we
will include the two prominent theories for the anisotropic cascade of
\Alfvenic turbulence in MHD plasmas, the Goldreich-Sridhar 1995
(\emph{GS95}) formulation for weak and strong MHD turbulence
\citep{Sridhar:1994,Goldreich:1995}, and the modification of this
theory by Boldyrev (\emph{B06}) that accounts for a weakening of the
nonlinearity by dynamic alignment \citep{Boldyrev:2006}.  Note that
since these are MHD theories of plasma turbulence---where strong
collisionality is assumed in the standard MHD approximation
\citep{Kulsrud:1983,Gurnett:2017}---the equilibrium species
temperatures are assumed to be anisotropic and equal, corresponding to
$\tau_\parallel=A_i=A_e=1$.  For the case of weakly collisional plasma
turbulence that is relevant to most space and astrophysical plasma
environments, these theories may need to be modified appropriately,
but such development of refined turbulence theories is beyond the
scope of the treatment here.

The particular scalings of the parallel wavenumber, intermediate
wavenumber, alignment angle, and magnetic field perturbation as a
function of the perpendicular wavenumber are given by
\begin{equation}
  k_\parallel \propto k_\perp^{2/(3+\alpha)}
  \label{eq:kpar}
\end{equation}
\begin{equation}
  k_i \propto k_\perp^{3/(3+\alpha)}
  \label{eq:ki}
\end{equation}
\begin{equation}
  \theta \propto  k_\perp^{-\alpha/(3+\alpha)}
  \label{eq:theta}
\end{equation}
\begin{equation}
\delta B_\perp \propto  k_\perp^{-1/(3+\alpha)}
  \label{eq:bperp}
\end{equation}
where $\alpha =0$ corresponds to the \emph{GS95} theory and $\alpha
=1$ corresponds to the \emph{B06} theory.  Here the turbulent
fluctuations as a function of scale are characterized by the three
spatial components of the wave vector relative to the local magnetic
field direction: (i) the parallel wavenumber $k_\parallel$; and (ii)
the intermediate wavenumber $k_i$ and (iii) the perpendicular
wavenumber $k_\perp$, which together describe the anisotropy of the
fluctuations in the plane perpendicular to the local magnetic field,
$k_i \ll k_\perp$, arising from the dynamic alignment of the
fluctuations \citep{Boldyrev:2006,Mason:2006,Boldyrev:2009,Perez:2012}
and predicting the development of current sheets at small scales.
These scaling theories predict the scaling of $k_\parallel$, $k_i$,
the angle of dynamic alignment $\theta$ (in radians), and the amplitude of the
perpendicular magnetic field fluctuations $\delta B_\perp$ as a
function of the perpendicular wavenumber $k_\perp$.

These theories yield specific predictions for the perpendicular
magnetic energy spectrum of turbulence in the MHD regime, $k_\perp
\rho_i \gg 1$, and for the scale-dependent anisotropy of the
turbulence, a qualitative feature that is well-established in MHD
turbulence
\citep{Cho:2000,Maron:2001,Cho:2002,Cho:2003,Mason:2006,Boldyrev:2009,Narita:2010,Sahraoui:2010b,Perez:2012,Roberts:2015b}. The
perpendicular magnetic energy spectrum for the two theories is given
by
\begin{equation}
  E_{B_\perp}(k_\perp) \propto \frac{(\delta B_\perp)^2}{k_\perp} \propto k_\perp^{-(5+\alpha)/(3+\alpha)} = \left\{\begin{array}{cc} k_\perp^{-5/3} & \alpha=0 \\ k_\perp^{-3/2} & \alpha=1 \\
     \end{array} \right. .
\end{equation}
The parallel-to-perpendicular wavevector anisotropy is given by
\begin{equation}
 \frac{k_\parallel}{k_\perp} \propto  k_\perp^{-(1+\alpha)/(3+\alpha)} = \left\{\begin{array}{cc} k_\perp^{-1/3} & \alpha=0 \\ k_\perp^{-1/2} & \alpha=1 \\
     \end{array} \right. ,
\end{equation}
demonstrating that both theories predict that the turbulent
fluctuations develop a scale-dependent anisotropy, whereby the
fluctuations become progressively more elongated along the magnetic
field as they cascade to smaller scale, eventually yielding
$k_\parallel/k_\perp \ll 1 $ at sufficiently small scales.  This
$k_\parallel/k_\perp$ anisotropy follows from the imposition of the condition of 
\emph{critical balance} between the linear and nonlinear timescales of
the turbulence \citep{Goldreich:1995,Mallet:2015}. The
intermediate-to-perpendicular wavevector anisotropy is given by
\begin{equation}
 \frac{k_i}{k_\perp} \propto  k_\perp^{-\alpha/(3+\alpha)} = \left\{\begin{array}{cc} k_\perp^{0} & \alpha=0 \\ k_\perp^{-1/4} & \alpha=1 \\
     \end{array} \right. .
\end{equation}
Here, the \emph{GS95} theory predicts no development of anisotropy in the perpendicular plane, at odds with the findings of current sheets at small perpendicular scales in numerical simulations \citep{Matthaeus:1980,Meneguzzi:1981,Uritsky:2010,Zhdankin:2012,Zhdankin:2013} and spacecraft observations \citep{Borovsky:2008,Osman:2011}. The \emph{B06} theory predicts the development of anisotropy in the perpendicular plane, corresponding to the development of current sheets with  $k_i/k_\perp \ll 1$ at sufficiently small scales, potentially triggering tearing instabilities \citep{Furth:1963,Zocco:2011} that enable magnetic reconnection to disrupt the turbulent cascade \citep{Boldyrev:2017,Loureiro:2017a,Loureiro:2017b,Mallet:2017b,Mallet:2017c,Walker:2018}.

It is worthwhile emphasizing that the idea of critical balance in the
turbulent cascade does \emph{not} imply that all of the turbulent
energy is concentrated at the parallel wavenumber $k_\parallel^{cb}$
determined by critical balance. Rather, $k_\parallel^{cb}$ is the
upper limit of parallel wavenumber, and thus generally also implies an
upper limit on the linear wave frequency (since $\omega \propto
k_\parallel$ for $\omega\ll \Omega_i$). Rather, the turbulent
fluctuation power extends over the range $0 \le k_\parallel \le
k_\parallel^{cb}$ \citep{Goldreich:1995,Maron:2001}.  Accounting for
how the turbulent energy is distributed over this $k_\parallel$ range would lead
to order unity or less quantitative changes in the heating
rates. Since the goal of predictive turbulent heating models is
generally to calculate the lowest-order, order-of-magnitude quantities
for species heating rates, any effects due to the distribution of
turbulent energy over $0 \le k_\parallel \le k_\parallel^{cb}$ are
likely to yield only higher-order corrections to these predictions;
thus, the effect of this range of turbulent energy in $k_\parallel$
is neglected in the first generation of turbulent heating models
addressed here.

In closing, note that these theories for MHD turbulence are largely
focused on the transport of energy in \Alfvenic fluctuations, so if
significant energy is transferred nonlinearly into other
(non-Alfv\'enic) wave modes, this may significantly alter the energy
transport from the predictions shown here, with the potential for
qualitatively different predictions for the damping mechanisms of
these non-\Alfvenic turbulent fluctuations.

\subsection{Exploiting Turbulent Scaling Theories to Reduce the Number of Fundamental Dimensionless Parameters}
\label{sec:reducing}
Given the large number of fundamental dimensionless parameters that
characterize weakly collisional plasma turbulence, shown in
\tabref{tab:params}, it is imperative to reduce the number of
important parameters to make progress in developing predictive models
of turbulent heating.  Adopting idealized approximations, such as
maintaining isotropic equilibrium velocity distributions for the
plasma, enables one to eliminate some of the parameters in
\tabref{tab:params}.  Another approach is to exploit the MHD turbulent
scaling theories described in \secref{sec:mhdtheory} to combine
several parameters into a new single parameter, an approach presented
here.


\begin{figure}
  \resizebox{4.25in}{!}{\includegraphics*[20pt,260pt][564pt,712pt]{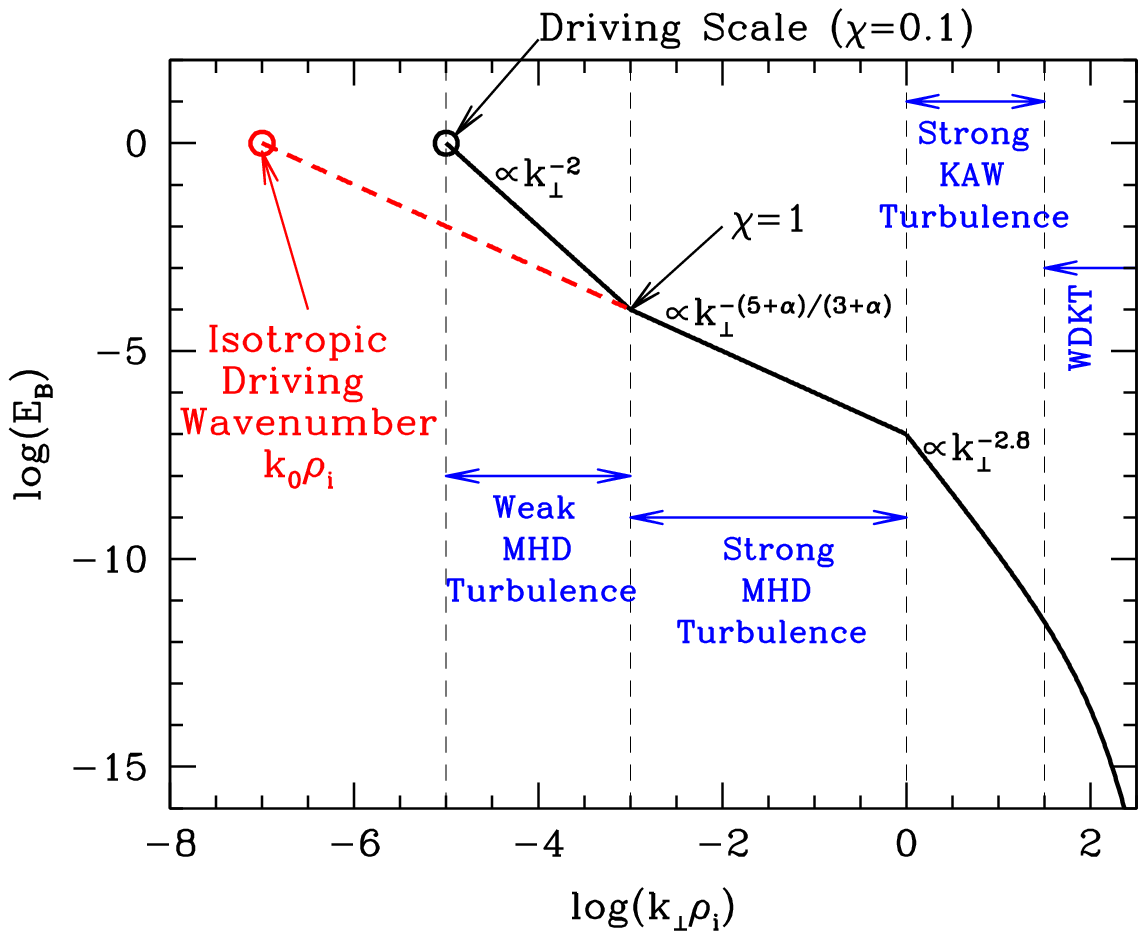}}\\
\resizebox{4.25in}{!}{\includegraphics*[20pt,260pt][564pt,690pt]{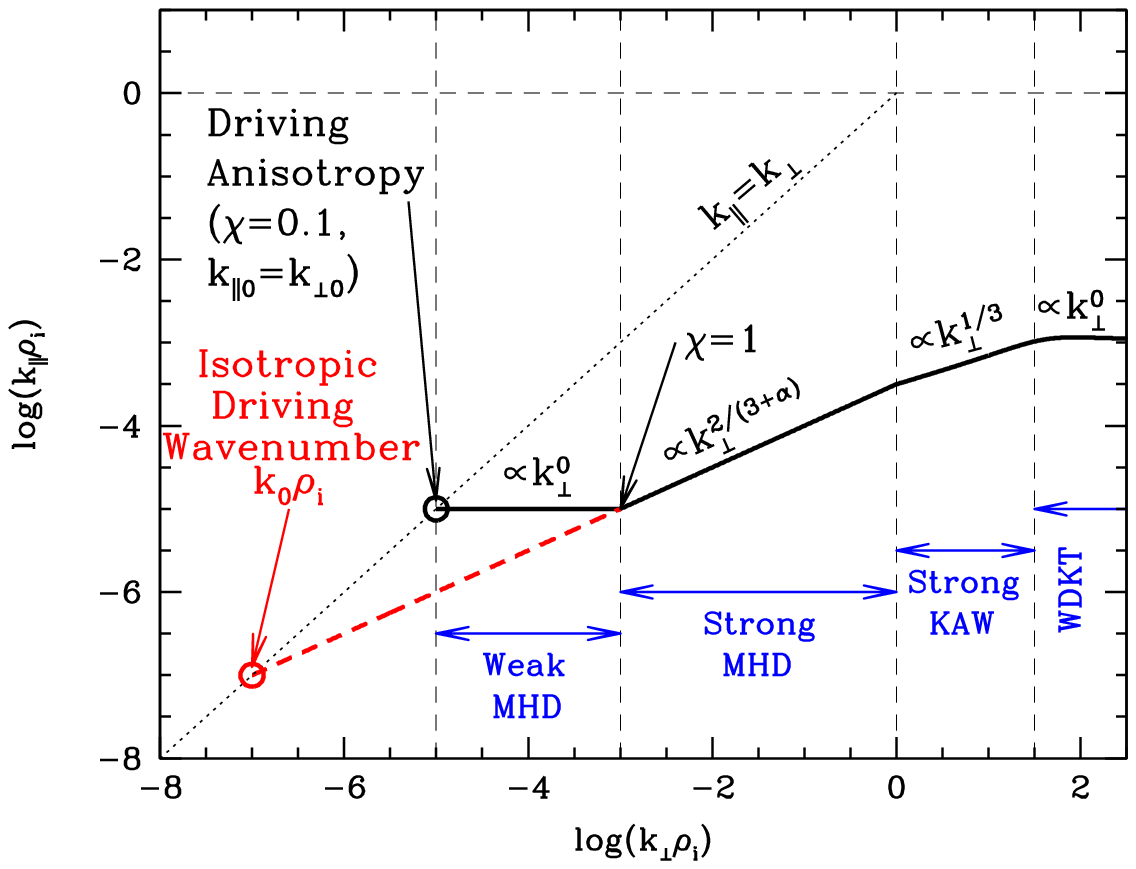}}
\caption{Diagram illustrating how to determine the isotropic driving
  wavenumber $k_0\rho_i$ in terms of the perpendicular driving
  wavenumber $k_{\perp 0} \rho_i$, the driving anisotropy
  $k_{\parallel 0}/k_{\perp 0}$, and the nonlinearity parameter of the
  driving, $\chi_0=k_{\perp 0} \delta B_\perp \theta_0/(k_{\parallel 0}
  B_0)$. 
\label{fig:k0def}}
\end{figure}

In general, the driving of the plasma turbulence depends on at least
three independent parameters: (i) the perpendicular wavenumber at which
the turbulence is driven $k_{\perp 0} \rho_i$, (ii) the anisotropy of the
driven fluctuations with respect to the local magnetic field direction
$k_{\parallel 0}/k_{\perp 0}$, and (iii) the nonlinearity parameter
which characterizes the strength of the driving, $\chi_0 \equiv k_{\perp 0} \delta
B_{\perp 0} \theta_0/(k_{\parallel 0} B_0)$.  Recall that quantities evaluated at the driving scale are given the subscript ``0''. Fortunately, for turbulence
driven at large scales (relative to the ion kinetic scales), as is typical in 
most space or astrophysical plasma environments, MHD scaling theory can be used to 
predict the properties of the turbulence 
upon reaching the kinetic scales, where kinetic plasma processes act to 
dissipate the turbulence.  In fact, as detailed here, 
it is possible to combine these three dimensionless parameters into 
a new, single dimensionless parameter: the \emph{isotropic driving wavenumber}, 
$k_0 \rho_i$.

MHD turbulent scaling theories can be used, for a given set of driving
parameters ($k_{\perp 0} \rho_i$, $k_{\parallel 0}/k_{\perp 0}$,
$\chi_0$), to predict the properties of the turbulent fluctuations
(\emph{e.g.}, anisotropy, amplitude) upon reaching the ion kinetic
scales, here taken to be the perpendicular wavenumber $k_\perp \rho_i
= 1$.  The isotropic driving wavenumber $k_0 \rho_i$ is the unique
wavenumber that produces the same conditions at $k_\perp \rho_i = 1$
as if the turbulence were driven isotropically at that wavenumber with
$k_{\perp 0} = k_{\parallel 0} \equiv k_0$ and in a condition of
critical balance $\chi_0=1$.

The key procedure for determining 
the isotropic driving wavenumber $k_0 \rho_i$ is illustrated in \figref{fig:k0def}.
Panel (a) shows the perpendicular wavenumber magnetic energy spectrum $E_{B_\perp}(k_\perp)$  
driven weakly with $\chi_0=0.1$ at perpendicular wavenumber $k_{\perp 0} \rho_i=10^{-5}$, 
where a weak turbulent cascade leads to a steep weak MHD turbulence spectrum 
with  $E_{B_\perp} \propto  k_\perp^{-2}$ \citep{Sridhar:1994,Galtier:2000,Boldyrev:2009b}, followed by a flatter strong 
MHD turbulence spectrum with  $E_{B_\perp} \propto  k_\perp^{-(5+\alpha)/(3 +\alpha)}$ \citep{Goldreich:1995,Boldyrev:2006}.
The cascade reaches the perpendicular ion kinetic scale  $k_\perp \rho_i = 1$, and 
at higher wavenumbers the cascade continues as strong kinetic \Alfven wave 
(KAW) turbulence with a 
significantly steeper magnetic energy spectrum with  
$E_{B_\perp} \propto  k_\perp^{-2.8}$  \citep{Sahraoui:2009,Kiyani:2009,Alexandrova:2009,Chen:2010b,Howes:2011a,Sahraoui:2013b}.  When the collisionless damping of the 
KAW turbulent cascade begins to diminish the rate of nonlinear energy cascade, 
the spectrum begins to fall off exponentially in the weakly dissipating KAW 
turbulence (WDKT) regime \citep{Howes:2011b}, ultimately terminating the turbulent 
cascade at scales $k_\perp \rho_e \gtrsim 1$.

In  \figref{fig:k0def}(b) is presented the characteristic path of the anisotropic cascade of energy through  $(k_\perp,k_\parallel)$ wavevector space for the different regimes of the turbulent cascade. There is no parallel cascade in the weak MHD turbulence regime  \citep{Sridhar:1994,Galtier:2000}, followed by a scale-dependent anisotropy given by $k_\parallel \propto k_\perp^{2/(3+\alpha)}$ in the strong MHD turbulence regime \citep{Goldreich:1995,Boldyrev:2006}.  The turbulence transitions to the strong KAW regime at $k_\perp \rho_i \sim 1$,  with  the anisotropy becoming stronger at  $k_\perp \rho_i > 1$, scaling as  $k_\parallel \propto k_\perp^{1/3}$ \citep{Cho:2004,Howes:2008b,Howes:2011b}. As the turbulence weakens due to collisionless damping of turbulent fluctuations in the WDKT regime, the parallel cascade is predicted to cease once again  \citep{Howes:2011b}.

The diagrams in \figref{fig:k0def} provide guidance to understand how MHD turbulence scaling theories can be used to combine  ($k_{\perp 0} \rho_i$, $k_{\parallel 0}/k_{\perp 0}$, $\chi_0$) into a single new parameter,  the isotropic driving wavenumber, 
$k_0 \rho_i$. If the turbulence is driven at critical balance ($\chi_0=1$) and 
isotropically ($k_{\parallel 0}/k_{\perp 0}=1$), then the isotropic driving 
wavenumber is simply the same as the perpendicular driving wavenumber,  
$k_0 \rho_i=k_{\perp 0} \rho_i$.  Therefore, it is necessary to 
use MHD turbulence scaling theory to determine $k_0 \rho_i$ only in the case that
the turbulence is driven weakly $\chi_0<1$ or anisotropically $k_{\parallel 0}/k_{\perp 0}\ne 1$.

For weakly driven turbulence with $\chi_0<1$, MHD turbulence theory predicts that 
the parallel cascade is inhibited and that the perpendicular wavenumber magnetic energy 
 spectrum of the turbulence scales as 
$E_{B_\perp} \propto (\delta B_\perp)^2 /k_\perp \propto k_\perp^{-2}$.
Therefore, in terms of the driving amplitude $\delta B_{\perp 0}$ 
and perpendicular driving wavenumber $k_{\perp 0}$, the amplitude of the 
turbulence as a function of $k_\perp$ is given by 
$\delta B_\perp(k_\perp)=\delta B_{\perp 0} (k_{\perp 0}/k_\perp)^{1/2} $.  
Given the lack of parallel cascade in weak MHD turbulence, 
we fix $k_\parallel = k_{\parallel 0}$, and obtain the following scaling\footnote{Here we have assumed that the dynamic alignment angle $\theta_0$ also does not evolve as a function of $k_\perp$ in weak MHD turbulence, as there is no prediction in the literature for the scaling  of dynamic alignment in weak turbulence.}  for the nonlinearity parameter
as a function of $k_\perp$,
\begin{equation}
    \chi (k_\perp) = \chi_0 \left(\frac{k_{\perp} \rho_i}{k_{\perp 0} \rho_i}\right)^{1/2}.
\end{equation}
MHD turbulence theory predicts that the nonlinearity parameter will increase through 
the weak turbulent cascade until it reaches a value of unity, $\chi \sim 1$, at which point the 
turbulence has reached a state of critical balance in the strong turbulence regime.
From that point, theory predicts that an anisotropic cascade will ensue (transferring 
energy to smaller scales more rapidly in the perpendicular than the parallel direction), with the parallel and perpendicular cascades balanced to maintain a nonlinearity parameter of unity, $\chi \sim 1$.  
Therefore, the key point in the cascade is the perpendicular wavenumber at which the turbulence first reaches
$\chi =1$, given by 
\begin{equation}
   \left. k_{\perp} \rho_i\right|_{\chi=1}= \frac{1}{\chi_0^2} k_{\perp 0} \rho_i
   \label{eq:kperpchi1}
\end{equation}
An example of this weakly turbulent, strictly perpendicular cascade in $(k_\perp,k_\parallel)$ space is depicted in 
\figref{fig:k0def}(b), where turbulence driven weakly with $\chi_0 =0.1$, cascades to 
$k_{\perp} \rho_i= 100 k_{\perp 0} \rho_i$, at which point the turbulence becomes strong, with $\chi=1$.

To obtain the isotropic driving wavenumber $k_0 \rho_i$ from this 
point in wavevector space $(\left. k_{\perp} \rho_i\right|_{\chi=1}, k_{\parallel 0})$, we need to extend the 
cascade backward along the critical balance line (red dashed lines) until it 
intersects with isotropy $k_\parallel=k_\perp$ (dotted line).  From the MHD scaling theories 
in \secref{sec:mhdtheory},  for a critically balanced, strong turbulent cascade that is driven strongly and isotropically at $k_0\rho_i$, the anisotropy as a function of $k_\perp$ scales
as 
\begin{equation}
\frac{k_\parallel}{k_\perp} = \left(\frac{k_0 \rho_i}{k_\perp \rho_i}\right)^{\frac{1+\alpha}{3 +\alpha}}.
\label{eq:anisocritbal}
\end{equation} 
The anisotropy at the point $(\left. k_{\perp} \rho_i\right|_{\chi=1},
k_{\parallel 0})$ is given by $k_\parallel/k_\perp = (k_{\parallel
  0}/k_{\perp 0}) \chi_0^2$.  Substituting this expression for
$k_\parallel/k_\perp$ on the left-hand side of \eqref{eq:anisocritbal}
and substituting \eqref{eq:kperpchi1} into the denominator on the
right-hand side of \eqref{eq:anisocritbal}, we can solve for the
isotropic driving wavenumber $k_0 \rho_i$ , obtaining the result
\begin{equation}
    k_0\rho_i = k_{\perp 0} \rho_i 
    \left(\frac{k_{\parallel 0}}{k_{\perp 0}}\right)^{\frac{3+\alpha}{1+\alpha}}
    \chi_0^{\frac{4}{1+\alpha}}
    \label{eq:isodrive}
\end{equation}
In \figref{fig:k0def}(b), the isotropic driving wavenumber $k_0
\rho_i$ (red circle) is the shown where the critical balance line
(solid black and red dashed lines) intersects the isotropy line,
$k_\parallel=k_\perp$ (dotted line).  Note that this reduction of the
three driving parameters to $k_0\rho_i$ is valid only in the limit
that the turbulent cascade becomes strong before it reaches the ion
kinetic scales at $k_\perp\rho_i \sim 1$.


The relation \eqref{eq:isodrive} provides the means to combine the
three turbulence driving parameters ($k_{\perp 0} \rho_i$,
$k_{\parallel 0}/k_{\perp 0}$, $\chi_0$) into the single new effective
parameter, the isotropic driving wavenumber $k_0 \rho_i$.  The
properties of the turbulent fluctuations at perpendicular scales
smaller than the point where the turbulent cascade becomes strong, or
$k_\perp \rho_i> \left. k_{\perp} \rho_i\right|_{\chi=1}$, will be the
same for the original case with driving parameterized by ($k_{\perp 0}
\rho_i$, $k_{\parallel 0}/k_{\perp 0}$, $\chi_0$) as for turbulence
driven strongly and isotropically at $k_0 \rho_i$.  In the development
of a turbulent heating model, it is the properties of the turbulent
fluctuations at the small, kinetic length scales that will determine
the kinetic damping mechanism and the resulting particle energization,
so reducing three turbulence driving parameters to a single parameter
substantially reduces the dimensionality of the parameter space to be
modeled.

It is worthwhile noting that the evolution of the dynamic alignment
angle with perpendicular wavenumber $\theta(k_\perp)$ is not well
modeled by $k_0 \rho_i$.  The development of current sheets at very
small scales due to dynamic alignment, and the possibility for the
interruption of the turbulent cascade by magnetic reconnection
\citep{Boldyrev:2017,Loureiro:2017a,Loureiro:2017b,Mallet:2017b,Mallet:2017c,Walker:2018},
will depend on this scaling of $\theta(k_\perp)$.  But, from a
practical point of view, this issue in modeling $\theta(k_\perp)$
while reducing ($k_{\perp 0} \rho_i$, $k_{\parallel 0}/k_{\perp 0}$,
$\chi_0$) to $k_0 \rho_i$ is likely to be a problem only in rare
cases, for the following reasons.  First, the scaling of dynamic
alignment angle $\theta \propto k_\perp^{-1/4}$ in \emph{B06} is
rather weak, scaling as the $-1/4$ power, so only when the MHD
inertial range spans many orders of magnitude in the perpendicular
length scale will the development of current sheets likely have a
significant impact on the turbulent dynamics. Second, dynamic
alignment appears to arise only in strong MHD turbulence, so it seems
reasonable to take $\theta_0 \sim 1$~radian (roughly equivalent to
isotropy in the perpendicular plane) at the perpendicular scale $
\left. k_{\perp} \rho_i\right|_{\chi=1}$ where the turbulence becomes
strong.  One may then estimate the evolution of that dynamic alignment
angle $\theta(k_\perp)$ from that point for the cases in which it
impacts the evolution. Finally, although the development of current
sheets in plasma turbulence is well established
\citep{Matthaeus:1980,Meneguzzi:1981,Borovsky:2008,Uritsky:2010,Osman:2011,Zhdankin:2012,Zhdankin:2013},
it is not clear that the scaling theory in \emph{B06} indeed properly
describes this development; an alternative explanation for the
development of current sheets arising self-consistently from \Alfven
wave collisions exists \citep{Howes:2015b,Howes:2016b,Verniero:2018a},
although a specific scaling for this alternative mechanism has not yet
been proposed.

\subsection{Efficiently Modeling the Small-Scale End of the Turbulent Inertial Range}
\label{sec:end_inertial}
The multiscale physics of astrophysical plasma turbulence presents a
particular challenge for numerical modeling that covers the entire
dynamic range of the turbulent cascade while simultaneously capturing
the kinetic mechanisms that govern the damping of the turbulence under
weakly collisional plasma conditions, in particular since the physics of
plasma turbulence is inherently three-dimensional
\citep{Howes:2015a}. For the development of turbulent heating models,
it is critical to investigate the physical mechanisms that remove
energy from the turbulence and energize the plasma particles.
Fortunately, since the inertial range of turbulence is defined as the
range of scales over which the turbulent energy cascades to smaller
scales but the dissipation is negligible, as depicted in
\figref{fig:cascade}(a), it is not strictly necessary to model the the
entire inertial range in order to capture the physics of the turbulent
dissipation.

\begin{figure}
  \resizebox{4.25in}{!}{\includegraphics*[20pt,260pt][564pt,712pt]{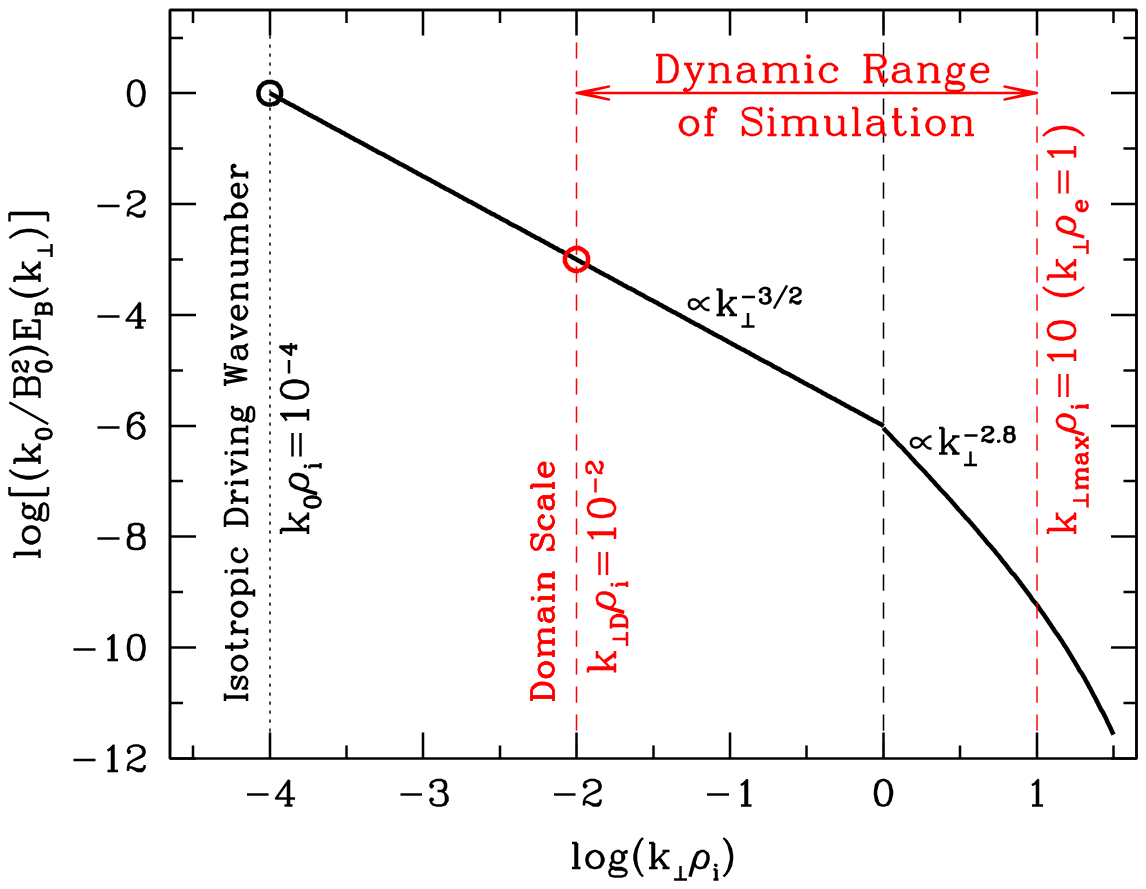}}\\
\resizebox{4.25in}{!}{\includegraphics*[20pt,260pt][564pt,690pt]{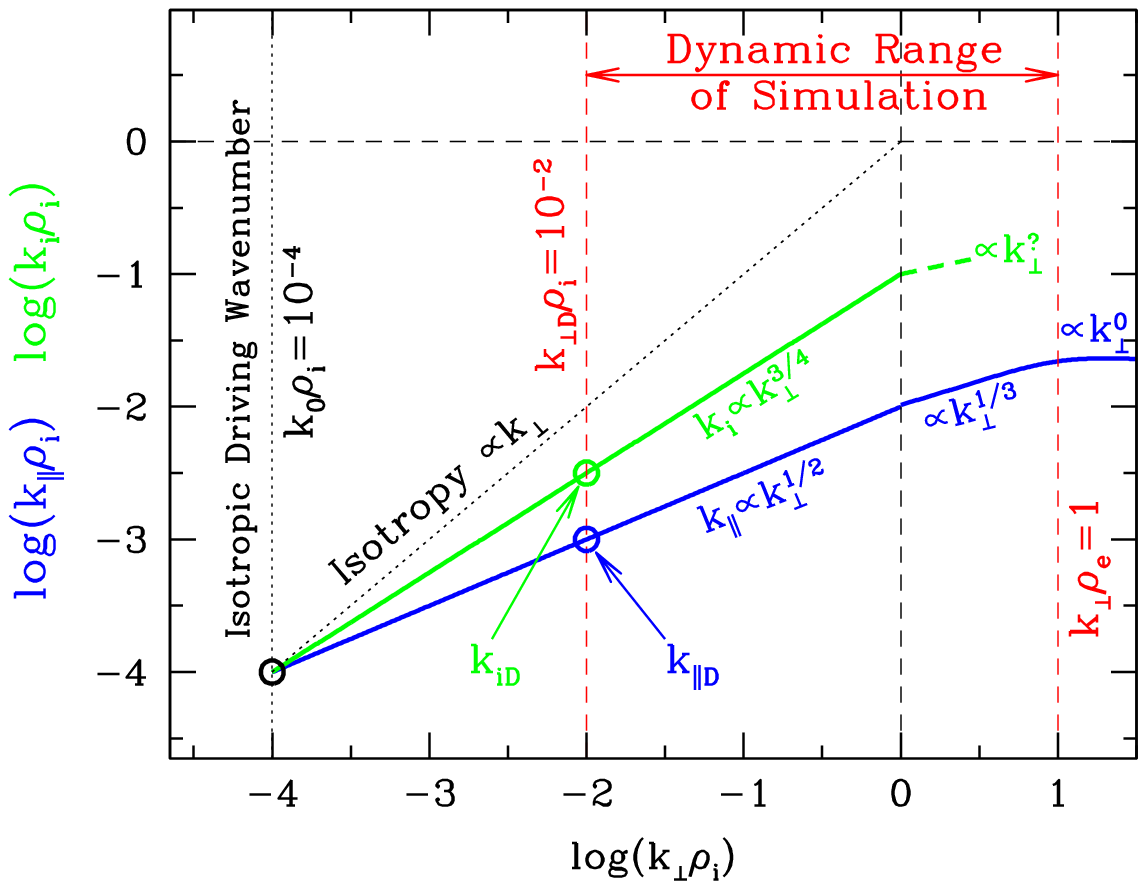}}
\caption{Modeling the  small-scale end of the turbulent inertial range using MHD scaling relations, choosing \emph{B06} scaling with $\alpha=1$, $m_i/m_e=100$, $k_0\rho_i =10^{-4}$, $k_{\perp D} \rho_i = 10^{-2}$.
\label{fig:turbdiss_sim}}
\end{figure}

One less computationally costly approach to capture numerically the
mechanisms of turbulent damping is to choose to model directly only
the small-scale end of the inertial range, preserving as much of the
resolved dynamic range of the simulation as possible to capture the dissipation
mechanisms.  To do so, one takes advantage of the scaling relations
for turbulence in the MHD regime $k_\perp \rho_i \ll 1$, summarized in
\secref{sec:mhdtheory}, to ``drive'' the turbulence at the domain
scale of the simulation with turbulent fluctuations that have the
properties dictated by the scaling theories.

As an example of how to set up simulations in this way, consider
turbulence physically driven strongly and isotropically at an isotropic
driving wavenumber of $k_0\rho_i =10^{-4}$, as depicted in
\figref{fig:turbdiss_sim}.  For a hydrogenic plasma of protons and
electrons with a realistic mass ratio $m_i/m_e=1836$, the turbulent
cascade extends at least down to the perpendicular scales of the
electron Larmor radius $k_\perp \rho_e \sim 1$
\citep{Sahraoui:2010b,Howes:2011a,TenBarge:2013b,Kiyani:2015}.  Thus,
a dynamic range of $L_0/\rho_e =k_{\perp \mbox{max}}/k_0 = [(k_{\perp
  \mbox{max}} \rho_e)/(k_0 \rho_i)] (\rho_i/\rho_e) \propto(k_0
\rho_i)^{-1}(T_i/T_e)^{1/2}(m_i/m_e)^{1/2} \sim 4.3 \times 10^5$ in
each of the three spatial dimensions is necessary when $T_i/T_e \sim
1$, well beyond the capabilities of any kinetic simulation code.  If
the simulation code can resolve a dynamic range of $10^3$ in each
spatial dimension, we can choose a perpendicular domain scale in the
middle of the physical inertial range at $k_{\perp D} \rho_i =
10^{-2}$ and use a reduced mass ratio $m_i/m_e=100$ so that $k_\perp
\rho_e=1$ corresponds to $k_\perp \rho_i=10$ for $T_i/T_e=1$.  Thus,
the resolved perpendicular length scales of the simulation span
$10^{-2} \le k_\perp \rho_i \le 10$, corresponding to $10^{-3} \le
k_\perp \rho_e \le 1$ in terms of the electron Larmor radius $\rho_e$,
as shown in \figref{fig:turbdiss_sim}.

To drive the turbulence in the middle of the turbulent inertial range
at $k_{\perp D} \rho_i = 10^{-2}$, we use the scaling relations
\eqref{eq:kpar}--\eqref{eq:bperp} to determine the appropriate turbulent
scale-dependent anisotropy and amplitude of the fluctuations at the
numerical domain scale for a \emph{physical} inertial range driven
strongly ($\chi_0=1$) and isotropically ($k_{\perp 0}= k_{\parallel
  0}=k_{i 0}\equiv k_0$ and $\theta_0=1$~rad) at $k_0\rho_i =10^{-4}$. For
this example, we will adopt the \emph{B06} scaling, choosing
$\alpha=1$.  The wavevector anisotropy and amplitude of the
fluctuations at the domain scale can be expressed using scaling
relations \eqref{eq:kpar}--\eqref{eq:bperp} to yield
\begin{equation}
  k_{\parallel D} \rho_i = k_0 \rho_i \left( \frac{k_{\perp D}\rho_i}{k_0 \rho_i}\right)^{1/2}
  \label{eq:kpar_d}
\end{equation}
\begin{equation}
  k_{iD}\rho_i= k_0 \rho_i \left( \frac{k_{\perp D}\rho_i}{k_0 \rho_i}\right)^{3/4}
  \label{eq:ki_d}
\end{equation}
\begin{equation}
  \theta_D =  \theta_0  \left( \frac{k_{\perp D}\rho_i}{k_0 \rho_i}\right)^{-1/4}
  \label{eq:theta_d}
\end{equation}
\begin{equation}
\frac{\delta B_{\perp D} }{B_0} =  \left( \frac{k_{\perp D}\rho_i}{k_0 \rho_i}\right)^{-1/4}
  \label{eq:bperp_d}
\end{equation}

The properties of the domain scale fluctuations at $k_{\perp D} \rho_i
= 10^{-2}$ determined by these scaling relations are illustrated in
\figref{fig:turbdiss_sim} for a turbulent cascade parameterized by
dimensionless parameters $k_0\rho_i =10^{-4}$ and $\theta_0=1$~rad.
The normalized value of energy spectrum at $k_{\perp D} \rho_i =
10^{-2}$ is $(k_0/B_0)^2 E_{B_\perp}(k_\perp) =10^{-3}$, consistent
with the solution $\delta B_{\perp D}/B_0=0.32$ of \eqref{eq:bperp_d}
at the domain scale, given by the red circle in
\figref{fig:turbdiss_sim}(a).  The three-dimensional anisotropy of the
wavevector of the turbulent fluctuations at the perpendicular domain
scale $k_{\perp D} \rho_i = 10^{-2}$ is characterized by: (i)
$k_{\parallel D} \rho_i =10^{-3}$ from \eqref{eq:kpar_d}, given by the
blue circle in \figref{fig:turbdiss_sim}(b); and (ii) $k_{i D} \rho_i
=3.2 \times 10^{-3}$ from \eqref{eq:ki_d}, given by the green circle
in \figref{fig:turbdiss_sim}(b).  The dynamic alignment of the
velocity and magnetic field fluctuations at the domain scale is given
by $\theta_D=0.32$~rad from \eqref{eq:theta_d}.  Note that for these
parameters, the scalings lead to a constant nonlinearity parameter as
a function of perpendicular wavenumber $\chi(k_\perp) = \chi_0=1$ at
all scales $k_\perp\rho_i<1$, as dictated by the conjecture of
critical balance in strong turbulence \citep{Goldreich:1995,Mallet:2015}.  These
scaling calculations determine the nature of the turbulent
fluctuations at the numerical domain scale, in the middle of the
physical inertial range being modeled.

Since the large-scale of the resolved dynamic range of the simulation
at $k_{\perp D} \rho_i$ is constantly being fed turbulent cascade
energy mediated by nonlinear interactions among turbulent fluctuations
at scales slightly larger than the domain scale with $k_\perp <
k_{\perp D}$, it is imperative that these turbulence simulations are
driven, rather than simply simulating the decay of the energy
turbulent fluctuations initialized at the domain scale. The nonlinear
interactions among perpendicularly polarized, counterpropagating
\Alfven waves drive the cascade of energy in \Alfvenic plasma
turbulence
\citep{Kraichnan:1965,Sridhar:1994,Goldreich:1995,Ng:1996,Galtier:2000,Howes:2012b,Howes:2013a,Nielson:2013a,Howes:2013b,Drake:2013,Howes:2016b,Verniero:2018a,Verniero:2018b}.
Exploiting this physical insight, driving counterpropagating \Alfven
waves, polarized in both dimensions perpendicular to the equilibrium
magnetic field, has been shown to generate successfully a steady-state
plasma turbulent cascade
\citep{Howes:2008a,Howes:2011a,TenBarge:2013b,Told:2015,Verniero:2018a,Verniero:2018b}.
A finite-time correlated driving, such as an antenna driven by a
Langevin equation \citep{TenBarge:2014a}, mimics the unsteady nature
of the turbulent cascade from larger scales expected of turbulent
driving at a chosen scale within the turbulent inertial range.

To set up the simulation to describe optimally the domain scale
fluctuations, it has become common practice to utilize a numerical
domain that is elongated along the direction of the equilibrium
magnetic field, consistent with the anisotropic scaling of turbulent
fluctuations at small scales, $k_\parallel/k_\perp \ll 1$. For the
example shown in \figref{fig:turbdiss_sim}, individual \Alfvenic
fluctuations are predicted to have a perpendicular plane anisotropy
given by $k_{iD}/k_{\perp D}=0.32$.  But since these \Alfvenic
fluctuations can be polarized in either of the directions in the
perpendicular plane, it is necessary for both perpendicular directions
to be large enough to resolve the intermediate scale (the larger of
the two perpendicular scales since $k_{iD}/k_{\perp D} < 1$).  Thus,
the simulation domain should be defined by $L_\parallel \times L_i^2$,
where $L_\parallel = 2 \pi \rho_i (k_{\parallel D} \rho_i)^{-1}$ and
$L_i = 2 \pi \rho_i (k_{i D} \rho_i)^{-1}$. Within this domain,
counterpropagating \Alfvenic fluctuations should be driven, polarized
in both perpendicular directions, with wavevectors characterized by
$(k_{\perp D},k_{i D},k_{\parallel D})$, with dynamic alignment
characterized by $\theta_D$, and with amplitudes $\delta B_{\perp D}$.
This approach enables the efficient simulation of the small-scale end
of the inertial range consistent with the scaling properties of MHD
turbulence, enabling kinetic simulations to focus on the physical
mechanisms damping the turbulent fluctuations.

\begin{table}
  \begin{center}
  \begin{tabular}{|l|c|}
    \hline
    \multicolumn{2} {|l|} {Plasma Parameters}\\
    \hline
    Parameter & Symbol and Definition \\
    \hline
    Ion Plasma Beta & $\beta_i=8 \pi n_i T_{i}/B^2$ \\
    Ion-to-Electron Temperature Ratio &  $\tau=T_{i}/T_{e}$\\
    Ion-to-Electron Mass Ratio & $\mu = m_i/m_e$\\
    \hline
    \multicolumn{2} {|l|} {Turbulence Parameters}\\
    \hline
    Parameter & Symbol and Definition \\
    \hline
    Isotropic Driving Wavenumber & $k_{0} \rho_i$\\
    \hline
    \end{tabular}
  \end{center}
  \caption{\emph{Isotropic Temperature Case}: Reduced set of fundamental dimensionless parameters for isotropic Maxwellian equilibrium velocity distributions with $A_i=A_e=1$, collisionless conditions $k_{\parallel 0} \lambda_{mfp,e} \gg 1$, strong turbulence assumed at the ion kinetic scales, balanced turbulence $Z_0^+/Z_0^-\sim 1$, and negligible energy in compressible fluctuations $E_{comp}/E_{inc} \ll 1$.  This results in a reduced set of plasma and turbulence parameters $(\beta_i,\tau; k_{0} \rho_i)$. \label{tab:reduced_params} }
  \end{table}
\subsection{A Reduced Parameter Space for Kinetic Plasma Turbulence}
Although the physics of plasma turbulence, its dissipation, and the
resulting heating of the plasma species formally depends on all ten of
the general plasma and turbulence parameters
$(\beta_{\parallel,i},\tau_\parallel,A_i,A_e,k_{\parallel 0} \lambda_{mfp,e}; k_{\perp 0}
\rho_i,k_{\parallel 0}/k_{\perp
  0},\chi_0,Z_0^+/Z_0^-,E_{comp}/E_{inc})$ listed in
\tabref{tab:params}, focusing initially on the development of
completely general turbulent heating models on a ten-dimensional
parameter space is unlikely to be successful.  Instead, first
developing predictive turbulent heating models over a reduced
parameter space is almost certainly a better strategy. As an initial
exploration of \Alfvenic turbulence and its kinetic damping
mechanisms,  we choose to focus on the simplified limit of
isotropic temperatures, modifying our approach as needed for cases
that violate these conditions.


For many of the calculations in this study, a reduced isotropic
temperature case is adopted, assuming that the equilibrium velocity
distribution for each species remains well described as an isotropic
Maxwellian distribution.  Note that kinetic temperature anisotropy
instabilities generally bound the species temperature anisotropies in
the vicinity of isotropy $A_s \sim 1$
\citep{Kasper:2002,Hellinger:2006,Stverak:2008,Bale:2009,Stverak:2009},
although if these instabilities are triggered,  particularly at $\beta_i \gg 1$, they can impact both
the linear wave physics at play in the turbulent dynamics of
critically balanced turbulence
\citep{Squire:2016,Squire:2017b,Squire:2017a} and mediate the nonlocal
transfer of energy directly from the large driving scales to the
kinetic length scales \citep{Arzamasskiy:2023}.


To define a reduced parameter space, we first restrict the temperature
anisotropies for ions and electrons to unity, $A_i=1$ and $A_e=1$, so
that $T_{\perp,i}=T_{\parallel,i}\equiv T_i$ and
$T_{\perp,e}=T_{\parallel,e}\equiv T_e$.  With this idealization, the
ion plasma beta and ion-to-electron temperature ratio are redefined in
terms of these isotropic temperatures $T_i$ and $T_e$, yielding the isotropic
parameters $\beta_i=8 \pi n_i T_{i}/B^2$ and $\tau=T_{i}/T_{e}$.  In
addition, for this reduced parameter space, we will also assume that
the turbulent inertial range is sufficiently large, or the driving is
sufficiently strong, that the dynamics at the perpendicular ion
kinetic length scales $k_\perp \rho_i \sim 1$ are in a state of strong
plasma turbulence with a nonlinearity parameter $\chi \sim 1$.  Under
these conditions, we can replace the three turbulent driving
parameters $(k_{\perp 0} \rho_i,k_{\parallel 0}/k_{\perp 0},\chi_0)$
by the single isotropic driving wavenumber $k_0\rho_i$, as detailed in
\secref{sec:reducing}.

We will also make two other simplifying assumptions for this reduced,  isotropic
temperature case. We will assume that the turbulence is balanced, with
approximately equal wave energy fluxes up and down the local magnetic
field $Z_0^+/Z_0^- \sim 1$, which is a necessary condition to apply
the MHD turbulence scaling relations given by
\eqref{eq:kpar}--\eqref{eq:bperp}.  Additionally, based on direct
spacecraft measurements of turbulence in the heliosphere which find
that incompressible \Alfvenic fluctuations appear to dominate
turbulent dynamics in space plasmas
\citep{Belcher:1971,Tu:1995,Alexandrova:2008,Bruno:2013}, as well as
theoretical considerations that suggest that the dynamics of \Alfven
waves dominate the turbulent cascade
\citep{Lithwick:2001,Maron:2001,Cho:2003,Schekochihin:2009,Howes:2015b},
we will also assume a negligible amount of energy in compressible
turbulent fluctuations at the driving scale, $E_{comp}/E_{inc} \ll 1$.

Therefore, we propose the \emph{isotropic temperature case}, a reduced
parameter space for weakly collisional plasma turbulence in the limit
of strong, balanced, \Alfvenic turbulence, with a governing set
of three dimensionless parameters, $(\beta_i,\tau; k_{0} \rho_i)$, as
shown in \tabref{tab:reduced_params}.  Although the isotropic
temperature case eliminates a number of important physical effects
such as temperature anisotropy instabilities, imbalanced turbulent
wave energy fluxes, and compressible fluctuations, it nonetheless
represents an excellent starting point for the development of refined
turbulent heating models.  Such models can produce theoretical
predictions that can be compared to kinetic numerical simulations and
spacecraft observations of weakly collisional plasma turbulence,
exemplified by the phase diagram for turbulent damping mechanisms
shown in \figref{fig:phase_k0bi}.

In terms of the fundamental parameters of the isotropic temperature
case, the MHD turbulent scaling relations
\eqref{eq:kpar}--\eqref{eq:bperp} simplify to the following easy-to-use forms
\begin{equation}
  k_\parallel \rho_i =  (k_0 \rho_i)^{(1+\alpha)/(3+\alpha)} (k_\perp \rho_i)^{2/(3+\alpha)}
  \label{eq:kpar_iso}
\end{equation}
\begin{equation}
  k_i  \rho_i =   (k_0 \rho_i)^{\alpha/(3+\alpha)} (k_\perp \rho_i)^{3/(3+\alpha)}
  \label{eq:ki_iso}
\end{equation}
\begin{equation}
  \theta =   \theta_0\left(\frac{k_0 \rho_i}{k_\perp \rho_i} \right)^{\alpha/(3+\alpha)}
  \label{eq:theta_iso}
\end{equation}
\begin{equation}
\frac{\delta B_{\perp } }{B_0} =   \left(\frac{k_0 \rho_i}{k_\perp \rho_i} \right)^{1/(3+\alpha)}
  \label{eq:bperp_iso}
\end{equation}

\section{Review of Previous Turbulent Heating Models}
\label{sec:review}
Several previous studies have constructed turbulent heating models
that attempt to quantify how the energy removed from the turbulent
cascade is partitioned between protons and electrons and between
parallel and perpendicular degrees of freedom. Here we briefly
summarize these existing turbulent heating models, including the
dissipation mechanisms considered, the assumptions of how that energy
is removed from the turbulence, and that stated limitations of each
model.

One of the earliest efforts to estimate the proton and electron
energization rates by turbulence due to collisionless damping
mechanisms was carried out by \citet{Quataert:1999} (hereafter
\emph{QG99}), following earlier results by each of these authors to
determine the collisionless damping rates under astrophysically
relevant conditions \citep{Quataert:1998,Gruzinov:1998}.  This study
adopted the \emph{GS95} scaling for dominantly \Alfvenic turbulence at
perpendicular scales restricted to the MHD limit, $k_\perp\rho_i
\lesssim 1$.  The turbulent fluctuations upon reaching the end of the
MHD limit at the perpendicular scale of the ion Larmor radius
$k_\perp\rho_i \sim 1$ are predicted to be significantly anisotropic
with $k_\perp \gg k_\parallel$; since the MHD \Alfven wave frequency $
\omega = k_\parallel v_A$ is proportional to $k_\parallel$, this
yields turbulent fluctuations at this perpendicular ion scale with low
frequencies relative to the ion cyclotron frequency, $\omega \ll
\Omega_i$.  Thus, all collisionless damping is provided by the $n=0$
Landau resonance, specifically Landau damping and transit-time
damping. In addition, this model assumed that all turbulent energy at
a given perpendicular wavenumber $k_\perp\rho_i$ was concentrated at
the parallel wavenumber given by the conjecture of critical balance
\citep{Goldreich:1995,Quataert:1999}.  They constructed a simple
cascade model that balanced the collisionless damping by ions and
electrons given by the Vlasov-Maxwell linear dispersion relation
\citep{Stix:1992} over the MHD range of scales, $k_\perp\rho_i
\lesssim 1$.  Beyond these scales at $k_\perp\rho_i \gtrsim 1$ in the
kinetic regime, uncertainty about the nature of the turbulence at
these sub-ion length scales led them to assume simply that any energy
reaching $k_\perp\rho_i \gtrsim 1$ ultimately heated the electrons.
They calculated the fraction of the turbulent energy that heats the
electrons, $Q_e/Q$, as function of total plasma beta
$\beta=\beta_i+\beta_e= \beta_i(1+T_i/T_e)$, yielding a numerical
result for $Q_e/Q(\beta)$.  Their resulting energy partition depended
strongly on a constant assumed in the model that characterized the
balance of the nonlinear turbulent cascade rate to the linear
collisionless damping rate; uncertainty in the value of this constant
(by a factor from $1/4$ to $4$) led the resulting prediction of
$Q_e/Q(\beta)$ to vary by more than one order-of-magnitude over the
range of $3 \lesssim \beta \lesssim 100$.

The Howes 2010 (hereafter \emph{H10}) turbulent heating model
\citep{Howes:2010d} was based on a turbulent cascade model
\citep{Howes:2008b} that extended the \emph{GS95} turbulence scaling
from MHD scales at $k_\perp\rho_i \lesssim 1$ into the kinetic regime
at $k_\perp\rho_i \gtrsim 1$, generalizing the conjecture of critical
balance to account for the dispersive nature of kinetic \Alfven waves
at these sub-ion length scales
\citep{Biskamp:1999,Cho:2004,Krishan:2004,Shaikh:2005,Schekochihin:2009}.
The \emph{H10} model was based on three primary assumptions: (i) the
Kolmogorov hypothesis that the energy cascade is determined by local
interactions \citep{Kolmogorov:1941}; (ii) the conjecture that
turbulence maintains a state of critical balance at all scales
\citep{Goldreich:1995}; and (iii) the speculation that linear
collisionless damping rates are applicable even in the presence of the
strong nonlinear interactions that mediate the turbulent cascade to
small scales.  The cascade model used by \emph{H10} balanced at each
perpendicular wavenumber $k_\perp \rho_i$ the nonlinear energy cascade
rate given by the extended turbulent scaling theories with the linear
collisionless damping rates at that value of $k_\perp \rho_i$.
Similar to \emph{QG99}, it assumed all turbulent energy at a given
value of $k_\perp \rho_i$ resided at the parallel wavenumber governed
by the extended critical balance. The \emph{H10} model assumed
turbulence under nonrelativistic conditions $v_{te}/c \ll 1$ with
isotropic ion and electron temperatures, strictly \Alfvenic turbulence
with $E_{comp}/E_{inc}=0$, balance turbulence with $Z_0^+/Z_0^-=1$,
and strongly and isotropically driven turbulence with $\chi_0 =1$ and
$k_{\parallel 0}/k_{\perp 0}=1$ (the driving was therefore
characterized by the isotropic driving wavenumber, $k_{0} \rho_i$).
Furthermore, adopting the same argument as the \emph{QG99} model, the
model assumed that the inertial range of the turbulence is
sufficiently large that the anisotropic fluctuations predicted by the
turbulent scaling relations remain at a low enough frequency relative
to the ion cyclotron frequency, $\omega \ll \Omega_i$, that the
cyclotron resonance plays a negligible role; the \emph{H10} model
calculated an estimate of the maximum value of $k_0\rho_i$ as a
function of $(\beta_i,T_i/T_e)$ for this assumption to be satisfied.
Thus, the \emph{H10} turbulent heating model predicted the damping of
the turbulent cascade due to Landau damping and transit-time damping
over the full range of the turbulent cascade, including MHD scales at
$k_\perp\rho_i \lesssim 1$ and kinetic scales at $k_\perp\rho_i
\gtrsim 1$.  The \emph{H10} model predicted the partitioning of energy
between ions and electrons as a function of the ion plasma beta
$\beta_i$ and the ion-to-electron temperature ratio $T_i/T_e$,
providing a simple analytical fit to the numerically calculated
predictions, $Q_i/Q_e(\beta_i,T_i/T_e)$. In short, the model found
that $Q_i/Q_e$ is a monotonically increasing function of $\beta_i$
with weak dependence on $T_i/T_e$, where the transition from
$Q_i/Q_e<1$ to $Q_i/Q_e>1$ happens at approximately $\beta_i \sim 1$.

The Chandran 2011 (hereafter \emph{C11}) turbulent heating model
\citep{Chandran:2011} looked to incorporate the recently quantified
damping rate due to nonlinear ion stochastic heating
\citep{Chandran:2010a,Chandran:2010b} into a model that also accounted
for ion and electron Landau damping and transit-time damping. Similar
to \emph{QG99} and \emph{H10}, it assumed a sufficiently large
inertial range that no significant cyclotron damping would occur since
the anisotropic cascade would lead to turbulent fluctuations with low
frequencies, $\omega \ll \Omega_i$. Focusing also on \Alfvenic
turbulence that transitions from an MHD \Alfven wave cascade at
$k_\perp\rho_i \lesssim 1$ and to a kinetic \Alfven wave cascade at
$k_\perp\rho_i \gtrsim 1$, the \emph{C11} model assumed that
dissipation occurred dominantly in two distinct wavenumber ranges: (i)
at $k_\perp\rho_i \sim 1$, where collisionless damping via the Landau
($n=0$) resonance can lead to parallel energization of the ions and
electrons and also where nonlinear stochastic heating can lead to
perpendicular energization of the ions; and (ii) at $k_\perp \rho_i
\gg 1$, where all remaining energy in the turbulent cascade is
ultimately deposited with the electrons.  The model used fitting forms
for the Landau-resonant collisionless damping rates for the ions and
electrons at $k_\perp \rho_i= 1$ over $10^{-3} < \beta_i < 10$ and $1
< T_i/T_e < 5$.  It also included an empirically derived formula for
the ion stochastic heating rate at $k_\perp \rho_i= 1$
\citep{Chandran:2010a}. This empirical prescription for the ion
stochastic heating depended on two dimensionless constants: one
governed the scaling of the stochastic heating rate relative to the
turbulent frequencies; and the other, which appeared as a factor in
the argument of the exponential function, effectively established a
threshold amplitude below which ion stochastic heating was negligible.
Thus, the \emph{C11} model calculated how the turbulent cascade rate
at $k_\perp \rho_i= 1$ is partitioned among ion Landau and
transit-time damping, electron Landau and transit-time damping, ion
stochastic heating, and nonlinear transfer to smaller scales with
$k_\perp \rho_i > 1$.  The model provided a partitioning of the total
turbulent heating rate into parallel ion heating by ion Landau and
transit-time damping $Q_{\parallel,i}/Q$, perpendicular ion stochastic
heating $Q_{\perp,i}/Q$, and electron heating $Q_{e}/Q$, each as a
function of ion plasma beta $\beta_i$, the ion-to-electron temperature
ratio $T_i/T_e$, and the amplitude of the turbulent fluctuations at
the ion Larmor radius scale determined using the \emph{B06} scaling
for the amplitude in terms of the driving scale, as given here by
\eqref{eq:bperp_iso}, indicating that the amplitude dependence can be
characterized by $k_0\rho_i$.

Motivated by the desire to understand the partitioning of dissipated
turbulent energy between ions and electrons in hot collisionless
accretion flows, the Rowan 2017 (hereafter \emph{R17}) heating model
\citep{Rowan:2017} took a fundamentally different approach from that
adopted in the \emph{QG99}, \emph{H10}, and \emph{C11} turbulent
heating models by focusing instead on the heating arising from
transrelativistic magnetic reconnection.  In this case, the protons
are nonrelativistic, but the electrons can be
ultrarelativistic. Utilizing a large suite of 2D particle-in-cell
(PIC) simulations of anti-parallel reconnection, the \emph{R17} model
studied the irreversible particle energization over a range of ion
plasma beta $10^{-4} \lesssim \beta_i \le 2$ and magnetization $0.1
\le \sigma_w \le 10$, where $\sigma_w$ is defined as the ratio of the
magnetic energy density to the enthalpy density. Starting with a
Harris current sheet \citep{Harris:1962} of the antiparallel magnetic
field (given by a $\tanh$ profile) with width $a \sim 40 d_e$, they
carefully separated the irreversible heating associated with an
increase in the entropy from the reversible heating associated with
adiabatic compression.  From their synthesis of the results of a large
suite of simulations varying different parameters, they fit a formula
to their numerical results that provided to electron-to-total heating
ratio $Q_e/Q$ as a function of the ion plasma beta $\beta_i$ and
magnetization $\sigma_w$.

The numerical study by \citet{Kawazura:2019} (hereafter \emph{K19})
used a large suite of hybrid gyrokinetic simulations to determine
directly the ion-to-electron heating ratio $Q_i/Q_e$ over a broad
range of ion plasma beta $0.1 \le \beta_i \le 100$ and ion-to-electron
temperature ratio $0.05 \le T_i/T_e \le 100$.  Using a hybrid code
\citep{Kawazura:2018} that evolves the ions gyrokinetically and the
electrons as an isothermal fluid, this study adopted many of the same
assumptions as the \emph{H10} model: (i) the anisotropic ($k_\perp \gg
k_\parallel$) \Alfvenic turbulence was driven in a statistically
balanced manner with $Z_0^+/Z_0^-=1$; (ii) the equilibrium ion and
electron temperatures were assumed to be isotropic; (iii) the
gyrokinetic approximation eliminated the physics of the ion cyclotron
resonances, leaving the Landau resonances and collisionless magnetic
reconnection in the large-guide-field limit as potential damping
mechanisms; and (iv) the strategy explained in
\secref{sec:end_inertial} was exploited to model numerically only the
small-scale end of the inertial range, where the isotropic driving
wavenumber $k_0 \rho_i$, characterizing the physical driving scale, was
assumed to be sufficiently small that the amplitudes of the turbulent
fluctuations at $k_\perp \rho_i \sim 1$ were small enough that ion
stochastic heating was inhibited \citep{Chandran:2010a,Chandran:2011}.
The \emph{K19} model provides a numerically validated prescription for
$Q_i/Q_e$ as a function of $\beta_i$ and $T_i/T_e$, presenting a
simple analytical formula that fit the numerical results well. The
numerical results largely validated the qualitative predictions of the
\emph{H10} turbulent heating model that $Q_i/Q_e$ is a monotonic
function of $\beta_i$ with little dependence on $T_i/T_e$, with a
couple of minor quantitative differences: the \emph{K19} model found a
ceiling of $Q_i/Q_e \simeq 30$ at $\beta_i \gg 1$, cutting off at
lower values than the \emph{H10} model; and the
numerical results did not find the drop in the heating rate $Q_i/Q_e
\ll 0.1$ predicted by the \emph{H10} model  at $\beta_i \ll 1$.

The first of the two disagreements above between the \emph{H10} model
and the \emph{K19} model---where the numerical simulations found an
effective ceiling at $Q_i/Q_e \simeq 30$ at $\beta_i \gg 1$---was
recently addressed by \citet{Gorman:2024}. They examined in
detail the small range of perpendicular wavenumber scales around
$k_\perp \rho_i \sim 1$ where the linear dispersion relation predicts
that the \Alfven wave solution becomes non-propagating at $\beta_i \gg
1$.  For turbulent cascade models assuming critical balance and
strictly local (in scale space) interactions mediating the turbulent
cascade, this gap in the \Alfven wave propagation leads to the
turbulent cascade being terminated at the onset of the gap with the
bulk of the turbulent energy transferred to the ions, leading to the
potential overestimation of $Q_i/Q_e$ in the $\beta_i \gg 1$
limit. \citet{Gorman:2024} (hereafter \emph{GK24}) tested the local
cascade model underlying the \emph{H10} model against a more refined
cascade model, the \emph{weakened cascade model}
\citep{Howes:2011b}. The weakened cascade model is a nonlocal cascade
model that accounts both for the weakening of the local nonlinear
interactions as damping taps the turbulent energy and for nonlocal
interactions by large-scale shearing and small-scale diffusion. It was
found that the inclusion of nonlocal interactions to the nonlinear
energy transfer prevents the termination of the cascade at the ion
scales, leading to predictions for $Q_i/Q_e$ at $\beta_i \gg 1$
\citep{Gorman:2024} that agree much more closely with the results of
the suite of simulations by \citet{Kawazura:2019}: the resulting
\emph{GK24} model produced by this study includes updated coefficients
for the \emph{H10} formula for $Q_i/Q_e(\beta_i,T_i/T_e)$.

Relaxing the limitation of their simulations to driving by strictly \Alfvenic
turbulent fluctuations, \citet{Kawazura:2020} used the same hybrid
gyrokinetic code \citep{Kawazura:2018} to explore the effect of
compressible driving $E_{comp}/E_{inc} > 0$ on the ion-to-electron
heating ratio $Q_i/Q_e$ (hereafter \emph{K20}). The turbulence was
driven at MHD scales $k_\perp\rho_i \ll 1$ with an adjustable mixture
of incompressible \Alfvenic fluctuations and compressible slow
magnetosonic fluctuations [fast magnetosonic wave modes are ordered
out of gyrokinetic simulations by the perpendicular pressure balance
that is maintained under the gyrokinetic approximation
\citep{Howes:2006}].  At $\beta_i \ll 1$, they found that $Q_i/Q_e =
E_{comp}/E_{inc}$, confirming a previous theoretical prediction that
compressive energy strictly heats ions at low ion plasma beta $\beta_i
\ll 1$ \citep{Schekochihin:2019}.  Surprisingly, the \emph{K20} model
also found that $Q_i/Q_e = E_{comp}/E_{inc}$ when the compressible
driving dominates, $E_{comp}/E_{inc} \gg 1$.  A simple fitting formula
was constructed that simply added the term $E_{comp}/E_{inc}$ to the
previous prescription for $Q_i/Q_e$ due to \Alfvenic turbulence in the
\emph{K19} model.  Thus, the \emph{K20} model provided a simple
analytical formula for $Q_i/Q_e(\beta_i , T_i/T_e, E_{comp}/E_{inc})$.

The application of the \emph{K20} model to understand the
ion-to-electron heating ratio in accretion disks where the
magnetorotational instability (MRI)
\citep{Balbus:1991,Hawley:1991,Balbus:1998} drives the turbulence
requires a determination of the ratio of the
compressible-to-incompressible driving $E_{comp}/E_{inc}$.  The
application of a rotating reduced MHD (RRMHD) model to simulate the
collisional MRI turbulence threaded by a nearly azimuthal magnetic
field in a 3D pseudo-spectral, shearing reduced MHD code
\citep{Kawazura:2022b} found a ratio of compressible-to-incompressible
turbulent energy of $ 2 \lesssim E_{comp}/E_{inc} \lesssim 2.5$
\citep{Kawazura:2022}, providing a solid foundation for the
application of the \emph{K20} model to astrophysical accretion disks.


\section{Dependence of Dissipation Mechanisms on Fundamental Parameters}
\label{sec:damping}
The development of turbulent heating models for weakly collisional
space and astrophysical plasmas requires two fundamental steps: (i)
identifying the mechanisms that play a role in the damping of the
turbulence; and (ii) determining upon which of the fundamental
dimensionless parameters of turbulence each mechanism depends and
approximating a functional form for those dependencies. 

It is worthwhile first noting an important subtlety regarding irreversible
plasma heating in the thermodynamic picture of turbulent dissipation
in weakly collisional plasmas \citep{Howes:2017c}. In the strongly
collisional limit of fluid plasma turbulence, the dissipation of
turbulence is fundamentally collisional on a microscopic level. For
example, in an MHD plasma, the physical mechanisms of viscosity and
resistivity arise in the limit of a non-negligible collisional mean
free path of the plasma particles.  These two dissipative mechanisms may
be derived rigorously beginning with kinetic theory through a
hierarchy of moment equations derived in the limit of small mean free
path by the Chapman-Enskog procedure \citep{Chapman:1970} for neutral
gases, or an analogous procedure for plasma systems
\citep{Spitzer:1962,Grad:1963,Braginskii:1965}.  Because viscosity and
resistivity are ultimately collisional in nature, the resulting energy
transfer from the turbulent fluctuations to the particles is
irreversible, leading to an increase in the entropy of the system.

Under the weakly collisional conditions typical of many space and
astrophysical plasmas, the collisionless removal of energy from the
turbulent fluctuations and the ultimate conversion of that removed energy
into plasma heat is a two-step process
\citep{Howes:2006,Howes:2008c,Schekochihin:2009,Navarro:2016,Howes:2017c,Howes:2018a}.
First, collisionless damping mechanisms remove energy from the
turbulent fluctuations, transferring that energy into nonthermal internal
energy contained in non-Maxwellian fluctuations of the particle
velocity distributions, an energy transfer that is reversible in
principle. That free energy in the fluctuations of the velocity
distribution about the equilibrium may be transferred to smaller
scales in velocity space---\emph{e.g.}, via linear phase mixing by the
ballistic term of the Boltzmann equation \citep{Landau:1946,Hammett:1992,Snyder:1997,Schekochihin:2009} or via
nonlinear phase mixing, a process often denoted the \emph{entropy cascade}
\citep{Dorland:1993,Schekochihin:2009,Tatsuno:2009,Plunk:2010,Plunk:2011,Kawamori:2013}---until
it reaches sufficiently small velocity-space scales that an
arbitrarily weak collisionality is sufficient to smooth out those
fluctuations, providing an irreversible conversion of that energy to
plasma heat and thereby increasing the entropy of the system
\citep{Howes:2006,Howes:2008c,Schekochihin:2009}.

The subtleties of energy conversion and plasma heating in weakly
collisional plasmas that are not in a state of local thermodynamic
equilibrium remains a topic of vigorous investigation, with many
recent developments that will not be reviewed in detail here
\citep{Schekochihin:2016,Parker:2016,Yang:2016,Yang:2017b,Yang:2017a,Howes:2018a,Kawazura:2019,Meyrand:2019,Meyrand:2021,Barbhuiya:2022,Cassak:2022a,Cassak:2022b,Squire:2022,Cassak:2023}.
In this paper, we use the term \emph{turbulent dissipation} to mean
the entire process of removing energy from the turbulent fluctuations
and subsequently thermalizing that energy via collisions to realize
irreversible heating of the plasma and an increase in the entropy of
the system.  When we refer to only the first step of the process, the
collisionless removal of energy from the turbulent fluctuations, we
denote those specific processes as \emph{turbulent damping
mechanisms}.

A primary aim of this study is to provide a unified framework for
characterizing the dependence of the turbulence and its dissipation on
a common set of dimensionless parameters.  The first step pursued here
is to identify the fundamental dimensionless parameters upon which
each proposed turbulent damping mechanism primarily depends, where the
set of dimensionless parameters for the general case is presented in
\tabref{tab:params} and for the isotropic temperature case in
\tabref{tab:reduced_params}. The key parameter dependencies for each
proposed damping mechanism found here are summarized in
\tabref{tab:mechanisms}.  Although, for any single particular
mechanism, different choices of dimensionless parameters may be more
natural (\emph{e.g.}, taking $\beta_e$ instead of $\beta_i/\tau$ in the
case of magnetic reconnection), the particular sets of dimensionless
parameters in Tables~\ref{tab:params} and~\ref{tab:reduced_params} are
proposed here as a useful framework for describing \emph{all} of the
proposed turbulent damping mechanisms, thereby simplifying the task of
determining how different mechanisms compete with each other throughout
the full parameter space.

\begin{table}
    \begin{center}
    \begin{tabular}{|l|l|l|l|}
    \hline
    Category & Abbrev & Mechanism & Key Parameters  \\
    \hline
       Resonant &iLD& Ion Landau Damping  & $\beta_i$  \\
        & eLD & Electron Landau Damping   & $\beta_i,T_i/T_e$, ($\mu$)   \\
         & iTTD& Ion Transit-Time Damping   &$\beta_i$  \\
        &  eTTD & Electron Transit-Time Damping  & $\beta_i,T_i/T_e$, ($\mu$) \\
        & iCD  & Ion Cyclotron Damping & $k_0 \rho_i, \beta_i$    \\
          \hline
   Non-      & iSH   & Ion Stochastic Heating  & $k_0 \rho_i, \beta_i$  \\
   Resonant        &  iMP & Ion Magnetic Pumping   & $E_{comp}/E_{inc},\beta_i,T_i/T_e,k_{\parallel 0} \lambda_{mfp,e},$ \\
   & & &  \hspace*{0.25in}$k_{\parallel 0}/k_{\perp 0},\chi_0$  \\
   &eMP & Electron Magnetic Pumping  & $\beta_i,T_i/T_e,k_{\parallel 0} \lambda_{mfp,e},k_{\parallel 0}/k_{\perp 0},\chi_0,$ \\
   & & & \hspace*{0.25in}$E_{comp}/E_{inc}$,  ($\mu$)  \\
       & iVH & Ion Kinetic Viscous Heating & $\beta_i,k_{\perp
  0}/k_{\parallel 0}, \chi_0$\\
           \hline
           Coherent &iRXN &Ion Magnetic Reconnection    &
           $k_0 \rho_i, \beta_i,T_i/T_e$,  ($\mu$)  \\
        Structures&  eRXN & Electron Magnetic Reconnection  &  $k_0 \rho_i, \beta_i,T_i/T_e$,  ($\mu$)  \\
          \hline 
    \end{tabular}
    \end{center}
    \caption{Proposed turbulent damping mechanisms and their key
      dimensionless parameter dependencies for the case of a fully
      ionized hydrogenic plasma. The ion-to-electron mass ratio $\mu$ is
      listed in parentheses since it is not a physically variable
      parameter, but it is often adjusted in numerical
      studies for computational efficiency. \label{tab:mechanisms}}
\end{table}

\subsection{Proposed Damping Mechanisms for Kinetic Turbulence}
\label{sec:proposed}
In weakly collisional space and astrophysical plasmas, the proposed
kinetic damping mechanisms that remove energy from the turbulent
fluctuations and transfer it to the particle species fall into three
broad categories: (i) resonant wave-particle interactions, (ii)
non-resonant wave-particle interactions, and (iii) damping occurring
in coherent structures.

Resonant mechanisms for the damping of the turbulent fluctuations in a
magnetized plasma transfer energy to particles in regions of the
velocity distribution that satisfy the resonant condition
\begin{equation}
  \omega- k_\parallel v_\parallel - n \Omega_s=0,
  \label{eq:res_denom}
\end{equation}
where $\Omega_s$ is the cyclotron frequency for species $s$ and
$|n|>1$ indicates the harmonics of the cyclotron frequency.  The $n=0$
resonance is often denoted the \emph{Landau resonance} and includes
two damping mechanisms: (i) Landau damping is mediated by the parallel
component of the electric field $E_\parallel$ doing work on the
charged particles
\citep{Landau:1946,Leamon:1998a,Leamon:1998b,Quataert:1998,Leamon:1999,Quataert:1999,Leamon:2000,Howes:2008b,Schekochihin:2009,TenBarge:2013a,Howes:2015b,TCLi:2016,Chen:2019,Afshari:2021};
and (ii) transit-time damping, also known as Barnes damping, is
mediated by the magnetic mirror force associated with magnetic field
magnitude fluctuations doing work on the magnetic moment of the
charged particle gyromotion
\citep{Barnes:1966,Quataert:1998,Quataert:1999,Howes:2008b,Huang:2024,Howes:2024b}. Both of these Landau-resonant collisionless damping mechanisms
energize particles in the degree of freedom parallel to the local
magnetic field.  The $n \ne 0$ resonances yield cyclotron damping of
the turbulent fluctuations by $E_\perp$, energizing particles in the
two perpendicular degrees of freedom
\citep{Coleman:1968,Denskat:1983,Isenberg:1983,Goldstein:1994,Leamon:1998b,Gary:1999a,Isenberg:2001,Hollweg:2002,Isenberg:2019,Afshari:2023}.

Non-resonant damping mechanisms can remove energy from turbulent
fluctuations through wave-particle interactions that do not require
satisfying a particular resonance condition. Stochastic heating can
arise when sufficiently large amplitude turbulent electromagnetic
fluctuations can effectively scatter the otherwise organized
gyromotion of particles, leading to a heating of the particles in the
perpendicular degrees of freedom
\citep{Johnson:2001,Chen:2001,White:2002,Voitenko:2004b,Bourouaine:2008,Chandran:2010a,Chandran:2010b,Chandran:2011,Bourouaine:2013,Chandran:2013,Klein:2016b,Arzamasskiy:2019,Martinovic:2020,Cerri:2021}.
Magnetic pumping energizes particles through a combination of two
effects: (i) particles undergo the reversible, double adiabatic
evolution of their perpendicular and parallel velocities in a magnetic
field with a time varying magnitude; and (ii) those particles are
scattered by collisions during this evolution, introducing
irreversibility and leading to a small net transfer of energy to the
particles over a full oscillation cycle
\citep{Spitzer:1951,Berger:1958,Lichko:2017,Lichko:2020,Montag:2022}.  One final
proposed non-resonant damping mechanism is the recently identified kinetic
``viscous'' heating mediated by the effective collisionality
associated with temperature anisotropy instabilities driven by
large-scale fluctuations of the magnetic field magnitude in high beta plasmas
\citep{Arzamasskiy:2023}.

The final category involves the removal of turbulent fluctuation energy in
coherent structures, such as particle energization arising from
collisionless magnetic reconnection in current sheets that are found
to arise naturally in plasma turbulence
\citep{Ambrosiano:1988,Dmitruk:2004,Markovskii:2011,Matthaeus:2011,Osman:2011,Servidio:2011a,Osman:2012a,Osman:2012b,Wan:2012,Karimabadi:2013a,Zhdankin:2013,Dalena:2014,Osman:2014a,Osman:2014b,Zhdankin:2015a,Zhdankin:2015b,Mallet:2017b,Mallet:2017c,Loureiro:2017a}.

Each of the kinetic damping mechanisms in these three categories
(with helpful abbreviations) are summarized in
\tabref{tab:mechanisms}. Also included in \tabref{tab:mechanisms} are
the primary dimensionless parameters upon which each damping mechanism
depends, as explored in more detail in the following subsections.

\subsection{A Competition between Turbulent Cascade Rates and Damping Rates}
\label{sec:competition}
One of the most challenging aspects of modeling the damping of plasma
turbulence is that the net energy transfer to each species arises from
the competition between the turbulent cascade rate and the kinetic
damping rates as a function of scale, as depicted by the arrows in
\figref{fig:cascade}(a).  Nonlinear interactions among turbulent
fluctuations---denoted \Alfven wave collisions \citep{Howes:2013a} in
the case of \Alfvenic turbulence---mediate the transfer of energy from
large to small scales that is dominantly local in scale space
\citep{Howes:2011b,Told:2015}, where energy is transferred locally
from a wavenumber $k$ to a wavenumber $2k$, then on to $4k$, and so
on.  This is indicated by the black curved arrows in the figure.
Because the physics of the cascade is dominantly local in nature
\citep{Howes:2011b,Told:2015}, scaling theory implies that the energy
transfer rate at a given wavenumber $k$ is determined solely by the
conditions at that scale.  Note that, when analyzed in Fourier space,
the energy transfer to small scales in the turbulent cascade is the
sum of the nonlinear interactions among many triads of plane-wave
modes, with a wide spread of energy transfer rates about zero; the net
energy transfer given by this sum at a given scale is much smaller in
amplitude than the spread of all of the individual three-wave
interaction rates, and the sum typically indicates a forward cascade
of energy to small scales for turbulence in three spatial dimensions
\citep{Coburn:2014,Coburn:2015}. The turbulent cascade rate
$\varepsilon$ used here specifically refers to this net energy transfer rate
obtained by integrating over all contributing nonlinear interactions
involving that scale\footnote{Because the net energy transfer rate at
a given scale fundamentally involves the integration over all possible
nonlinear triads, it can in principle include a non-negligible
contribution from nonlocal interactions, such as those incorporated
 in the \emph{weakened cascade model} of the turbulent cascade \citep{Howes:2011b}.}.

\begin{figure}
  \begin{center}
    \resizebox{4.0in}{!}{\includegraphics{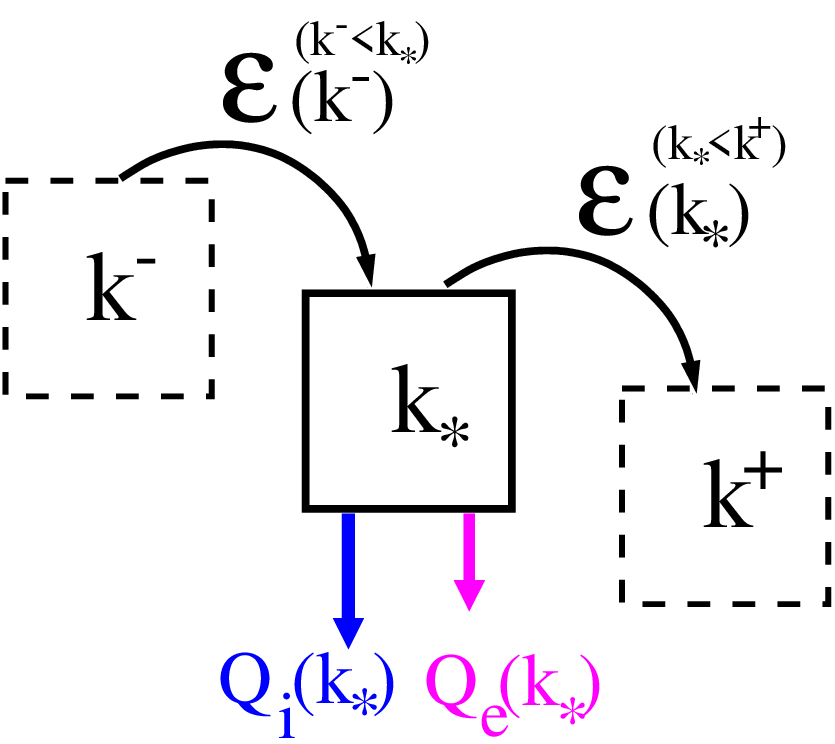}}
\end{center}
  \caption{Diagram of the balance of the energy transfer rates in a steady-state turbulent cascade at a local wavenumber $k_*$, where the following terms sum to zero:
    (i)  the turbulent nonlinear energy transfer rate from lower wavenumbers  $\varepsilon(k^-)^{(k^-<k_*)}$;  (ii)  the turbulent nonlinear energy transfer rate to higher wavenumbers $\varepsilon(k_*)^{(k_*<k^+)}$; and (iii) the local (in scale) kinetic damping rates of the turbulent energy, due to the sum of ion energization rate 
$Q_i(k_*)$ (blue) and  electron energization rate 
$Q_e(k_*)$ (magenta).
\label{fig:balance}}
\end{figure}

Competing with that turbulent cascade rate at each scale is the
damping rate due to collisionless interactions between the
electromagnetic fields and the plasma particles, as shown by the blue
arrows for the transfer of energy to ions and the magenta arrows for
the transfer of energy to electrons in \figref{fig:cascade}(a).  The
net damping rate of the entire turbulent cascade, resulting from the
sum of the energization rates for each particle species $s$, can be
computed by integration over all scales of the turbulent cascade.  In
a steady-state, at each wavenumber $k_*$, the sum of three terms must
balance to zero, as illustrated in \figref{fig:balance}: (i) the
energy transfer from lower wavenumbers $k^-<k_*$ (larger scales),
denoted by $\varepsilon(k^-)^{(k^-<k_*)}$; (ii) the energy transfer to
higher wavenumbers $k^+>k_*$ (smaller scales), denoted by
$\varepsilon(k_*)^{(k_*<k^+)}$; and (iii) the kinetic damping of the
fluctuations at wavenumber $k_*$, given by $Q(k_*)$, which is the sum of the local (in scale) ion
energization rate  $Q_i(k_*)$ (blue) and electron energization rate $Q_e(k_*)$ (magenta).
Complicating the determination of the partitioning of damped
turbulent energy between ions and electrons $Q_i/Q_e$ is the fact that the kinetic
damping rates due to ions and electrons may overlap as a function of
scale \citep{Howes:2011a,Howes:2011b,Told:2015}.

\begin{figure}
  \begin{center}
    \resizebox{\textwidth}{!}{\includegraphics*[20pt,185pt][566pt,746pt]{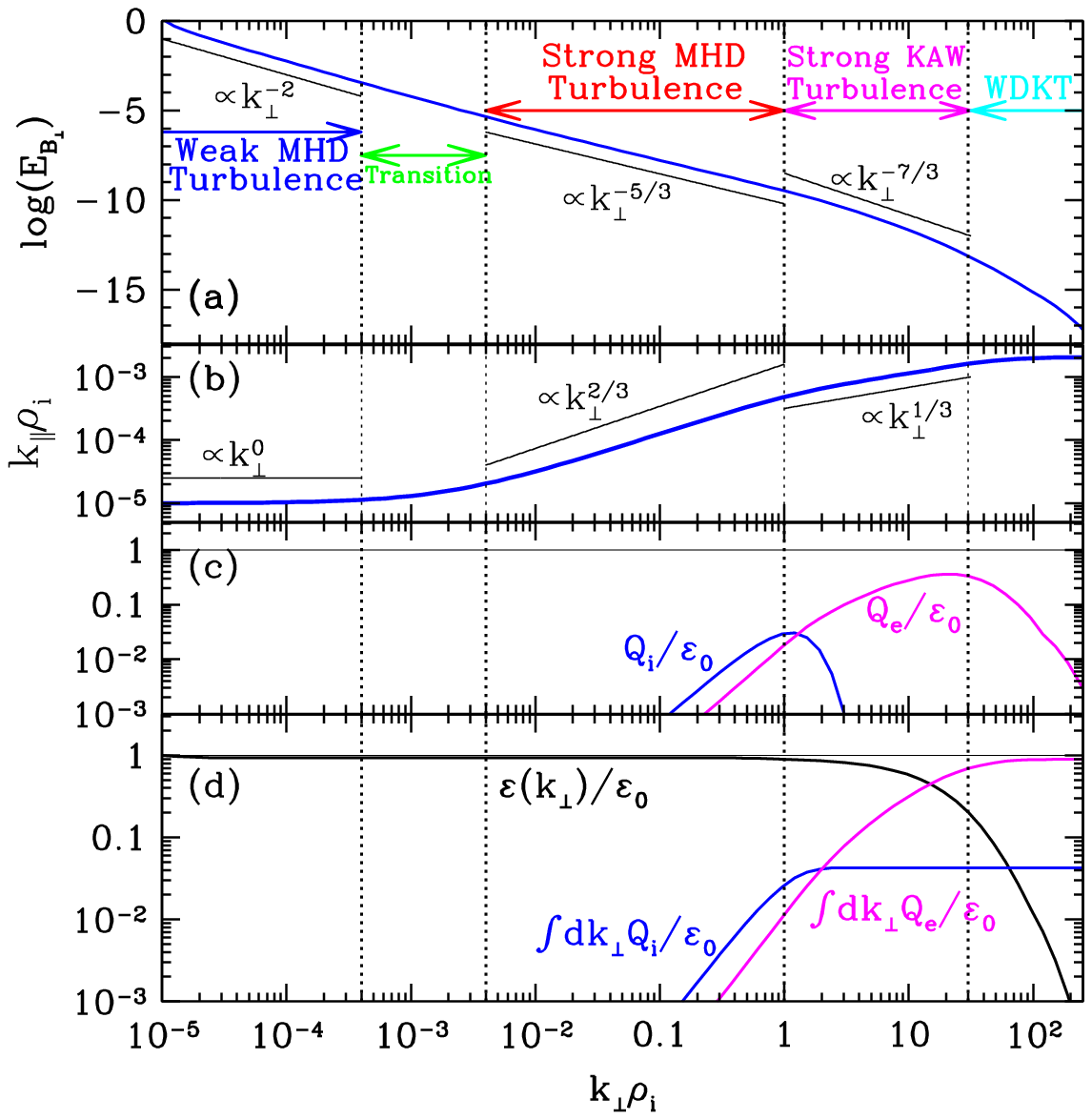}}
\end{center}
  \caption{Weakened cascade model results for a turbulent cascade
    driven weakly at $k_{\perp 0} \rho_i=k_{\parallel 0} \rho_i=10^{-5}$
    with nonlinearity parameter $\chi(k_{\perp 0})=0.1$ in a plasma
    with isotropic equilibrium Maxwellian velocity distributions with
    $T_i/T_e=9$ and $\beta_i=1$: (a) One-dimensional perpendicular
    magnetic energy spectrum $E_{B_\perp}(k_\perp)$, (b) anisotropic
    cascade through wave vector space given by ,
    $k_\parallel(k_\perp)$, (c) energy transfer rates to the ions
    $Q_i(k_\perp)$ (blue) and electrons $Q_e(k_\perp)$ (magenta), and
    (d) integrated energy damping rates by ions $\int_0^{k_\perp} d
    k'_\perp Q_i(k'_\perp)$ (blue) and by electrons $\int_0^{k_\perp}
    d k'_\perp Q_e(k'_\perp)$ (magenta), all plotted as a function of
    the perpendicular wavenumber $k_\perp \rho_i$.  The ranges of $k_\perp \rho_i$ of the full turbulent cascade include Weak MHD Turbulence (blue), the Transition from Weak to Strong MHD Turbulence (green), Strong MHD Turbulence (red), Strong Kinetic \Alfven Wave (KAW) Turbuence (magenta), and Weakly Dissipating KAW Turbulence (WKDT) (cyan) \citep{Howes:2011b}.
\label{fig:damping}}
\end{figure}

Determining this balance of cascade and damping rates as a function of
scale, and ultimately using the terms in that balance to determine the
partitioning of turbulent energy among species and degrees of
freedom---parameterized by $Q_i/Q_e$, $Q_{\perp,i}/Q_{i}$, and
$Q_{\perp,e}/Q_{e}$---has been attempted in empirical turbulent
cascade models
\citep{Pao:1965,Howes:2008b,Podesta:2010a,Howes:2011b,Zhao:2013,Schreiner:2017}.
An example of the modeling of the turbulent energy cascade and its
kinetic dissipation is presented in \figref{fig:damping} from the
results of the \emph{weakened cascade model} by
\citet{Howes:2011b}. In addition to balancing the physics of the local
cascade rate $\varepsilon(k_\perp)$ with the kinetic damping rates due
to ions $Q_i(k_\perp)$ and electrons $Q_e(k_\perp)$ via the Landau
resonance, this cascade model also includes the contributions to the
local cascade rate from nonlocal interactions by large-scale shearing
and small-scale diffusion.  The model results shown here in
\figref{fig:damping} are from the same calculation as presented in
Fig.~8 of \citet{Howes:2011b}, where we reproduce (a) the
one-dimensional perpendicular magnetic energy spectrum $E_{B_\perp}
(k_\perp)$ and (b) the anisotropic cascade through wave vector space
characterized by the characteristic parallel wavenumber as a function
of the perpendicular wavenumber, $k_\parallel(k_\perp)$.  This model
for a plasma with $\beta_i=1$ and isotropic equilibrium Maxwellian
velocity distributions with $T_i/T_e=9$ is driven weakly with
$\chi(k_{\perp 0})=0.1$ at $k_{\perp 0} \rho_i=k_\parallel
\rho_i=10^{-5}$, taking the model Kolmogorov constants to be $C_1=1.4$
and $C_2=1.0$.  Note that an important avenue of research is to use
kinetic numerical simulations or spacecraft observations to constrain
the values of these Kolmogorov constants, as recently done by
\citet{Shankarappa:2023} using observations in the inner heliosphere
by the \emph{Parker Solar Probe} \citep{Fox:2016}.

In \figref{fig:damping}(c) is plotted the rate of energy density transfer to
the ions $Q_i(k_\perp)$ (blue) and to the electrons $Q_e(k_\perp)$
(magenta) as a function of $k_\perp \rho_i$, both normalized by the
turbulent energy cascade rate at the driving scale $\varepsilon_0$.  This
shows that the ion kinetic damping occurs at the ion kinetic scales
$k_\perp\rho_i \sim 1$, whereas the electron damping rises
monotonically from the ion scales, peaking around $k_\perp\rho_i \sim
30$ for this case. To obtain the net partitioning of energy between
ions and electrons $Q_i/Q_e$, it is necessary to integrate these
energy damping rates over the entire cascade, as shown in
\figref{fig:damping}(d): the turbulent energy cascade rate
$\varepsilon(k_\perp)$ (black) is ultimately terminated at
$k_\perp\rho_i \gg 1$ by the integrated energy damping rates by ions
$\int_0^{k_\perp} d k'_\perp Q_i(k'_\perp)$ (blue) and by electrons
$\int_0^{k_\perp} d k'_\perp Q_e(k'_\perp)$ (magenta).  Note that this
turbulent cascade model determines the solutions presented in
\figref{fig:damping} by computing the balance of the turbulent cascade
rates with the local kinetic damping rates, as depicted in
\figref{fig:balance}.  Integration of those damping rates over the
entire cascade yields the predicted partitioning of dissipated
turbulent energy between ions and electrons $Q_i/Q_e$.

The energy density transfer rates $Q_i(k_\perp)$ and $Q_e(k_\perp)$
that remove energy from the turbulent fluctuations and transfer it to
the plasma particles in turbulent cascade models require as input the
kinetic damping rates due to ions $\gamma_i(k_\perp)$ and electrons
$\gamma_e(k_\perp)$.  In the case of the weakened cascade model shown
in \figref{fig:damping} \citep{Howes:2011b}, the damping rates used
are collisionless damping rates via the Landau ($n=0$) resonance for
the \Alfven wave mode (and its extension as a kinetic \Alfven wave at
scales $k_\perp\rho_i \gtrsim 1$) from the Vlasov-Maxwell linear
dispersion relation \citep{Stix:1992}.  Previous studies have called
into question whether resonant collisionless wave-particle
interactions, such as Landau damping, can effectively damp turbulent
fluctuations in the presence of the strong nonlinear interactions that
mediate the turbulent cascade \citep{Plunk:2013,Schekochihin:2016}.
Using the field-particle correlation technique
\citep{Klein:2016a,Howes:2017a,Klein:2017b}, several recent analyses
of kinetic simulations \citep{Klein:2017b,Howes:2018a,Klein:2020} and
spacecraft observations \citep{Chen:2019,Afshari:2021} of plasma
turbulence find clear velocity-space signatures of Landau damping.
These findings indicate indeed that resonant wave-particle
interactions do play a role, and possibly a dominant role
\citep{Afshari:2021}, in the damping of space and astrophysical plasma
turbulence, settling the question of whether Landau damping can
effectively damp strong plasma turbulence\footnote{Note that recent
work found that an anti-phase-mixing process can inhibit the process
of the linear phase mixing that transfers nonthermal internal energy
to small parallel velocity-space scales, thereby shutting down the
effect of Landau damping of compressible turbulent fluctuations within
the inertial range under sufficiently collisionless conditions
\citep{Schekochihin:2016,Parker:2016,Meyrand:2019}.}.

In quantifying how a particular damping mechanism competes with the
nonlinear turbulent cascade, the key dimensionless quantity to
evaluate is the ratio of linear damping rate due to nonlinear cascade
rate $\gamma/\omega_{nl}$ \citep{Howes:2008b,Howes:2011b}. Adopting
the conjecture of critical balance between linear and nonlinear
timescales in strong plasma turbulence $\omega \sim \omega_{nl}$
\citep{Goldreich:1995,Howes:2008b,Howes:2011b,Mallet:2015}, the key
dimensionless measure of the importance of any particular mechanism
is the ratio $\gamma/\omega$, which can be estimated using the linear
theory for each of the different damping mechanisms.

\subsection{The Relevance of Linear Physics to Strong Plasma Turbulence}
The concept of critical balance in plasma turbulence
\citep{Goldreich:1995,Howes:2008b,Howes:2011b,Mallet:2015}---which
suggests that the nonlinear timescales of the turbulent cascade remain
in approximate balance with the timescales of the linear wave physics,
$\omega_{nl} \sim \omega$---implies that the physics of the linear
plasma response remains valid even in strong plasma turbulence.  But,
perhaps in analogy with the case of incompressible hydrodynamic
turbulence in which no linear response exists \citep{Howes:2015b}, the
validity of linear physics has been called into question in the
presence of strong plasma turbulence \citep{Matthaeus:2014}.  Although
there exist numerous counterexamples in which linear physics
properties have been shown to be relevant to strong plasma turbulence
\citep{Maron:2001,Cho:2003,Alexandrova:2008,Howes:2008a,Howes:2008b,Svidzinski:2009,Sahraoui:2010b,Hunana:2011,Howes:2011a,TenBarge:2012b,Chen:2013a},
the question has persisted. A more recent careful study of kinetic
numerical simulations and spacecraft observations has shown clearly
that both small-amplitude and large-amplitude fluctuations, including
variations interpreted to be coherent structures, preserve the linear
properties of \Alfvenic fluctuations \citep{Groselj:2019}, hopefully
providing sufficient evidence to establish definitively the relevance
of linear physics to strong plasma turbulence.

The key to predicting qualitatively which kinetic damping mechanisms
are likely to dominate for a given instance of turbulence is to
determine how the various collisionless damping mechanisms in
\secref{sec:proposed} depend on the fundamental dimensionless plasma
and turbulence parameters.  For a fully ionized proton and electron
plasma with anisotropic (bi-Maxwellian) equilibrium velocity
distribution functions
\citep{Stix:1992,Quataert:1998,Quataert:1999,Swanson:2003,Klein:2015a},
the complex eigenfrequencies determined by the linear Vlasov-Maxwell
dispersion relation depend on the dimensionless parameters that
characterize the nature of the turbulent fluctuations through the
typical wave vector components in a plane-wave decomposition of the
turbulent fluctuations $k_\perp \rho_i$ and $k_\parallel \rho_i$ and
the plasma parameters through the parallel ion plasma beta $\beta_{\parallel,i}$, the
parallel ion-to-electron temperature ratio $\tau_\parallel=T_{\parallel,
  i}/T_{\parallel, e}$, the temperature anisotropies of each species
$A_i=T_{\perp,i}/T_{\parallel,i}$ and
$A_e=T_{\perp,e}/T_{\parallel,e}$, and the ratio $v_{t\parallel,i}/c$.  Thus,
these dependencies can be expressed through the general relation
\begin{equation}
 \omega =\omega_{VM}^{aniso}(k_\perp \rho_i, k_\parallel \rho_i,
 \beta_{\parallel,i},\tau_\parallel,A_i,A_e,v_{ti}/c).
 \label{eq:vm_aniso}
\end{equation}
In the limit of isotropic equilibrium velocity distributions for the
ions and electrons---leading to the simplifications $A_i=1$, $A_e=1$, and $\tau_\parallel= \tau \equiv T_{i}/T_{e}$---this
set of dependencies is reduced to
\begin{equation}
 \omega =\omega_{VM}(k_\perp \rho_i, k_\parallel \rho_i,
 \beta_i,\tau,v_{ti}/c).
 \label{eq:vm_iso}
\end{equation}
For both the isotropic and anisotropic equilibrium temperatures, in
the non-relativistic limit  $v_{ti}/c \ll 1$ typical of many space and astrophysical
plasmas of interest, the linear Vlasov-Maxwell
dispersion relation is practically independent of $v_{ti}/c$, so this parameter
may effectively be dropped for the case of non-relativistic plasma turbulence.

In the limit of a sufficiently large inertial range of turbulence, the
MHD turbulence scaling parameters given by
\eqref{eq:kpar}--\eqref{eq:bperp} predict that turbulent fluctuations at
the small-scale end of the inertial range---where most of the kinetic damping
mechanisms are believed to become significant---will be anisotropic
with $k_\parallel \ll k_\perp$ and small amplitude relative to the
equilibrium magnetic field $|\delta \V{B}| \ll B_0$
\citep{Howes:2006,Howes:2008b,Schekochihin:2009}.  In this limit, the
turbulent frequencies of the incompressible \Alfvenic fluctuations
directly observed to dominate turbulence in space plasmas
\citep{Belcher:1971,Tu:1995,Alexandrova:2008,Howes:2012a,Bruno:2013}
have frequencies much smaller than the ion cyclotron frequency,
$\omega/\Omega_i = k_\parallel d_i \ll 1$, where
$d_i=v_A/\Omega_i=\rho_i/\sqrt{\beta_i}$ is the ion inertial length.
In this limit, the turbulent fluctuations at the small-scale end of
the inertial range satisfy the conditions for the applicability of
gyrokinetic theory
\citep{Rutherford:1968,Taylor:1968,Catto:1978,Antonsen:1980,Catto:1981,Frieman:1982,Dubin:1983,Hahm:1988,Brizard:1992,Howes:2006,Brizard:2007,Schekochihin:2009}.
Gyrokinetic theory can be rigorously derived from plasma kinetic
theory under the limits of low frequency ($\omega/\Omega_i \ll 1$) and 
anisotropic ($k_\parallel \ll k_\perp$) fluctuations in the
non-relativistic limit ($v_{ti}/c \ll 1$). In the gyrokinetic
approximation, the physics of the the cyclotron resonances and of the
fast magnetosonic fluctuations---and their kinetic extension to
whistler fluctuations \citep{Howes:2014a}---are ordered out of the
system.  Gyrokinetics retains the physics of the $\V{E} \times \V{B}$
nonlinearity that mediates the turbulent cascade to small scales, full
finite-Larmor-radius effects, and the collisionless damping associated
with the $n=0$ Landau resonance.  In this reduced limit, the complex
eigenfrequencies of the linear gyrokinetic dispersion relation for
isotropic equilibrium velocity distributions are linearly dependent on
the parallel wavenumber, and have the significantly reduced
dimensionless parameter dependency given by \citep{Howes:2006}
\begin{equation}
  \overline{\omega} \equiv \frac{\omega}{k_\parallel v_A}
  =\overline{\omega}_{GK}(k_\perp \rho_i,   \beta_i,\tau).
 \label{eq:gk}
\end{equation}

Although the ion-to-electron mass ratio $\mu=m_i/m_e$ is not a
parameter that varies in the physical world, many numerical studies of
plasma turbulence choose to employ a reduced mass ratio for
computational efficiency.  Therefore, we choose to include the $\mu$
dependence of different proposed turbulent damping mechanisms in our
calculations below to determine how the use of a reduced mass ratio
may impact the partitioning of turbulent energy removed by different
damping mechanisms. Note also that it is important to employ a
sufficiently large mass ratio to achieve a separation between the ion
and electron scales, with a previous study suggesting $m_i/m_e\ge 32$
is necessary to model with qualitative fidelity the distinct ion and
electron responses to the electromagnetic fluctuations due to this
scale separation \citep{Howes:2018a}.


\subsection{Landau Damping and Transit-Time Damping}
\label{sec:landaures}
To explore the two Landau-resonant collisionless damping
mechansims---Landau damping and transit-time damping---we take $n=0$
in the resonance condition in \eqref{eq:res_denom}, obtaining the
condition that particles with parallel velocities equal to the
parallel phase velocity of the waves, $v_\parallel =
\omega/k_\parallel$, may resonantly exchange energy with the
electromagnetic waves.

For simplicity in this analysis, we will assume the isotropic
temperature case for a proton-electron plasma with linear wave
properties given by the isotropic linear Vlasov-Maxwell dispersion
relation \eqref{eq:vm_iso}; extending these calculations to allow for
anisotropic temperatures can be done simply following the same logical
procedure and using the linear Vlasov-Maxwell dispersion relation for
anisotropic temperatures given by \eqref{eq:vm_aniso}.  Furthermore,
justified by observations in solar wind turbulence that incompessible
\Alfvenic fluctuations energetically dominate the turbulence
\citep{Belcher:1971,Tu:1995,Alexandrova:2008,Bruno:2013}, we will
focus on the linear collisionless damping rates of the \Alfvenic
fluctautions, including both \Alfven waves in the MHD limit at
$k_\perp \rho_i \ll 1$, and kinetic \Alfven waves at $k_\perp \rho_i
\gtrsim 1$. Consideration of the damping of compressible turbulent
fluctuations, specifically fast and slow magnetosonic fluctuations at
$k_\perp \rho_i \ll 1$ \citep{Howes:2012a}, is easily performed using
the same procedure with the linear dispersion relations for these
alternative wave modes\footnote{Although, it is important to keep in mind
that the recently identified nonlinear effect of anti-phase-mixing in
turbulent plasmas can suppress the collisionless damping predicted by
linear kinetic theory for the compressible wave modes in the inertial
range of scales at $k_\perp \rho_i \ll 1$
\citep{Schekochihin:2016,Parker:2016,Meyrand:2019}.}.

Normalizing the resonant parallel velocity by the thermal velocity for
both species in turn and converting the result into our dimensionless
parameters for the isotropic temperature case, we obtain the key
relations
\begin{equation}
  \frac{v_\parallel}{v_{ti}} =  \frac{\omega}{k_\parallel v_{ti}} = \overline{\omega}\beta_i^{-1/2}, \quad \quad  \quad \quad 
  \frac{v_\parallel}{v_{te}} =  \frac{\omega}{k_\parallel v_{te}} = \overline{\omega} \left(\frac{\tau}{\beta_i \mu}\right)^{1/2},
  \label{eq:ldresvel}
\end{equation}
where we define the  dimensionless  wave frequency normalized the the \Alfven wave frequency in the MHD limit,
\begin{equation}
  \overline{\omega} \equiv \frac{\omega}{k_\parallel v_A}
  \label{eq:ombar}
\end{equation}
For \Alfven waves, this normalized frequency
$\overline{\omega}$ is well approximated by \citep{Howes:2014a}
\begin{equation}
  \overline{\omega}(k_\perp\rho_i,\beta_i, \tau)  =
  \sqrt{ 1 + \frac{(k_\perp\rho_i)^2}{\beta_i + 2/(1+1/\tau)}},
  \label{eq:LDres}
\end{equation}
an expression that is valid in the limits that the ion plasma beta is
not very low $\beta_i \gg \mu^{-1}$ and that the wave frequency
remains smaller than the ion cyclotron frequency $\omega /\Omega_i \ll
1$, equivalent to the condition $k_\parallel d_i \ll 1$, or
$k_\parallel \rho_i \ll \beta_i^{1/2}$.

\begin{figure}
  \begin{center}
    \resizebox{4.0in}{!}{\includegraphics*[18pt,308pt][566pt,700pt]{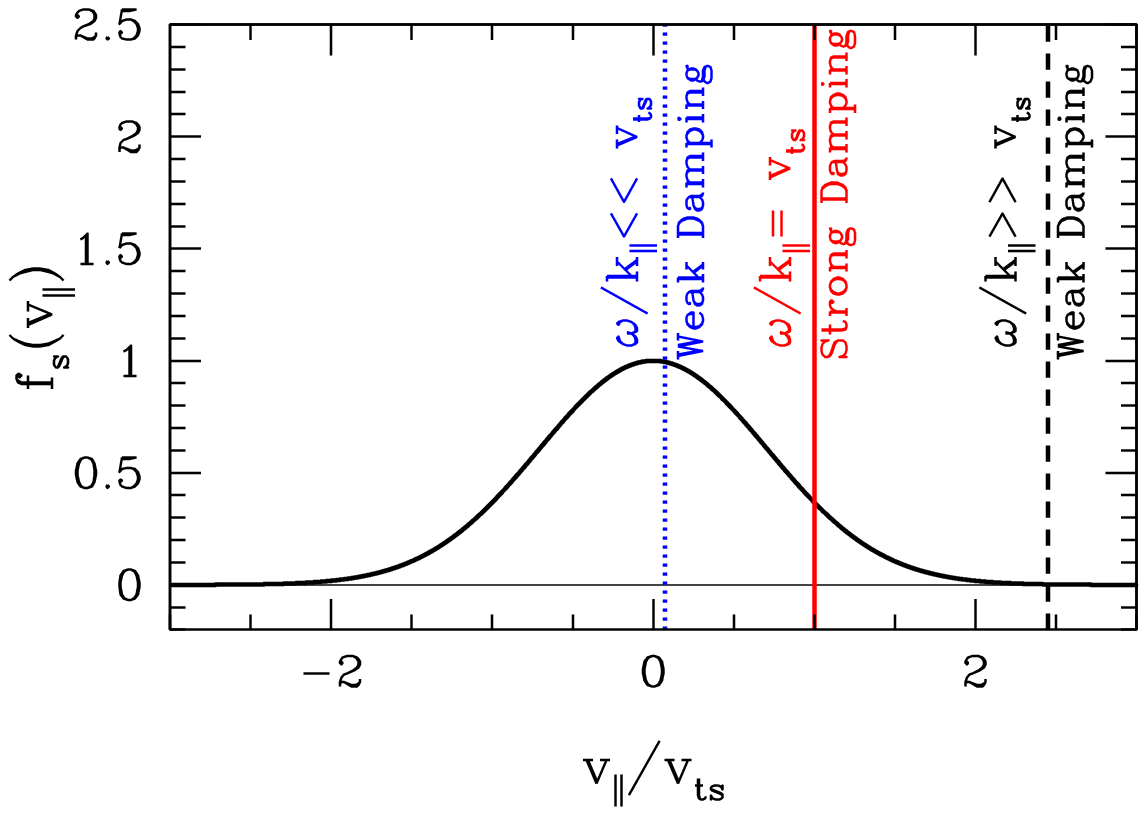}}
\end{center}
  \caption{Diagram demonstrating how Landau resonant damping qualitatively depends on where the parallel phase velocity $\omega/k_\parallel$ of the wave falls within the velocity distribution of particle species $s$, with weak damping for $\omega/k_\parallel \ll v_{ts}$ or  $\omega/k_\parallel \gg v_{ts}$, and strong damping for  $\omega/k_\parallel \sim v_{ts}$.
\label{fig:landau resonant}}
\end{figure}

Whether damping by the Landau-resonant mechanisms is weak or strong
primarily depends on where the parallel wave phase velocity falls
within the equilibrium particle velocity distribution, as quantified
by the expressions in \eqref{eq:ldresvel}.  The application of the
weak-growth-rate approximation to the kinetic theory for Landau
damping \citep{Krall:1973} indicates that the linear damping rate is
proportional to the slope of the distribution at the resonant
velocity, as depicted in \figref{fig:landau resonant}.  For phase
velocities that fall deep in the core of the velocity distribution
$\omega/k_\parallel \ll v_{ts}$ (blue dotted) where the slope of
$f_s(v_\parallel)$ is relatively flat, the damping is weak because
nearly equal numbers of particles with $v_\parallel <
\omega/k_\parallel$ are accelerated by the wave as particles with
$v_\parallel > \omega/k_\parallel$ are decelerated by the wave,
leading to little net energy exchange.  Similarly, for phase
velocities far out in the tail of the velocity distribution
$\omega/k_\parallel \gg v_{ts}$ (black dashed), there are few
particles to resonate with the waves and the slope of
$f_s(v_\parallel)$ is also quite flat, so the damping is weak.  Only
if the parallel phase velocity falls in the steep region of the
velocity distribution at $\omega/k_\parallel \sim v_{ts}$ (red solid)
is there a significant net energy transfer to the particles: more
particles with $v_\parallel < \omega/k_\parallel$ are accelerated by
the wave than those particles with $v_\parallel > \omega/k_\parallel$
are decelerated by the wave, so that particles gain net energy at the
expense of the wave.  Note that this qualitative picture of
collisionless damping by the Landau resonance being controlled by
where the parallel wave phase velocity falls within the distribution
function is independent of the form of the equilibrium velocity
distribution; thus, one can apply the same reasoning to estimate the
strength of the Landau-resonant damping for other non-Maxwellian forms
of the equilibrium velocity distribution, such as the kappa
distributions often applied to model \emph{in situ} measurements in
space plasmas
\citep{Vasyliunas:1968,Abraham-Shrauner:1977,Gosling:1981,Lui:1981,Armstrong:1983,Summers:1991,Thorne:1991,Summers:1994,Livadiotis:2013,Livadiotis:2018}. 

The ultimate goal is to characterize the turbulent plasma heating---
specifically $Q_i/Q_e$, $Q_{\perp,i}/Q_{i}$,and
$Q_{\perp,e}/Q_{e}$---due to Landau damping and transit-time damping
as a function of the fundamental dimensionless parameters of
turbulence.  To accomplish this goal, it is necessary to integrate the
damping rates over the full range of the turbulent cascade, as
discussed in \secref{sec:competition} and illustrated in
Figures~\ref{fig:balance} and~\ref{fig:damping}.  This integration
over all scales effectively eliminates the dependence of the damping
mechanisms on the components of the wavevector $k_\perp \rho_i$ and
$k_\parallel \rho_i$, leaving only the desired dependence on the
plasma and turbulence parameters enumerated in Tables~\ref{tab:params}
and~\ref{tab:reduced_params}.

With the scale dependence on $k_\perp \rho_i$ eliminated, the ion
Landau damping (iLD) and ion transit-time damping (iTTD) rates depend
primarily on $\beta_i$, and the electron Landau damping (eLD) and
electron transit-time damping (eTTD) rates depend on $\beta_i$,
$\tau$, and $\mu$, as shown by \eqref{eq:ldresvel}.  Note that the
dependence of eLD and eTTD on $\tau$ and $\mu$ primarily arises from
those two parameters controlling the scale separation of the ion and
electron Larmor radius scales, $\rho_i/\rho_e = (\tau \mu)^{1/2}$.

It is worthwhile pointing out that the physics of collisionless
damping via the Landau resonance is fully captured in the gyrokinetic
approximation, with the linear wave frequencies and damping rates both
linearly proportional to $k_\parallel$, as shown in \eqref{eq:gk}.
Thus, the ratio of the damping-to-cascade rate\footnote{Here we have
adopted the condition of critical balance for strong plasma
turbulence $\omega_{nl} \simeq\omega$
\citep{Goldreich:1995,Mallet:2015}.}, $\gamma/\omega_{nl} \simeq
\gamma/\omega$, is formally independent of $k_\parallel$, leaving the
expected parameter dependence on the plasma parameters $\beta_i$ and
$\tau$, along with the mass ratio $\mu$, once the cascade is
integrated over the perpendicular scales of the cascade.

In addition to the fact that the dependence of the linear gyrokinetic dispersion
relation on $(k_\perp \rho_i, \beta_i,\tau)$ in \eqref{eq:gk} is
consistent with the results of our analysis of the parameter
dependence of the Landau-resonant damping mechanisms, the linear
gyrokinetic dispersion relation also determines the amplitudes and
phases of the key electromagnetic fields $E_\parallel$ and $\delta
B_\parallel$ that mediate the particle energization by Landau damping
and transit-time damping.  As done in the \emph{H10} turbulent heating
model (see \secref{sec:review}), the linear dispersion relation can
simply be used to calculate directly the linear collisionless damping
rates as a function of $(k_\perp \rho_i, \beta_i,\tau)$. It is
possible, of course, that these linear damping rates may be modified
under strongly turbulent plasma conditions, but modest quantitative
changes to the effective damping rates are typically absorbed into the
adjustable constant typically used in cascade models to specify the
balance between the nonlinear cascade rate and collisionless damping
rate \citep{Howes:2008b,Howes:2011b}. However, field-particle
correlation analyses of numerical simulations
\citep{Klein:2017b,Klein:2020,Horvath:2020,Conley:2023,Huang:2024} and
spacecraft observations \citep{Chen:2019,Afshari:2021} of plasma
turbulence have clearly shown that Landau damping and transit-time
damping do indeed play a role in the damping of strong plasma
turbulence.

\begin{figure}
  \begin{center}
    \resizebox{\textwidth}{!}{\includegraphics*[18pt,288pt][576pt,690pt]{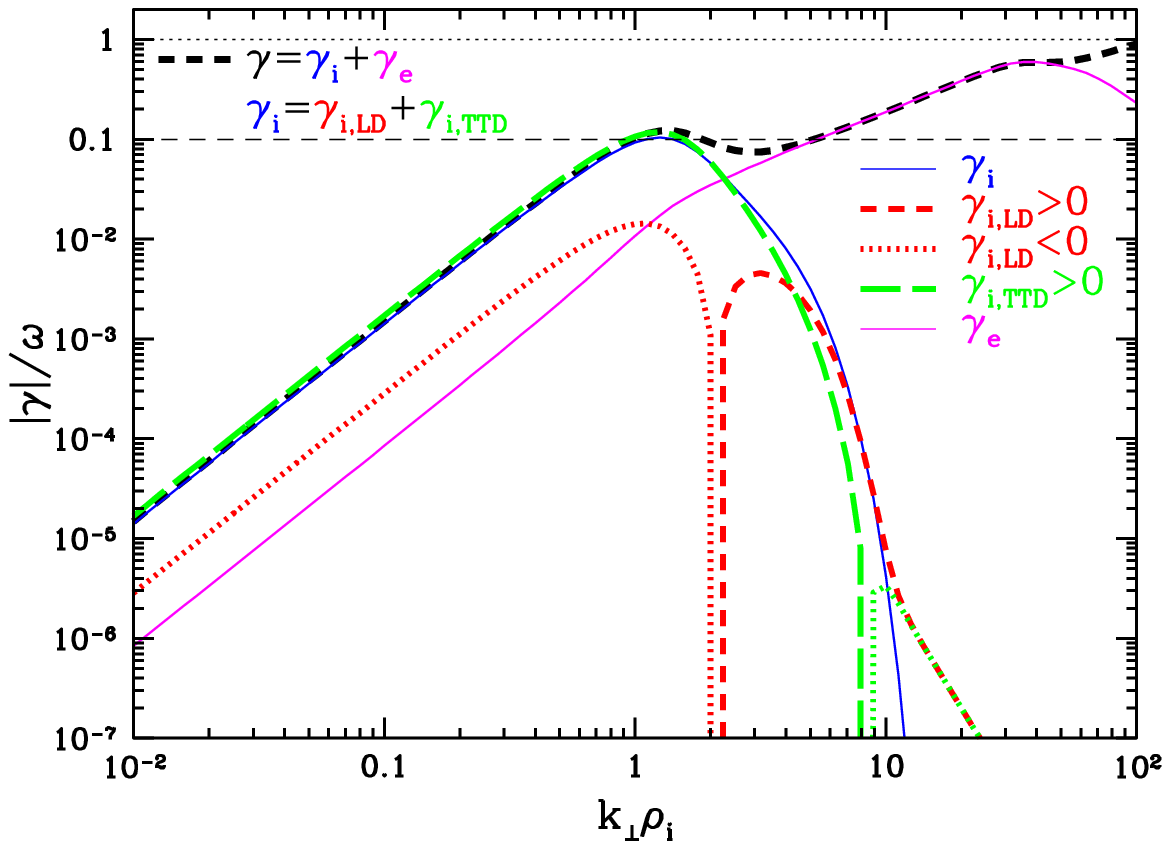}}
\end{center}
  \caption{Plot of the normalized damping rate $\gamma/\omega$ (black
    dashed) for the \Alfven and kinetic \Alfven wave over the
    perpendicular wavenumber range $10^{-2} \le k_\perp \rho_i \le
    10^2$ for a plasma with $\beta_i=3$, $T_i/T_e=1$, and
    $m_i/m_e=1836$, where the complex eigenfrequency is given by
    $\omega_c=\omega-i\gamma$, so that $\gamma>0$ corresponds to wave
    damping and $\gamma<0$ to wave growth. The contributions from the
    ion damping rate $\gamma_i$ (thin blue) and electron damping rate
    $\gamma_e$ (thin magenta) are separately plotted, with the
    decomposition by species failing at strong damping rates with $\gamma/\omega
    \gtrsim 0.5$, occurring for $ k_\perp \rho_i \gtrsim 40$.  Ion
    damping due to Landau damping $\gamma_{i,LD}>0$ (red dashed) and
    transit-time damping $\gamma_{i,TTD}>0$ (green long dashed) are
    separated, showing a region of net energy transfer \emph{from}
    ions \emph{to} the wave  by the Landau resonance with the ions  yielding
    $\gamma_{i,LD}<0$ (red dotted) at $k_\perp \rho_i \le 2$.
\label{fig:ld_ttd_disp}}
\end{figure}

One final step is to estimate how the ion-to-electron heating ratio
$Q_i/Q_e$ depends on the remaining key parameters $\beta_i$, $\tau$,
and $\mu$.  As illustrated in \figref{fig:damping}(c) and
\figref{fig:ld_ttd_disp}, the ion damping rate $\gamma_i$ peaks at
scales $k_\perp \rho_i \sim 1$, and numerical solutions of the
Vlasov-Maxwell dispersion relation (not presented here) show that this
Landau-resonant damping onto the ions is essentially independent of
the ion-to-electron temperature ratio.  The electron damping rate
$\gamma_e$, on the other hand, increases monotonically as the
perpendicular wavevector increases to the electron Larmor radius scale
$k_\perp \rho_e \rightarrow 1$, eventually yielding a normalized
collisionless damping rate that is sufficiently strong to terminate
the turbulent cascade with $\gamma_e/\omega \rightarrow 1$, as shown
in \figref{fig:ld_ttd_disp}.  Thus, any energy in the turbulent
cascade that passes beyond the ion kinetic scales at $k_\perp \rho_i
\sim 1$ will ultimately be transferred to the electrons.  Since the
ion damping rate depends dominantly on $\beta_i$ with very weak
dependence on $\tau$, yielding $Q_i(\beta_i)$ to lowest order, this
means that the ratio $Q_i/Q_e$ is effectively determined by the ion
plasma beta alone.  This simplification arises because the total
turbulent damping rate must eventually balance the total cascade rate,
$Q =Q_i+ Q_e \sim \varepsilon_0$, so it can be shown that $Q_e/Q_i=
\varepsilon_0/Q_i(\beta_i) - 1 $.  This dominant dependence of the
Landau-resonant collisionless damping mechanisms on $\beta_i$ is
consistent with existing turbulent heating models for \Alfvenic
turbulence \citep{Howes:2010d,Kawazura:2019}.

One final unexpected physical behavior of Landau damping and
transit-time damping is also illustrated in \figref{fig:ld_ttd_disp}.
For isotropic Maxwellian velocity distributions of the ions and
electrons, the imaginary component of the complex frequency, $\omega_c
\equiv \omega-i \gamma$, determined by the linear dispersion relation
is always negative (solid black curve), corresponding to a damping of
the wave with damping rate $\gamma >0$.  However, for the particular
plasma parameter choices in \figref{fig:ld_ttd_disp} of $\beta_i=3$,
$T_i/T_e=1$, and $m_i/m_e=1836$, the individual energy transfer rates
to the ions through Landau damping\footnote{The separation of the
contributions of Landau damping and transit-time to the total
collisionless damping rate for the Vlasov-Maxwell dispersion relation
is detailed in \citet{Huang:2024}.}  mediated by $E_\parallel$ are
actually negative over the range $k_\perp \rho_i \lesssim 2$ (red
dotted), indicating a net transfer of energy from the ions to the wave
fields when averaged over the full period (or, alternatively, over
$2\pi$ in phase) of the \Alfven wave.  But the contribution from
transit-time damping mediated by $E_\perp$
\citep{Huang:2024,Howes:2024b} is positive and much larger (green long
dashed), leading to a net damping of the wave ($\gamma_i >0$, thin
blue) by the sum of the Landau-resonant ($n=0$) interactions of ions
with the \Alfven wave.  Whether the wave-period-averaged energy
transfer to ions is positive or negative depends on the relative
phases of $E_\parallel$ and $E_\perp$ to the self-consistently
determined components of the ion current density $\V{j}_i$.  For the
case of $\beta_i=3$ pictured in \figref{fig:ld_ttd_disp}, at scales
$k_\perp \rho_i \lesssim 2$ (red dotted), this relative phase $\delta
\phi$ between $E_\parallel$ and $j_{\parallel,i}$ must fall in the
range $\pi/2 \le \delta \phi\le 3 \pi/2$ (corresponding to wave growth
at the expense of ion energy), while the relative phase between
$E_\perp$ and $j_{\perp,i}$ must fall in the range $-\pi/2 \le \delta
\phi\le \pi/2$ (corresponding to wave damping and ion energy gain).
Because each individual mechanism can possibly lead to negative energy
transfer while the sum always leads to damping for an isotropic
Maxwellian distribution, it suggests that perhaps a separation of the
particle energization rates by Landau damping and transit-time damping
is not particularly important in unraveling the turbulent heating of
the plasma (especially since both of these mechanisms ultimately lead
to energization of the parallel degree of freedom of the particles),
so that only the summed effect of the Landau-resonant collisionless
damping rates is important.

Summarizing our results for Landau-resonant collisionless damping of
the turbulent cascade, the resulting ion-to-electron heating is
predicted to be dominantly a function of the ion plasma beta,
$Q_i/Q_e(\beta_i)$, with a weak dependence on the ion-to-electron
temperature ratio $\tau$ and mass ratio $\mu$.  Both Landau damping,
mediated by the parallel component of the electric field
$E_\parallel$, and transit-time damping, mediated by the perpendicular
component of the electric field $E_\perp$, lead to energization of the
degrees of freedom parallel to the magnetic field, so
$Q_{\perp,i}/Q_{i}=0$ and $Q_{\perp,e}/Q_{e}=0$.

\subsection{Cyclotron Damping }
\label{sec:icd}
For the fundamental ($n=1$) cyclotron resonance, the resonance
condition in \eqref{eq:res_denom} simplifies to $(\omega-k_\parallel
v_\parallel)/\Omega_s =1$.  The numerator involves the fluctuation
frequency modified by the Doppler shift of the particle's motion
parallel to the magnetic field. When that Doppler-shifted frequency
matches the ion cyclotron frequency, the perpendicular component of the
electric field $E_\perp$ can resonantly interact with the particles to
energize the particles in the directions perpendicular to the magnetic
field, yielding $Q_{\perp,s}/Q_s=1$.  To estimate the conditions
required for cyclotron resonant damping, it is often sufficient to
require the condition $\omega/\Omega_s \rightarrow 1$.

In typical space and astrophysical plasmas, the turbulent cascade can
reach the ion cyclotron frequency $\omega/\Omega_i \rightarrow 1$
under the conditions derived below, but since the electron cyclotron
frequency is higher by the mass ratio, $\Omega_e/\Omega_i = m_i/m_e$,
there is typically little energy remaining in the cascade at the very
high frequencies needed to transfer a significant amount of energy to
the electrons through their cyclotron resonance.  Thus, here we
consider only the physics of the ion cyclotron resonance as a possible
physical mechanism for the dissipation of plasma turbulence.

Evaluating the conditions needed to obtain $\omega/\Omega_i
\rightarrow 1$ in terms of the dimensionless parameters of turbulence
in the isotropic temperature case, we find
\begin{equation}
  \frac{\omega}{\Omega_i} = \overline {\omega} k_\parallel d_i
  = \overline {\omega} \frac{k_\parallel  \rho_i}{\beta_i^{1/2}},
\end{equation}
where $ \overline {\omega}$ is defined by \eqref{eq:ombar}.  Note that
the middle expression indicates that the most natural unit to
normalize the length scale parallel to the magnetic field is the ion
inertial length $d_i=c/\omega_{pi}=v_A/\Omega_i$, but to maintain
consistency with the length normalizations in Tables~\ref{tab:params}
and ~\ref{tab:reduced_params}, we substitute $d_i= \rho_i/
\beta_i^{1/2}$.  Utilizing the general scaling of the parallel wavenumber in terms of the isotropic driving wavenumber,
\begin{equation}
 k_\parallel \rho_i = (k_0 \rho_i)^{(1+\alpha)/(3+\alpha)}(k_\perp \rho_i)^{2/(3+\alpha)},
\end{equation}
to determine the frequency achieved when the cascade reaches $k_\perp
\rho_i=1$, we obtain the condition
\begin{equation}
  \frac{\omega}{\Omega_i} = \frac{\overline {\omega}} {\beta_i^{1/2}}  (k_0 \rho_i)^{(1+\alpha)/(3+\alpha)}.
  \label{eq:icdscale}
\end{equation}
This indicates clearly that ion cyclotron damping is more likely to
occur for small ion plasma beta $\beta_i \ll 1$ or a smaller turbulent
inertial range (meaning a value of $k_0 \rho_i$ that is not too
small). For the \emph{GS95} scaling with $\alpha=0$, we obtain a condition
for the onset of ion cyclotron damping of $k_0 \rho_i
=\beta_i^{3/2}/\overline {\omega}^3$; for the B06 scaling with
$\alpha=1$, we obtain $k_0 \rho_i=\beta_i/\overline {\omega}^2$.
For simplicity, note that at the small-scale end of the inertial range
with $k_\perp \rho_i \lesssim 1$, the normalized wave frequency $\overline
{\omega} \sim 1$ for \Alfven waves can be approximated in the limit $k_\parallel d_i \rightarrow 1$ by the expression \citep{DiMare:2024}
\begin{equation}
 \overline {\omega} \simeq \frac{1}{\sqrt{1+(k_\parallel d_i)^2}}.
  \label{eq:icw_approx}
\end{equation}

In summary, ion cyclotron damping is predicted to play a role in the
damping of plasma turbulence for conditions with a smaller inertial
range and low plasma beta values $\beta_i \ll 1$.  The nature of the
resulting plasma heating is strictly ion energization $Q_i/Q=1$ in the
perpendicular degrees of freedom $Q_{\perp,i}/Q_i=1$.

Note that recent theoretical and numerical studies of turbulence with
a sufficiently large amount of imbalance $Z_0^+/Z_0^- \gg 1$, such as
may be encountered in the inner heliosphere, has uncovered a new
phenomenon denoted the \emph{helicity barrier}, which can lead to the
generation of sufficiently high-frequency fluctuations, specifically
ion cyclotron waves, that may cause energization of the ions by the
cyclotron resonance \citep{Meyrand:2021,Squire:2022,Squire:2023}.

\subsection{Stochastic Heating }
\label{sec:iSH}
The stochastic heating of ions has long been proposed as a means to energize ions in turbulent plasmas \citep{Johnson:2001,Chen:2001,White:2002,Voitenko:2004b,Bourouaine:2008}, with more recent work providing significant theoretical and numerical evidence for ion stochastic heating as a significant mechanism for turbulent damping in space and astrophysical plasmas \citep{Chandran:2010a,Chandran:2010b,Chandran:2011,Bourouaine:2013,Chandran:2013,Klein:2016b,Vech:2017,Arzamasskiy:2019,Martinovic:2020,Cerri:2021}.

\citet{Chandran:2010a} used test particle simulations of randomly
phased \Alfven and kinetic \Alfven waves to develop an empirical
formula for the rate of ion stochastic heating by turbulent
fluctuations with frequencies much lower than the ion cyclotron
frequency $\omega/\Omega_i \ll 1$.  Following this,
\citet{Chandran:2011} developed a refined implementation to predict
the partitioning of turbulent energy among parallel ion heating,
perpendicular ion heating, and electron heating (the \emph{C11} model
in \secref{sec:review}), with a formula for the damping rate by ion
stochastic heating given by
\begin{equation}
  \gamma_{SH} = c_1 \left(\frac{\delta v_\perp^{(\rho_i)}}{v_{ti}}\right) \Omega_i e^{-c_2 v_{ti}/\delta v_\perp^{(\rho_i)}}
  \label{eq:gamsh}
\end{equation}
where $\delta v_\perp^{(\rho_i)}$ is the root-mean-square amplitude of
the perpendicular velocity fluctuations at the perpendicular scale of
the ion Larmor radius, $k_\perp \rho_i=1$.  Here $c_1$ and $c_2$ are
fitting constants with suggested values $c_1=0.18$ and $0.15 \lesssim
c_2 \lesssim 0.34$ \citep{Chandran:2011}. The key dimensionless
parameter controlling the onset of stochastic heating in this
formulation is the ratio of the perpendicular velocity fluctuation to
the thermal velocity, $\epsilon \equiv \delta
v_\perp^{(\rho_i)}/v_{ti}$; these studies found a critical threshold
$\epsilon_{crit} \simeq 0.19$ that determines the minimum fluctuation
amplitude required for the onset of stochastic heating
\citep{Chandran:2010a}.

Using the relation for \Alfven waves to connect the perpendicular
fluid velocity fluctuations to the perpendicular magnetic field
fluctuations $\delta v_\perp /v_A =\pm \overline{\omega} \delta
B_\perp/B_0$ \citep{Howes:2008b} and the MHD scaling of the turbulent
perpendicular magnetic field fluctuation amplitude given by
\eqref{eq:bperp} with the B06 scaling with $\alpha=1$, we obtain the
expression
\begin{equation}
\frac{\delta v_\perp}{v_{ti}} = \overline{\omega} \beta_i^{-1/2}
\left(\frac{k_0\rho_i}{k_{\perp} \rho_i} \right)^{1/4}
  \label{eq:dvperp}
\end{equation}
At  $k_\perp \rho_i \sim 1$,  we obtain the normalized frequency $\overline{\omega} \sim 1$ and  find the key relation for the amplitude  in terms of our fundamental turbulence parameters
\begin{equation}
\epsilon \equiv  \frac{\delta v_\perp^{(\rho_i)}}{v_{ti}} =  \beta_i^{-1/2}
\left(k_0\rho_i \right)^{1/4}
  \label{eq:dvperp_rhoi}
\end{equation}

To estimate the importance of damping by ion stochastic heating relative to the turbulent cascade rate, we can calculate $ \gamma_{SH}/\omega_{nl} \sim  \gamma_{SH}/\omega$ to obtain the final result
\begin{equation}
  \frac{ \gamma_{SH}}{\omega} = c_1\left(k_0\rho_i \right)^{-1/4}  e^{-c_2  \beta_i^{1/2} (k_0\rho_i )^{-1/4}}
  \label{eq:gwsh}
\end{equation}
where we have used the substitution $\Omega_i/\omega =
\beta_i^{1/2}/k_\parallel \rho_i$.

In summary, the proposed turbulent damping mechanism of ion stochastic
heating depends primarily on the fundamental parameters $\beta_i$ and
$k_0\rho_i$, where lower ion plasma beta and larger isotropic driving
wavenumber (which corresponds to a smaller inertial range) both
conspire to yield larger turbulent amplitude at the ion Larmor radius
scale $k_\perp \rho_i \sim 1$, enhancing stochastic heating of the
ions.  The exponential effectively provides a threshold turbulent
amplitude at the ion scales below which stochastic ion heating is
ineffective.  Note that the exponential dependence on the constant
$c_2$ makes its accurate determination critical to assess the
importance of ion stochastic heating in any particular case; its value
is likely to depend on the detailed nature of the turbulent
fluctuations at the perpendicular scale of the ion Larmor radius
\citep{Chandran:2011}.  This heating mechanism is predicted to lead to
strictly ion energization $Q_i/Q=1$ in the perpendicular degrees of
freedom $Q_{\perp,i}/Q_i=1$.

Note that, because the amplitude of the turbulent fluctuations is
typically found to decrease monotonically as the turbulent cascade
progress to smaller scales, it is unlikely that electrons will
experience significant stochastic heating, so we do not explore
electron stochastic heating here.  Further research will be necessary
to evaluate thoroughly the potential for electron stochastic
heating to damp turbulence at the perpendicular scale of the electron
Larmor radius, $k_\perp \rho_e \sim 1$.  Furthermore, a consideration
of the intermittency---which can enhance the amplitude of individual
fluctuations---of the turbulent fluctuations at the ion Larmor radius
scale suggests that the rate of ion stochastic heating may be
dramatically increased in the presence of significant intermittency
\citep{Mallet:2019}.

\subsection{Magnetic Pumping }
Magnetic pumping
\citep{Spitzer:1951,Berger:1958,Lichko:2017,Lichko:2020,Montag:2022}
is a particle energization mechanism in which variations of magnetic
field magnitude at low frequency $\omega$ lead to an oscillating
transfer of energy from parallel to perpendicular degrees of freedom
and back, obeying the double-adiabatic (Chew-Goldberger-Low, CGL)
equations of state \citep{Chew:1956}.  In the absence of any
collisions, the net change of particle energy over a full oscillation
is zero.  But, in the presence small but finite
collisionality\footnote{Note that the collisionality can be due to
Coulomb collisions or to an effective collisionality due to pitch-angle
scattering through collisionless wave-particle interactions with
small-scale electromagnetic fluctuations, such as those driven by
kinetic temperature anisotropy instabilities
\citep{Kunz:2018,Arzamasskiy:2023}.}, there can be a non-zero transfer
of energy from the oscillating electromagnetic fields to the particles
over the course of a full oscillation.

As a proposed mechanism for the damping of turbulence, magnetic
pumping necessarily requires turbulent fluctuations that involve
changes of the magnetic field magnitude from the equilibrium field
strength $B_0$, or $\delta B \equiv |\V{B}|-B_0 \ne 0$.  For \Alfvenic
turbulence, which is dominantly incompressible\footnote{Note, however,
that the magnetic field magnitude variations arising from strong
turbulent driving at the outer scale with $|\delta \V{B}| \sim B_0$ can lead
to the ``interruption'' of linearly polarized \Alfven waves by kinetic
temperature anisotropy instabilities at sufficient high $\beta_i \gg
1$ \citep{Squire:2016,Squire:2017b,Squire:2017a}} in the MHD regime at
$k_\perp \rho_i \ll 1$, significant fluctuations of the magnetic field
magnitude arise only within the kinetic \Alfven wave regime at
$k_\perp\rho_i \gtrsim 1$ under conditions of ion plasma beta  $\beta_i > 1$.  Alternatively, if the
turbulence has a significant component of compressible
magnetosonic fluctuations with $E_{comp}/E_{inc}>0$, because the
turbulent amplitudes typically decrease monotonically with decreasing
scale lengths, the largest amplitude compressible fluctuations occur at the
driving scale and may lead to magnetic pumping, directly energizing
particles from these large-scale fluctuations.

We determine first the parameters controlling magnetic pumping in
\Alfvenic turbulence, following a similar procedure subsequently to analyze
the magnetic pumping arising from compressible turbulence with
$E_{comp}/E_{inc}>0$.  The foundation of magnetic pumping is the
double-adiabatic evolution of the weakly collisional plasma which
conserves the particle magnetic moment $\mu_m \equiv m v_\perp^2/2B$
and the action integral of the parallel bounce motion $J_\parallel \equiv \oint
v_\parallel dl$ \citep{Lichko:2017,Montag:2018}.  Conservation of
these adiabatic invariants requires a scale separation of the perpendicular
wavenumber relative to the particle species Larmor radius $k_\perp
\rho_s \ll 1$ and of the wave parallel phase velocity relative to the particle
thermal velocity $\omega/k_\parallel v_{ts} \ll 1$.

Kinetic \Alfven waves give rise to non-negligible magnetic field
magnitude variations over the perpendicular wavenumber range $1
\lesssim k_\perp \rho_i \lesssim (\tau \mu)^{1/2}$ for values of
$\beta_i \gtrsim 1$.  This range of $\delta B$ fluctuations is
incompatible with the necessary condition $k_\perp \rho_i \ll 1$ for
the ions, so we expect that \emph{magnetic pumping by kinetic \Alfven
waves cannot energize the ions}.  For the electrons, on the other
hand, the requisite condition $k_\perp \rho_e \ll 1$ can be rewritten
as $k_\perp \rho_i \ll (\tau \mu)^{1/2}$, so we expect that kinetic
\Alfven waves may be able to damp turbulent fluctuations and energize
the electrons under ion plasma beta conditions with $\beta_i \gtrsim
1$.  The phase velocity condition for electrons can be expressed as
$\overline{\omega} \ll (\beta_i \mu/\tau)^{1/2}$.  Under the ion
plasma beta condition $\beta_i \gtrsim 1$, the equation for the
dimensionless frequency of kinetic \Alfven waves given by
\eqref{eq:LDres} can be approximated by $\overline{\omega} \simeq
k_\perp \rho_i/\beta_i^{1/2}$, so this requirement can be converted to
a condition on the perpendicular wavenumber given by $k_\perp \rho_i
\ll \beta_i (\mu/\tau)^{1/2}$, which can certainly be satisfied for
the perpendicular wavenumber range of kinetic \Alfven waves with
significant magnetic field magnitude fluctuations.  Therefore, the
necessary conditions for the magnetic pumping of electrons to play a
role in the dissipation of \Alfvenic plasma turbulence with $\beta_i
\gtrsim 1$ appear to be satisfied.

Previous studies have shown that the ratio of the damping rate due to
magnetic pumping $\gamma_{MP}$ to the pump wave frequency $\omega$ is
given by \citep{Lichko:2017,Montag:2022}
\begin{equation}
  \frac{\gamma_{MP}}{\omega} = \frac{\chi_{c,s}}{1+\chi_{c,s}^2}\left(\frac{\delta B}{B_0}\right)^2
  \label{eq:gmp}
\end{equation}
where $\chi_{c,s} \equiv 6 \nu_s/\omega$ is a dimensionless measure of
the collision frequency $\nu_s$ for species $s$ and $\delta B$ is the
amplitude of the magnetic field magnitude fluctuations.

To determine the dependence of the magnetic pumping of electrons on the plasma and
turbulence parameters in Tables~\ref{tab:params} and
~\ref{tab:reduced_params}, we assume isotropic driving $k_{\parallel
  0}/k_{\perp 0}=1$ for simplicity, write the electron-electron collision
frequency as $\nu_e = v_{te}/\lambda_{mfp,e}$, and use the general
parallel wavenumber scaling for arbitrary $\alpha$ given by
\eqref{eq:kpar} to obtain
\begin{equation}
\chi_{c,e} = \frac{6}{ \overline{\omega} (k_{\parallel 0}
  \lambda_{mfp,e})} \left(\frac{\beta_i \mu}{\tau}\right)^{1/2} \left(
\frac{k_0 \rho_i}{k_\perp \rho_i} \right)^{2/(3+\alpha)} .
\end{equation}
For a sufficiently large turbulent inertial range with $k_0\rho_i \ll
1$, kinetic \Alfven waves over the perpendicular wavenumber range $1
\lesssim k_\perp \rho_i \lesssim (\tau \mu)^{1/2}$ will have
relatively small turbulent amplitudes $\delta B_\perp/B_0 \ll 1$, so
the magnetic field magnitude fluctuations can be expressed to lowest
order as variations of the parallel component of the perturbed
magnetic field $\delta B \simeq \delta B_\parallel$
\citep{Huang:2024}.  The turbulent magnetic field magnitude variations
may then be expressed as
\begin{equation}
  \frac{\delta B}{B_0} = \left(\frac{\delta B_\parallel}{\delta B_\perp}\right)
   \left( \frac{k_0 \rho_i}{k_\perp \rho_i} \right)^{1/(3+\alpha)},
\end{equation}
where the ratio $\delta B_\parallel/\delta B_\perp$ for kinetic
\Alfven waves over the relevant range from the linear dispersion
relation is strictly a function of $\beta_i$, or $\delta
B_\parallel/\delta B_\perp=f(\beta_i)$.  Therefore, electron magnetic
pumping by \Alfvenic turbulent fluctuations at scales $k_\perp \rho_i
\gtrsim 1$ depends on the parameters $\beta_i$, $\tau$, $\mu$, $k_0
\rho_i$, and $k_{\parallel 0}\lambda_{mfp,e}$.

Next we consider that case that the turbulence is driven compressibly
with $E_{comp}/E_{inc}>0$.  In this situation, the magnetic field
magnitude fluctuations at the driving scale, which typically have the
largest amplitude of $\delta B/B_0$, may contribute to magnetic
pumping of both the ions and electrons. In this case of turbulent
damping at the driving scale, it is necessary to use the three driving
parameters $(k_{\perp 0} \rho_i,k_{\parallel 0}/k_{\perp 0},\chi_0)$
in \tabref{tab:params}, rather than the simplified parameterization
through the isotropic driving wavenumber $k_0 \rho_i$, to determine
the parameter dependence of the turbulent damping via magnetic
pumping.

First, the normalized fast and slow magnetosonic wave phase velocities
can be expressed by the linear dispersion relation \citep{Klein:2012}
\begin{equation}
\widetilde{\omega}^4 - \widetilde{\omega}^2(1 + \beta) + \beta \cos^2 \Theta=0,
\end{equation}
which has  four solutions given by 
\begin{equation}
  \widetilde{\omega}^2 = \frac{1}{2}\left[ 1+ \beta \pm \sqrt{(1 + \beta)^2 - 4 \beta \cos^2 \Theta}\right],
  \label{eq:fsvphase}
\end{equation}
where the plus (minus) sign corresponds to the two fast (slow)
magnetosonic wave modes.  Here we define the normalized wave frequency
$\widetilde{\omega} \equiv \omega/k v_A$, the total (MHD) plasma beta
$\beta = \beta_i (1 + 1/\tau)$, and the wavevector components
$(k_\parallel,k_\perp)$ are alternatively parameterized by
$(k,\Theta)$, where $k_\parallel = k \cos \Theta$, $k_\perp= k \sin
\Theta$, $k=(k_\parallel^2+k_\perp^2)^{1/2}$, and $\Theta
=\tan^{-1}(k_\perp/k_\parallel)$. The dimensionless form of the
solutions in \eqref{eq:fsvphase} makes clear that the fast and slow
magnetosonic wave phase velocities depend only on two dimensionless
parameters, $\widetilde{\omega}( \beta,\Theta)$.

We may determine the parameter dependence of the normalized ion
collision frequency by using the relation for the ion-ion collision
rate in terms of the electron-electron collision rate $\nu_i =
\mu^{-1/2} \tau^{-3/2} \nu_e$ and writing the pumping wave frequency
at the driving scale as $\omega = \widetilde{\omega}_0 k_0 v_A$ to
obtain
\begin{equation}
\chi_{c,i} = \frac{6}{ \widetilde{\omega}_0 (k_{\parallel 0}
  \lambda_{mfp,e}) \tau^2} \beta_i^{1/2} \cos \Theta_0.
\label{eq:chici}
\end{equation}
Note that the ion collisional mean free path can be expressed in terms
of the electron collisional mean free path by the relation $
\lambda_{mfp,i}= \lambda_{mfp,e} \tau^2$, so the denominator is
actually independent of the ion-to-electron temperature ratio $\tau$,
depending rather only on the normalized ion mean free path,
$k_{\parallel 0} \lambda_{mfp,i}$.

To evaluate the turbulent magnetic field magnitude variations for ion
magnetic pumping by compressible fluctuations at the driving scale, we
first note that the phenomenon of dynamic alignment, parameterized by
the alignment angle $\theta_0$, will not play a role: dynamic
alignment develops through the nonlinear interactions as the
turbulence cascades to smaller scales, but does not come into play at
the driving scale, so we set $\theta_0 =1$, thus reducing the
nonlinearity parameter at the driving scale to $\chi_0 = k_{\perp 0}
\delta B_{\perp 0}/(k_{\parallel 0} B_0)$.  If the magnetic
perturbations at the driving scale have $|\delta \V{B}|/B_0 \sim 1$,
the linear theory for waves breaks down, and the nonlinearity must be
taken into account in the lowest-order description of the wave
dynamics; in this case, one may directly take the measured magnetic
field magnitude fluctuations $\delta B /B_0$ as the parameter to
estimate the damping rate by magnetic pumping in \eqref{eq:gmp}.
However, if the magnetic perturbations at the driving scale have
$\epsilon \equiv |\delta \V{B}|/B_0 <1$, the binomial expansion may be
used to obtain the expression
\begin{equation}
\delta B /B_0 \simeq \delta B_\parallel /B_0 + (\delta B_\parallel
/B_0)^2/2 + (\delta B_\perp /B_0)^2/2,
\label{eq:dbamp}
\end{equation}
where the first term is of order $\epsilon$ and the next two terms are
of order $\epsilon^2$.  Thus, to lowest order, we may estimate $\delta
B \simeq \delta B_\parallel$, where the error in this estimation
scales as $\epsilon^2/2$, yielding a fractional error of 25\% for
$|\delta \V{B}|/B_0 =1/2$ and of 6\% for $ |\delta \V{B}|/B_0 =1/4$.

Using this approximation, we can derive an expression the normalized
fluctuations of the magnetic field magnitude by $\delta B /B_0 \simeq
\delta B_\parallel /B_0 = (\delta B_\perp/B_0) (\delta
B_\parallel/\delta B_\perp)$.  If we estimate $\delta
B_\parallel/\delta B_\perp \sim (E_{comp}/E_{inc})^{1/2}$, we can then
obtain the expression for the normalized amplitude of the magnetic field magnitude fluctuations
\begin{equation}
\frac{\delta B}{B_0} = \frac{k_{\parallel 0}}{k_{\perp 0}} \chi_0
\left(\frac{E_{comp}}{E_{inc}}\right)^{1/2}.
\label{eq:dbpari}
\end{equation}
Combining the dependencies of $\chi_{c,i}$ in \eqref{eq:chici} and of
$\delta B /B_0$ in \eqref{eq:dbpari}, ion magnetic pumping by
compressible turbulent fluctuations at the driving scale depends on the
parameters $\beta_i$, $k_{\parallel 0}\lambda_{mfp,i}= (k_{\parallel
  0}\lambda_{mfp,e}) \tau^2$, $k_{\parallel 0}/k_{\perp 0}$, $\chi_0$,
and $E_{comp}/E_{inc}$.

For magnetic pumping of electrons by compressible turbulent
fluctuations at the driving scale, we can follow an analogous
procedure to obtain the normalized electron collision frequency 
\begin{equation}
\chi_{c,e} = \frac{6}{ \widetilde{\omega}_0 (k_{\parallel 0}
  \lambda_{mfp,e})}  \left(\frac{\beta_i \mu}{\tau}\right)^{1/2}
 \cos \Theta_0.
\label{eq:chice}
\end{equation}
and the same expression in \eqref{eq:dbpari} for the the normalized
fluctuations of the magnetic field magnitude by $\delta B /B_0$, which
is independent of the species properties.  Thus, electron magnetic
pumping by compressible turbulent fluctuations at the driving scale
depends on the parameters $\beta_i$, $\tau$, $\mu$, $k_{\parallel
  0}\lambda_{mfp,e}$, $k_{\parallel 0}/k_{\perp 0}$, $\chi_0$, and
$E_{comp}/E_{inc}$.

In summary, \Alfvenic turbulent fluctuations can damp turbulence
and energize electrons through turbulent fluctuations at perpendicular
scales $1 \lesssim k_\perp \rho_i \lesssim (\tau \mu)^{1/2}$ for
plasma conditions with $\beta_i \gtrsim 1$, but ions are not expected
to be energized by \Alfvenic fluctuations.  The resulting magnetic
pumping of electrons by \Alfvenic turbulent fluctuations depends on
the parameters $\beta_i$, $\tau$, $\mu$, $k_0 \rho_i$, and
$k_{\parallel 0}\lambda_{mfp,e}$.  If the turbulence is driven
compressibly with $E_{comp}/E_{inc}>0$, we expect that the turbulent
fluctuations at the driving scales, which have the largest amplitude
of $\delta B/B_0$, may lead to damping of the turbulence through the
magnetic pumping of both the ions and electrons.  The resulting
magnetic pumping of ions by compressible turbulence at the driving
scales depends on the parameters $\beta_i$, $k_{\parallel
  0}\lambda_{mfp,i}= (k_{\parallel 0}\lambda_{mfp,e}) \tau^2$,
$k_{\parallel 0}/k_{\perp 0}$, $\chi_0$, and
$E_{comp}/E_{inc}$. Analogously, the magnetic pumping of electrons by
compressible turbulence depends on the parameters $\beta_i$, $\tau$,
$\mu$, $k_{\parallel 0}\lambda_{mfp,e}$, $k_{\parallel 0}/k_{\perp
  0}$, $\chi_0$, and $E_{comp}/E_{inc}$.  These parameter dependencies
for magnetic pumping of ions and electrons are summarized in
\tabref{tab:mechanisms}.  Note that magnetic pumping energizes particles in energy over all pitch angles   \citep{Montag:2022}, so that both the parallel and perpendicular degrees of freedom gain energy, yielding for either ion or electron species  $Q_{\perp,s}/Q_{\parallel,s} \sim 1$.

\subsection{Kinetic Viscous Heating}
\label{sec:iVH}

As discussed above for the case of magnetic pumping, low-frequency
turbulent fluctuations that generate variations in the magnetic field
magnitude will lead to a double-adiabatic (CGL) \citep{Chew:1956}
evolution of weakly collisional plasmas that preserves to lowest order
the particle magnetic moment $\mu_m$ and the action integral of the
parallel bounce motion $J_\parallel$
\citep{Lichko:2017,Montag:2018,Arzamasskiy:2023}.  This dynamics leads
to the development of anisotropies in the velocity distribution of the
plasma particles, yielding temperature and pressure anisotropies with
$p_\parallel \ne p_\perp$.  These deviations from local thermodynamic
equilibrium can trigger rapidly growing kinetic instabilities that can
disrupt the linear wave dynamics
\citep{Squire:2016,Squire:2017b,Squire:2017a} and mediate the nonlocal
transfer of turbulent energy directly from the large driving scales to
kinetic length scales \citep{Kunz:2018,Arzamasskiy:2023}, as
illustrated by the red arrow in \figref{fig:cascade}(b). The effect of
these kinetic instabilities on the turbulent dynamics has recently
been modeled as an effective ``viscosity'' \citep{Arzamasskiy:2023}.

Here we use the term ``kinetic viscous heating'' to describe the
transfer of energy by this effective viscosity mediated by temperature
anisotropy instabilities \citep{Arzamasskiy:2023} to avoid confusion
with the usual viscous heating arising from microscopic Coulomb
collisions.  We note that, although this effect can be grossly modeled
as an effective viscosity that leads to diffusion of the bulk fluid
velocity field, its properties are entirely distinct from the standard
viscosity in fluid dynamics mediated by microscopic collisions between
neutral particles \citep{Chapman:1970} or by microscopic Coulomb
collisions between charged particles in a plasma under strongly
collisional conditions \citep{Spitzer:1962,Grad:1963,Braginskii:1965}.
Standard viscosity is a linear mechanism that may be modeled
mathematically by a Laplacian differential operator in the fluid
momentum equation, and thus its effect increases monotonically with
$k^2$ in Fourier space; this kinetic viscosity is a nonlinear
mechanism operating only under weakly collisional conditions that
strongly depends on the value of the ion plasma beta $\beta_i$, and
its effect diminishes at higher wavenumber $k$ with the monotonically
decreasing amplitudes of the turbulent fluctuations.  Determining how
this kinetic viscosity may tap some fraction of the energy from the
turbulent fluctuations at the large, driving scales and nonlocally
transfer that energy directly to fluctuations at the small, ion
kinetic length scales remains an open line of investigation.

To properly describe the kinetic physics of turbulence in temperature
anisotropic plasmas, it is necessary to employ the full set of plasma
and turbulence parameters in \tabref{tab:params}.  The relevant
kinetic instabilities driven by anisotropies in the ion velocity
distribution are the four ion temperature anisotropy instabilities:
(i) the parallel (or whistler) firehose instability
\citep{Kennel:1966,Gary:1976} and (ii) the \Alfven (or oblique)
firehose instability \citep{Hellinger:2000}, which are both relevant
to plasmas with an ion temperature anisotropy sufficiently less than
unity $A_i=T_{\perp,i}/T_{\parallel,i}<1$ and with a parallel ion
plasma beta $\beta_{\parallel,i} >1$; and (iii) the mirror instability
\citep{Vedenov:1958,Tajiri:1967,Southwood:1993} and (iv) the ion
cyclotron instability \citep{Gary:1976}, which are both relevant to
plasmas with an ion temperature anisotropy sufficiently greater than
unity $A_i=T_{\perp,i}/T_{\parallel,i}>1$ for any value of the
parallel ion plasma beta $\beta_{\parallel,i}$.  When the plasma exceeds a
threshold value of the ion temperature anisotropy
\citep{Boris:1977,Matteini:2006,Hellinger:2008,Klein:2015a}, these
instabilities can tap the free energy associated with the anisotropic
ion temperature, driving electromagnetic fluctuations and
ultimately reducing the temperature anisotropy, thereby moving the
plasma back toward a state of marginal stability.

\begin{figure}
  \begin{center}
 \resizebox{4.0in}{!}{\includegraphics{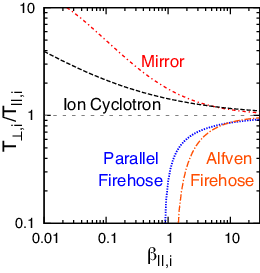}}
\end{center}
\caption[Marginal stability conditions for the four ion temperature anisotropy
instabilities.]
{Marginal linear 
stability thresholds for the four ion temperature 
anisotropy instabilities using \eqref{eq:unstableFit}
and the parameters in \tabref{tb:unstableFit}.
}
\label{fig:unstableFit}
\end{figure}

\citet{Hellinger:2006} has compiled values for the marginal stability
boundaries for these four ion temperature anisotropy instabilities
over the $(\beta_{\parallel,i}, T_{\perp,i}/T_{\parallel,i})$ plane.
Using a linear Vlasov-Maxwell dispersion relation solver with
bi-Maxwellian equilibrium ion velocity distributions in a fully ionized, hydrogenic (proton and electron) plasma, the marginal
stability boundary is determined by calculating the complex
eigenfrequency $\omega(\V{k})$ for a fixed $\beta_{\parallel,i}$ over
all possible wavevectors $\V{k}$.  The ion temperature anisotropy $
T_{\perp,i}/T_{\parallel,i}$ is then varied until the \emph{most}
unstable wavevector has a growth rate of $|\gamma|/\Omega_i=10^{-3}$,
thus establishing the instability criterion.  For the four ion temperature anisotropy
instabilities, the corresponding instability criteria on the
$(\beta_{\parallel,i}, T_{\perp,i}/T_{\parallel,i})$ plane are
generally well fit by an expression of the form
\begin{equation}
\frac{T_{\perp,i}}{T_{\parallel,i}}=1+\frac{a}{(\beta_{\parallel,i}-\beta_0)^b}
\label{eq:unstableFit}
\end{equation} 
where $a, \ b,$ and $\beta_0$ 
are unique values for each of the four instabilities, computed by 
 \citet{Hellinger:2006} and found here in \tabref{tb:unstableFit}.
These instability thresholds are plotted in \figref{fig:unstableFit}, and observational studies of measured intervals in the solar wind generally show that the measurements are more or less constrained by these threshold limits on the $(\beta_{\parallel,i}, T_{\perp,i}/T_{\parallel,i})$ plane \citep{Hellinger:2006,Bale:2009,Maruca:2011}.

\begin{table}
\begin{center}
\begin{tabular}{|c|c|c|c|}
\hline
Instability & $a$ & $b$ & $\beta_0$\\
\hline
Ion Cyclotron & 0.43 & 0.42 & -0.0004 \\
Parallel Firehose & -0.47 & 0.53 & 0.59 \\
\Alfven Firehose & -1.4 & 1.0 & -0.11 \\
Mirror & 0.77 & 0.76 & -0.016 \\
\hline
\end{tabular}
\end{center}
\caption{Instability Threshold Parameters for a maximum growth rate $|\gamma|/\Omega_i=10^{-3}$ 
from \citet{Hellinger:2006}. \label{tb:unstableFit}}
\end{table}

Recent theoretical and numerical studies have highlighted the fact
that, at sufficiently high parallel ion plasma beta $\beta_{\parallel,i} \gg 1$,
the marginal stability boundary for these kinetic ion temperature
anisotropy instabilities asymptotes towards isotropy, as shown in
\figref{fig:unstableFit}, so the double-adiabatic evolution of
turbulent fluctuations can easily exceed these boundaries and trigger
these rapidly growing instabilities.  Large-scale \Alfven waves in the
MHD limit at $k_\perp \rho_i \ll 1$ yield only perpendicular magnetic
field perturbations $\delta B_\perp \ne 0$ with $\delta
B_\parallel=0$, so the variation in the magnetic field magnitude
$\delta B$ is negligible in the limit of small amplitude waves $\delta
B_\perp/B_0 \ll 1$.  However, at sufficiently large amplitudes $\delta
B_\perp/B_0 \rightarrow 1$, these fluctuations can lead to significant
variation in the magnetic field magnitude, as made clear by
\eqref{eq:dbamp}.  Thus, at sufficiently high values of parallel ion
plasma beta $\beta_{\parallel,i} \gg 1$, the physics of linearly
polarized\footnote{Note that circularly polarized \Alfvenic
fluctuations, which have no magnetic field magnitude variation, do not
suffer this kinetic-instability-driven interruption
\citep{Squire:2016}.} \Alfven waves can be ``interrupted'' by these
kinetic ion temperature anisotropy instabilities
\citep{Squire:2016,Squire:2017b,Squire:2017a}.

Further examination of the effect of kinetic instabilities on the
turbulent cascade under conditions of high parallel ion plasma beta,
$\beta_{\parallel,i} \gg 1$, has found that these instabilities can
nonlocally transfer energy directly to kinetic length scales
\citep{Arzamasskiy:2023}, as illustrated by the red arrow in
\figref{fig:cascade}(b).  This new channel of energy transfer can
compete with the traditional local transfer of energy to smaller
scales of the turbulent cascade that is mediated by nonlinear
interactions, as illustrated by the black arrows in
\figref{fig:cascade}(a).  The effect of these kinetic instabilities on
the large-scale turbulent dynamics has been quantified as an effective
viscosity by \citet{Arzamasskiy:2023}.  Here we denote this
kinetic-instability-mediated viscosity as the \emph{kinetic
viscosity}\footnote{This \emph{kinetic viscosity} terminology for
kinetic-instability-mediated viscosity should not be confused with the
\emph{kinematic viscosity} defined in hydrodynamic fluids.}  to
distinguish it from the usual definition of viscosity that arises
through Coulomb collisions among the charged plasma particles.

These ion kinetic instabilities act to limit the ion pressure
anisotropy in the plasma to values corresponding to marginal
stability, given by $\Delta p_i/p_i \lesssim 1/\beta_{\parallel,i}$, where $\Delta
p_i \equiv p_{\perp i} - p_{\parallel i}$ and $p_i \equiv (2
p_{\perp i} + p_{\parallel i})/3$
\citep{Kunz:2014b,Melville:2016,Arzamasskiy:2023}. Using this
marginal stability scaling with $\beta_{\parallel,i}$, one can estimate an
effective collision frequency  \citep{Arzamasskiy:2023}
\begin{equation}
\nu_{\mbox{eff}} \sim \beta_{\parallel,i} \omega (\delta B_\perp/B_0)^2,
\end{equation}
where $\omega$ is the linear wave frequency at the scale where the
dynamics generate the largest pressure anisotropy, thus leading to
the maximum instability growth rate. Since the turbulent fluctuation
amplitudes generally decrease monotonically with scale, the kinetic
temperature anisotropy instabilities will be driven most strongly by
turbulent fluctuations at the driving scale.  As with the case for
magnetic pumping by compressible fluctuations that is dominated by the
driving scale dynamics, here we must use the three driving parameters
$(k_{\perp 0} \rho_i,k_{\parallel 0}/k_{\perp 0},\chi_0)$ in
\tabref{tab:params}, rather than $k_0 \rho_i$, to estimate the
turbulent damping rate by the kinetic viscosity.

Using the effective collision frequency $\nu_{\mbox{eff}}$ above as an estimate of the damping rate by this kinetic viscous heating $\gamma_{VH}$, we can
calculate the ratio of the damping rate arising from kinetic viscous heating
to the wave frequency at the driving scale, obtaining
\begin{equation}
  \frac{\gamma_{VH}}{\omega}\sim \beta_{\parallel,i} \left(\chi_0 \frac{k_{\parallel 0}}{k_{\perp 0}}
  \right)^2,
    \label{eq:gvh}
\end{equation}
where the definition of the nonlinearity parameter $\chi_0 = (k_{\perp
  0}/k_{\parallel 0})( \delta B_{\perp 0}/B_0) \theta_0$ with
$\theta_0=1$ at the driving scale is used to replace the amplitude of
the driving-scale turbulent fluctuations $ \delta B_{\perp 0}/B_0$.

In summary, the kinetic viscous heating due to ion temperature
anisotropy instabilities arising from the nonthermal velocity
distributions driven by the large-scale turbulent motions occurs in
weakly collisional turbulent plasmas with parallel ion plasma beta
$\beta_{\parallel,i} \gg 1$. The kinetic viscous damping rate depends
on the parallel ion plasma beta $\beta_{\parallel,i}$, the
nonlinearity parameter $\chi_0$, and the anisotropy of the turbulent
driving $k_{\perp 0}/k_{\parallel 0}$.  It is important to keep in
mind that the scaling used to derive the form of the effective
collision frequency $\nu_{\mbox{eff}}$ depends on the scaling of the
marginal stability boundaries that are generally calculated in
quiescent plasmas in the absence of turbulence.  The scaling of the
marginal stability boundaries---such as those quantified by
\eqref{eq:unstableFit} and the coefficients in
\tabref{tb:unstableFit}---may differ in the presence of strong plasma
turbulence, although evidence from spacecraft observations
\citep{Hellinger:2006,Bale:2009,Maruca:2011} and from hybrid kinetic
ion and fluid electron simulations of temperature anisotropic plasma
turbulence \citep{Kunz:2014b,Bott:2021,Arzamasskiy:2023} appear to
suggest that any modifications of these boundaries due to the presence
of turbulence are relatively small.

\subsection{Collisionless Magnetic Reconnection}
\label{sec:rxn}
Numerical simulations of plasma turbulence demonstrate the ubiquitous
development of current sheets at small scales
\citep{Matthaeus:1980,Meneguzzi:1981,Biskamp:1989,Spangler:1998,Spangler:1999,Biskamp:2000,Maron:2001,Merrifield:2005,Greco:2008},
and it has been found that the dissipation of the turbulence is
largely concentrated in the vicinity of these current sheets
\citep{Uritsky:2010,Wan:2012,Karimabadi:2013a,TenBarge:2013a,Wu:2013a,Zhdankin:2013},
giving rise to the proposal that magnetic reconnection may play a role
in the damping of plasma turbulence.  Motivated by these findings,
statistical analyses of observations of turbulence in the solar wind
have sought evidence for such spatially localized heating
\citep{Osman:2011,Borovsky:2011,Osman:2012a,Osman:2012b,Perri:2012a,Wang:2013,Wu:2013a,Osman:2014b}.
Yet the physical mechanisms that lead to the development of these
current sheets in plasma turbulence is yet to be fully understood.

\begin{figure}
\begin{center}\resizebox{4.0in}{!}{\includegraphics*[50pt,498pt][565pt,702pt]{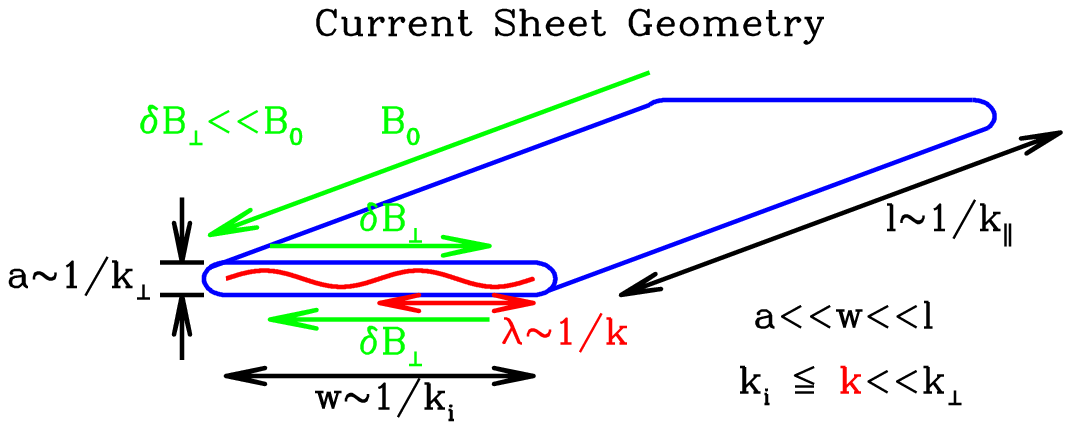}}
\end{center}
\caption{Diagram of the current sheet geometry generated
  self-consistently in plasma turbulence in the MHD regime at
  $k_\perp\rho_i \lesssim 1$, with the scalings of
  $(k_\perp,k_i,k_\parallel)$ given by the \emph{B06} scaling.  The
  current sheet has thickness $a\sim 1/k_\perp$ and width $w\sim 1/k_i$
  in the perpendicular plane and length $l \sim 1/k_\parallel$ along the
  equilibrium magnetic field $\V{B}_0$.  The wavelength of the mode
  unstable to the collisionless tearing instability is given by
  $\lambda \sim 1/k$ (red), and must fall within the bounds determined by the current sheet scaling, yielding an ordering $k_i \lesssim k \ll k_\perp$.
\label{fig:geom}}
\end{figure}

Under the \emph{B06} theory for the scaling of MHD turbulence, the
phenomenon of dynamic alignment \citep{Boldyrev:2006} is predicted to
lead to the development of current sheets with a width $w\sim 1/k_i$
to thickness $a\sim 1/k_\perp$ ratio in the plane perpendicular to the
magnetic field given by $w/a = k_\perp/k_i = [(k_\perp
  \rho_i)/(k_0\rho_i)] ^{1/4}$,  according
to the scaling in \eqref{eq:ki} for $\alpha=1$. The geometry of the
current sheets expected to develop in plasma turbulence is illustrated
in \figref{fig:geom}, where for sufficiently small perpendicular
scales relative to the driving scale, $k_\perp \rho_i \gg k_0 \rho_i$,
the current sheets have a parallel extent along the mean magnetic
field $\V{B}_0$, given by $l\sim 1/k_\parallel$, and a scale
separation of thickness to width to length obeying $a \ll w \ll l$;
the corresponding three components of the wave vector have the
complementary scaling $k_\parallel \ll k_i \ll k_\perp$.  In  \figref{fig:geom}, the
wavelength of a mode that is unstable to the collisionless tearing
instability is illustrated by $\lambda \sim 1/k$ (red). Note that the
minimum value of this tearing unstable wavenumber $k$ is equal to the
intermediate wavenumber $k_i$ corresponding to the width of the
current sheet, $w \sim 1/k_i$. The maximum value of  $k$ assumed in calculations of the linear collisionless tearing instability \citep{Loureiro:2017b,Mallet:2017b}  must be much smaller than the
perpendicular wavenumber $k_\perp$ characterizing  the  current sheet  thickness ($a \sim 1/k_\perp$), thereby yielding $k/k_\perp = k a \ll 1$.  Thus, the range of the unstable wavenumber $k$ is given
by
\begin{equation}
  k_i \rho_i = (k_0 \rho_i)^{1/4}(k_\perp \rho_i)^{3/4} \lesssim k \rho_i \ll k_\perp \rho_i \lesssim 1,
  \label{eq:klimits}
\end{equation}
where the final constraint $k_\perp \rho_i \lesssim 1$ is needed for
the tearing instability to arise for a current sheet thickness within
the MHD regime, otherwise different scalings for electron-only
reconnection \citep{Phan:2018} in the regime $k_\perp \rho_i \gtrsim
1$ would need to be incorporated.

Note that, in turbulence obeying the \emph{B06} scaling, the
self-consistent generation of current sheets arises only at
sufficiently small perpendicular scales relative to the driving scale,
$k_\perp \rho_i \gg k_0 \rho_i$. Therefore, the ratio of the
perpendicular magnetic field perturbation to the mean magnetic field
will be small, $\delta B_\perp/B_0 = (k_0 \rho_i/k_\perp \rho_i)^{1/4}
\ll 1$, so the reconnection that arises in plasma turbulence
\emph{necessarily} falls in the limit of strong-guide-field magnetic
reconnection, as illustrated in \figref{fig:geom} (green). An
implication of this fact is that, if numerical simulations are
employed to assess the importance of magnetic reconnection in
turbulence, it is critical that those simulations be modeled in three
spatial dimensions (3D).  Previous studies of 2D vs.~3D plasma
turbulence have shown that 3D simulations exhibit a significantly
faster turbulent cascade rate \citep{TCLi:2016}, so enabling full 3D
evolution is likely to be essential to determine the competition
between the turbulent cascade and magnetic reconnection.

The tearing instability growth rate typically increases with an
increasing perpendicular aspect ratio $w/a$ ratio of the current
sheets. If this ratio reaches a sufficiently large value, it was
recently recognized that the growth rate of the tearing instability
could exceed the turbulent cascade rate at that scale,
$\gamma/\omega_{nl}\gtrsim 1$, triggering a reconnection flow that
would disrupt the turbulent dynamics, deviating from the scaling
relations summarized in \secref{sec:mhdtheory}.  Such changes in the scaling of the turbulent fluctuations may potentially lead
to differences in which physical mechanisms dominate the damping of the
turbulent cascade at small scales
\citep{Boldyrev:2017,Loureiro:2017a,Loureiro:2017b,Mallet:2017b,Mallet:2017c,Walker:2018}.

It is worth noting here that an alternative explanation for the development
of current sheets in plasma turbulence due to the nonlinear
interactions between counterpropagating \Alfven waves---a phenomenon
often denoted \emph{\Alfven wave collisions} \citep{Howes:2013a}---was proposed by
\citet{Howes:2016b}, but no scaling for the resulting aspect ratio of
the current sheets was provided.  If the scaling of this alternative
mechanism differs from that of the \emph{B06} theory, than the scaling
predictions for the onset of reconnection would also change.

Here we adopt the \emph{B06} scaling for the development of current
sheets to determine the dependence of magnetic reconnection as a
turbulent damping mechanism on the fundamental plasma and turbulence
parameters for the isotropic temperature case in
\tabref{tab:reduced_params}.  This calculation follows previous
investigations that assess the conditions under which collisionless
magnetic reconnection may disrupt the turbulent cascade and thereby
impact how the turbulent energy is channeled into particle energy
\citep{Loureiro:2017b,Mallet:2017b}.  Our goal here is to express the
results from these previous studies in terms of the dimensionless parameters
for the isotropic temperature case proposed in
\tabref{tab:reduced_params}.

The growth rate of the collisionless tearing instability, which is the
instability that initiates the process of magnetic reconnection under
the weakly collisional conditions typical of space and astrophysical
plasmas, has been derived in the limit of very low electron plasma
beta $\beta_e \sim \mu^{-1}$ \citep{Zocco:2011}, where we convert
$\beta_e = \beta_i/\tau$.  The physics of the collisionless tearing
instability depends on three length scales: (i) the thickness of the
current sheet $a \sim 1/k_\perp$, assumed here to occur on MHD scales with  $k_\perp \rho_i \ll 1$; (ii) the ion sound Larmor radius $\rho_s
\equiv \rho_i/(2 \tau)^{1/2}$ where two fluid effects begin to lead to
differences between the ion and electron dynamics; and (iii) the inner
boundary layer scale $\delta_{in}$ where the magnetic flux can be
unfrozen from the electron fluid flow.  We will restrict our analysis
to $\beta_e \gg \mu^{-1}$, such that the flux unfreezing arises from
electron inertia \citep{Zocco:2011,Loureiro:2017b,Mallet:2017b},
occurring approximately at the electron inertial length scale $d_e
\equiv c/\omega_{pe} = \rho_i/(\beta_i \mu)^{1/2}$. These scales are
ordered by $\delta_{in} \sim d_e \ll \rho_s \ll a$.  Together, these
restrictions limit us to moderately small plasma beta,
\begin{equation}
\mu^{-1} \ll
\beta_i/\tau \ll 1.
\label{eq:lowbeta}
\end{equation}

The tearing instability depends on the tearing mode instability
parameter $\Delta'$, determined by the outer-region solution (at MHD
scales $k_\perp \rho_i \ll 1$) \citep{Zocco:2011}.  This parameter can
be expressed over two limiting regimes in a dimensionless form
\citep{Loureiro:2017b} as
\begin{equation}
  \Delta' a = (ka)^{-n} \quad \quad \mbox{for} \quad \quad 1 <  n \le 2,
  \label{eq:deltaprime}
\end{equation}
where $k$ is the unstable wavenumber along the width of the current layer
governing the reconnection flow, as illustrated in \figref{fig:geom}
by the red sinusoidal curve. Here $n=1$ corresponds to a Harris
current sheet (a hyperbolic tangent profile of the reconnecting magnetic field
across the current sheet thickness) \citep{Harris:1962}, while $n=2$
corresponds to a sinusoidal variation of the reconnection magnetic
field \citep{Loureiro:2017b}, which may be the more likely case for
the onset of reconnection at current sheets that self-consistently
arise in a turbulent plasma.  The tearing instability growth rates are
solved in two limits of
$\Delta'$ \citep{Zocco:2011,Loureiro:2017b,Mallet:2017b}: (i) $\Delta' \delta_{in} \ll 1$,
\begin{equation}
  \gamma \sim k v_{A \perp} \frac{d_e \rho_s \Delta'}{a} \quad \quad \mbox{with}
  \quad \quad  \delta_{in} \sim d_e \rho_s^{1/2} \Delta'^{1/2},
  \label{eq:dprimesmall}
\end{equation}
and (ii) $\Delta' \delta_{in} \gtrsim 1$,
\begin{equation}
  \gamma \sim k v_{A \perp} \frac{d_e^{1/3} \rho_s^{2/3}}{a} \quad \quad \mbox{with}
   \quad \quad \delta_{in} \sim d_e^{2/3} \rho_s^{1/3},
   \label{eq:dprimelarge}
\end{equation}
where $v_{A \perp} \equiv \delta B_\perp/(4 \pi n_i m_i)^{1/2}$ is the
\Alfven speed based on the magnitude of \emph{only} the reconnecting
component of the magnetic field $\delta B_\perp$.  Since the growth
rates above are calculated in the limit $k_\perp \rho_i \ll 1$, we
will focus our assessment on whether magnetic reconnection will
disrupt the turbulent cascade at MHD scales $k_\perp \rho_i \lesssim
1$, which is the same approach as taken in previous studies
\citep{Loureiro:2017b,Mallet:2017b}.

Using \eqref{eq:deltaprime} and either \eqref{eq:dprimesmall} or
\eqref{eq:dprimelarge} for $\delta_{in}$, the normalized tearing mode instability parameter
$\Delta' \delta_{in}$ can be expressed in terms of the dimensionless
quantities in \tabref{tab:reduced_params} by
\begin{equation}
  \Delta' \delta_{in} = \frac{k_\perp \rho_i}{(2 \tau \beta_i^2 \mu^2)^{1/6}}
 \left(\frac{k_\perp \rho_i}{k \rho_i} \right)^n.
  \label{eq:dprime}
\end{equation}
This equation shows that $\Delta' \delta_{in}$ increases as the
perpendicular wavenumber $k_\perp \rho_i$ increases, corresponding to
a thinner current sheet, or as the unstable wavenumber $k\rho_i$
decreases, requiring to a wider current sheet.  Since the width
$w\sim 1/k_i$ of the current sheets in \emph{B06} scaling theory
governed by \eqref{eq:ki} bounds the unstable wavenumber $k$ from
below by the intermediate wavenumber, $k>k_i$, the maximum value of
instability parameter $\Delta' \delta_{in}$ increases with the
perpendicular aspect ratio of the current sheets $w/a = k_\perp/k_i
\gg 1$. Note also that $\Delta' \delta_{in} \propto \beta_i^{-1/3}$,
so that lower values of ion plasma beta also lead to a larger instability
parameter.

Note that the value of the unstable wavenumber $k \rho_i$ at the
transition between the limiting regimes of the tearing mode
instability parameter, where $\Delta' \delta_{in}=1$, is given by
\begin{equation}
  k \rho_i = k_t \rho_i \equiv  \frac{(k_\perp \rho_i)^{(n+1)/n}}{(2 \tau \beta_i^2 \mu^2)^{1/6n}},
  \label{eq:dprime1}
\end{equation}
which provides a bound on the value of $k \rho_i$ for the validity of
the tearing growth rate formulas in \eqref{eq:dprimesmall} and
\eqref{eq:dprimelarge}.


To assess whether the collisionless tearing instability grows fast
enough to enable the onset of magnetic reconnection to disrupt the
turbulent cascade, we will assess $\gamma_{RXN}/\omega_{nl} \sim
\gamma_{RXN}/\omega$, assuming critical balance of linear and nonlinear
timescales $\omega_{nl}\sim \omega$
\citep{Goldreich:1995,Mallet:2015}.  Taking the linear wave frequency
for MHD \Alfven waves $\omega = k_\parallel v_A = (k_\parallel
\rho_i)^{1/2} (k_\perp \rho_i)^{1/2} v_A/\rho_i$ using \eqref{eq:kpar}
with $\alpha=1$ for the \emph{B06} scaling of $k_\parallel$, and
substituting $v_{A \perp} = v_A (\delta B_\perp/B_0) = v_A [(k_0
  \rho_i)/(k_\perp \rho_i)]^{1/4}$ using \eqref{eq:bperp}, we find
that the growth rate in the $\Delta' \delta_{in} \ll 1$ limit, which
corresponds to the large unstable wavenumber $k$ limit by inspection of
\eqref{eq:dprime}, can be expressed as
\begin{equation}
  \frac{\gamma_{RXN}}{\omega} \sim \frac{ (k_\perp \rho_i)^{(5+4n)/4}}
  { (k \rho_i)^{n-1}(k_0 \rho_i)^{1/4}(2 \tau \beta_i \mu)^{1/2} },
  \label{eq:dpsmall1}
\end{equation}
where this expression is formally valid in the MHD limit with $k_\perp
\rho_i \ll 1$.  Combining the limitations on the possible unstable
wavenumber from \eqref{eq:klimits} and the requirement $\Delta'
\delta_{in} \lesssim 1$ using \eqref{eq:dprime1}, the formula in
\eqref{eq:dpsmall1} is valid over the range of normalized unstable
wavenumbers
\begin{equation} 
  \max\left[ (k_0 \rho_i)^{1/4}(k_\perp \rho_i)^{3/4},
    \frac{(k_\perp \rho_i)^{(n+1)/n}}{(2 \tau \beta_i^2 \mu^2)^{1/6n}} \right]
  \lesssim k \rho_i \ll k_\perp \rho_i \lesssim 1.
  \label{eq:krangedpsmall}
\end{equation}
Note that this valid unstable wavenumber range can be used to show
that no range with $\Delta' \delta_{in} \lesssim 1$ exists if $\beta_i
\lesssim ( 2 \tau \mu^2)^{-1/2}$. For typical values of $\tau \sim 1$,
however, the very low values of $\beta_i$ required to violate this
condition would already have violated the assumed moderate beta limit
given by \eqref{eq:lowbeta}, so this limitation is of no practical
concern.

The tearing growth rate in \eqref{eq:dpsmall1} demonstrates that, in
the large $k$ limit, the normalized growth rate is independent of $k$
with $ \gamma_{RXN}/\omega \propto k^0$ for Harris-type current sheets
with $n=1$, and is inversely proportional to $k$ with $
\gamma_{RXN}/\omega \propto k^{-1}$ for the $n=2$ case of sinusoidal
variations of the perpendicular magnetic field component that may be
expected to occur in plasma turbulence.

The maximum growth rate $\gamma_{RXN}/\omega$ from \eqref{eq:dpsmall1}
increases with $k_\perp \rho_i$, but also requires the limit $k_\perp
\rho_i \ll 1$ for the validity of this linear tearing growth rate
calculation.  Therefore, to estimate quantitatively the maximum
normalized tearing growth rate, we take $k_\perp \rho_i \rightarrow
\mathcal{F}$, where $\mathcal{F} \ll 1$ is the maximum value of the
normalized perpendicular wavenumber for which \eqref{eq:dpsmall1}
remains valid; numerical simulations of collisionless magnetic
reconnection will be invaluable to provide estimates of the maximum
value of $\mathcal{F}$.  The maximum growth rate $\gamma_{RXN}/\omega$
also occurs for the minimum unstable wavenumber $k\rho_i$ allowed by
the range in \eqref{eq:krangedpsmall}.  Thus, taking these two limits
and requiring a collisionless tearing instability growth rate sufficient to
disrupt the turbulent cascade, $\gamma_{RXN}/\omega \gtrsim 1$,
yields the constraint
\begin{equation}
  k_0 \rho_i \lesssim \frac{\mathcal{F}^{(4+5n)/n}}
  {  (2 \tau )^{(4n+2)/3n}  (\beta_i \mu)^{(2n+4)/3n}}
  \label{eq:dpsmall_lim1}
\end{equation}
if the minimum $k\rho_i$ is constrained by the condition $\Delta'
\delta_{in} \lesssim 1$.  Alternatively, if the minimum $k\rho_i$ is
constrained by the width of the current sheet given by
\eqref{eq:klimits}, then the condition $\gamma_{RXN}/\omega \gtrsim 1$
yields the constraint
\begin{equation}
  k_0 \rho_i \lesssim \frac{\mathcal{F}^{(5+4n)/n}}{ (2 \tau \beta_i \mu)^{2/n}}.
  \label{eq:dpsmall_lim2}
\end{equation}

\begin{figure}
\begin{center}\resizebox{4.0in}{!}{\includegraphics{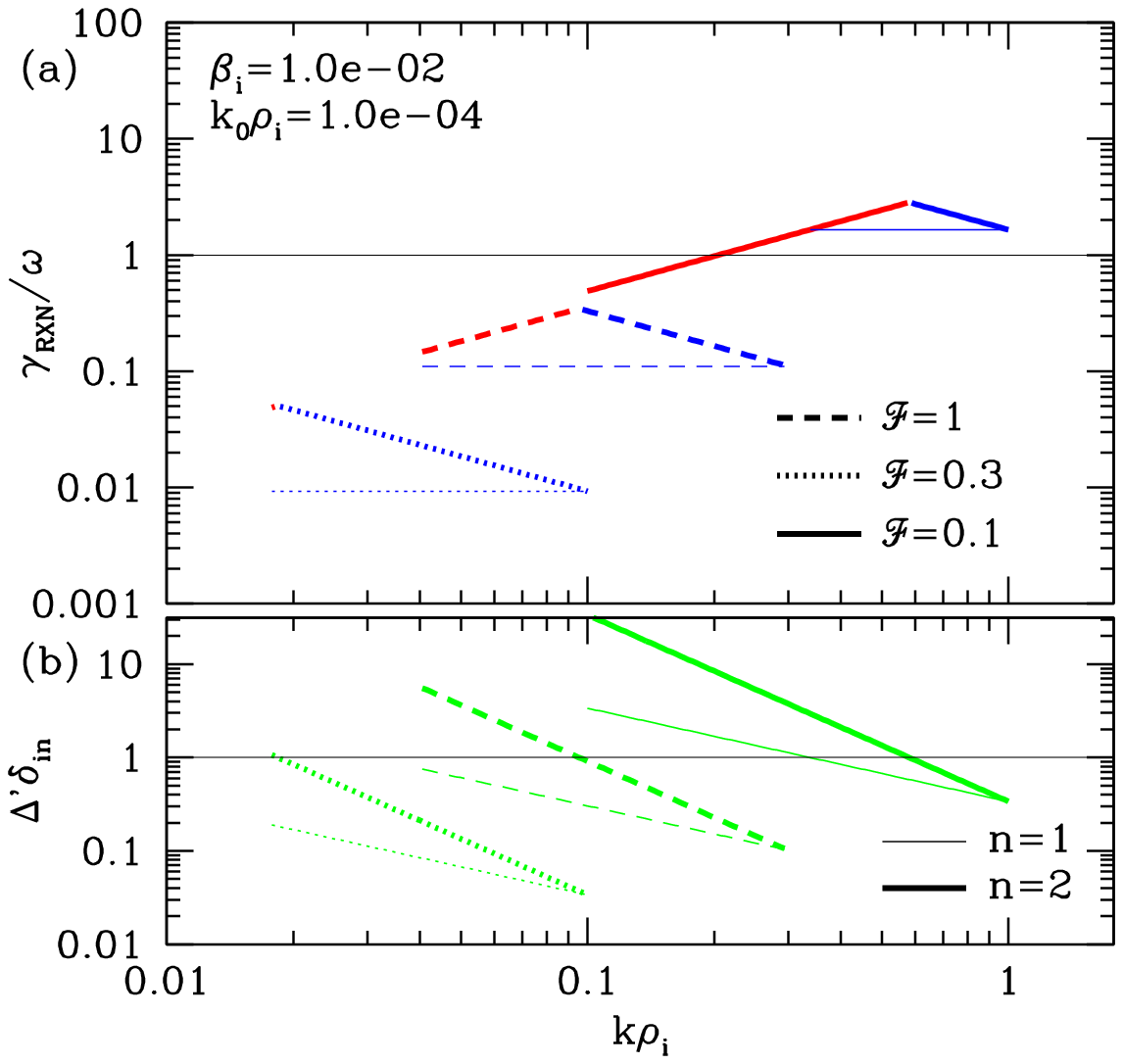}}
\end{center}
\caption{(a) Scaling the normalized collisionless tearing instability
  growth rate $ \gamma_{RXN}/\omega$ for a turbulent plasma with
  $\beta_i=0.01$, $\tau=1$, $\mu=1836$, and $k_0\rho_i=10^{-4}$ as a
  function of the normalized unstable wavenumber $k \rho_i$. Colors
  indicate the solutions in the low $k$ limit (corresponding to
  $\Delta' \delta_{in} \gtrsim 1$) (red) and in the high $k$ limit
  (corresponding to $\Delta' \delta_{in} \ll 1$) (blue). Line
  thickness indicates the either a sinusoidal current profile with
  $n=2$ (thick) or a Harris-like hyperbolic tangent current profile
  with $n=1$ (thin).  Line styles indicate different values of maximum
  perpendicular wavenumber $\mathcal{F} = 1.0$ (solid), $0.3$
  (dashed), and $0.1$ (dotted). (b) The corresponding normalized
  tearing mode instability parameter $\Delta' \delta_{in}$ vs.~$k
  \rho_i$ for each case.
\label{fig:gwrxn}}
\end{figure}

The normalized growth rates in the $\Delta' \delta_{in} \lesssim 1$
limit vs.~normalized unstable wavenumber $k \rho_i$ are illustrated in
\figref{fig:gwrxn}(a) for $n=1$ (thin blue) and $n=2$ (thick blue) for
turbulent parameters $\beta_i=0.01$, $\tau=1$, $\mu=1836$, and
$k_0\rho_i=10^{-4}$ and for three values of the maximum  of
$k_\perp \rho_i$ given by $\mathcal{F}=1.0$ (solid), $0.3$ (dashed),
and $0.1$ (dotted).  The corresponding values of $\Delta' \delta_{in}$
vs.~$k \rho_i$ are presented in \figref{fig:gwrxn}(b) for each of the
$n$ and $k_\perp \rho_i$ cases.  For the turbulence parameters
specified in this example, we find $\gamma_{RXN}/\omega \gtrsim 1$
only for the $k_\perp \rho_i=1.0$ case over the unstable wavenumber
range $0.3 \lesssim k\rho_i \lesssim 1$ for the $n=1$ case (thin solid
blue) and over $0.6 \lesssim k\rho_i \lesssim 1$ for the $n=2$ case
(thick solid blue).


Next we assess the initiation of collisionless tearing in the $\Delta'
\delta_{in} \gtrsim 1$ limit with the growth rate given by
\eqref{eq:dprimelarge}.  Using similar substitutions for $\omega$ and
$v_{A \perp}$ (as detailed above for the $\Delta' \delta_{in} \ll 1$
limit), we find that the growth rate in the $\Delta' \delta_{in}
\gtrsim 1$ limit, which is the small unstable wavenumber $k$ limit as
shown by \eqref{eq:dprime}, can be expressed as
\begin{equation}
  \frac{\gamma_{RXN}}{\omega} \sim \frac{ (k \rho_i) (k_\perp \rho_i)^{1/4}}
  { (k_0 \rho_i)^{1/4}(4 \tau^2 \beta_i \mu)^{1/6} },
  \label{eq:dplarge1}
\end{equation}
where this expression is formally valid in the MHD limit with $k_\perp
\rho_i \ll 1$. 
Combining the limitations on the possible unstable wavenumber from
\eqref{eq:klimits} and the requirement $\Delta' \delta_{in}
\gtrsim 1$ using \eqref{eq:dprime1}, the formula in
\eqref{eq:dplarge1} is valid over the range of normalized unstable
wavenumbers
\begin{equation}
  (k_0 \rho_i)^{1/4}(k_\perp \rho_i)^{3/4} \lesssim  k \rho_i \ll
  \max\left[ k_\perp \rho_i,
    \frac{(k_\perp \rho_i)^{(n+1)/n}}{(2 \tau \beta_i^2 \mu^2)^{1/6n}} \right]
  \lesssim 1.
  \label{eq:krangedplarge}
\end{equation}
This valid unstable wavenumber range can be used to show
that no range with $\Delta' \delta_{in} \gtrsim 1$ exists if
\begin{equation}
  k_\perp \rho_i \lesssim
  (k_0 \rho_i)^{n/(n+4)}(2 \tau \beta_i^2 \mu^2)^{2/3(n+4)}.
  \label{eq:nodphigh}
\end{equation}
This finding is consistent with the fact the that value of $\Delta'
\delta_{in}$ decreases with decreasing $k_\perp \rho_i$ in
\eqref{eq:dprime}, so by decreasing $k_\perp \rho_i$ enough, one will
always reach a point where there is no range of $k\rho_i$ with
$\Delta' \delta_{in} \gtrsim 1$.

The tearing growth rate in \eqref{eq:dplarge1} shows that, in the
small $k$ limit, the normalized growth rate increases linearly with
the unstable wavenumber $k$ and increases weakly with the
perpendicular wavenumber $k_\perp \rho_i$, independent of whether the
turbulently generated current sheets are Harris-like or sinusoidal
(thus, independent of $n$). 

Once again the maximum growth rate $\gamma_{RXN}/\omega$ from
\eqref{eq:dplarge1} increases with $k_\perp \rho_i$ (although weakly
with the $1/4$ power), so once again we take the limit $k_\perp \rho_i
\rightarrow \mathcal{F}$, where  $\mathcal{F} \ll 1$ is the maximum value of the
normalized perpendicular wavenumber for which \eqref{eq:dplarge1}
remains valid. The maximum growth rate will also occur for the
maximum unstable wavenumber $k\rho_i$ given by
\eqref{eq:krangedplarge}.  Thus, taking these two limits and requiring
a collisionless tearing instability growth rate sufficient to disrupt the
turbulent cascade, $\gamma_{RXN}/\omega \gtrsim 1$, yields the
constraint
\begin{equation}
  k_0 \rho_i \lesssim \frac{\mathcal{F}^{(4+5n)/n}}
  {  (2 \tau )^{(4n+2)/3n}  (\beta_i \mu)^{(2n+4)/3n}}.
  \label{eq:dplarge_lim1}
\end{equation}
if the maximum $k\rho_i$ is constrained by the condition $\Delta'
\delta_{in} \gtrsim 1$.  Note that this condition is the same as
\eqref{eq:dpsmall_lim1}, as it must be since this maximum growth rate
occurs at $\Delta' \delta_{in} = 1$, where the growth rates for the
$\Delta' \delta_{in} \lesssim 1$ and $\Delta' \delta_{in} \gtrsim 1$
limits cross.  Alternatively, if the maximum $k\rho_i$ is constrained
by the maximum $k_\perp \rho_i \rightarrow \mathcal{F}$, then the condition
$\gamma_{RXN}/\omega \gtrsim 1$ yields the constraint
\begin{equation}
  k_0 \rho_i \lesssim  \frac{\mathcal{F}^{(5+4n)/n}}{ (4 \tau^2 \beta_i \mu)^{2/3}}
  \label{eq:dplarge_lim2}
\end{equation}

The normalized growth rates in the $\Delta' \delta_{in} \gtrsim 1$
limit vs.~normalized unstable wavenumber $k \rho_i$ are illustrated in
\figref{fig:gwrxn}(a) (thick red) for turbulent parameters
$\beta_i=0.01$, $\tau=1$, $\mu=1836$, and $k_0\rho_i=10^{-4}$ and for
three values of  the maximum  of
$k_\perp \rho_i$ given by $\mathcal{F}=1.0$ (solid), $0.3$ (dashed), and
$0.1$ (dotted).  The corresponding values of $\Delta' \delta_{in}$
vs.~$k \rho_i$ are presented in \figref{fig:gwrxn}(b) for each of the
$n$ and $k_\perp \rho_i$ cases.  For the turbulence parameters
specified in this example, we find $\gamma_{RXN}/\omega \gtrsim 1$
only for the $k_\perp \rho_i=1.0$ case over the unstable wavenumber
range $0.2 \lesssim k\rho_i \lesssim 0.3$ for the $n=1$ case (thick
solid red) and over $0.2 \lesssim k\rho_i \lesssim 0.6$ for the $n=2$
case (thick solid red).

It is necessary to highlight here a few important caveats in using the
linear collisionless tearing mode growth rates and MHD turbulence
scaling theory to predict when magnetic reconnection may disrupt the
dynamics of the turbulent cascade.  First, the linear tearing mode
growth rates used here are rigorously derived in asymptotic limits,
but we employ these same rates up to the extreme boundary of those
limits.  For example, the linear growth rates formally require
$k_\perp \rho_i \ll 1$, but we take $k_\perp \rho_i \rightarrow 1$ to estimate the maximum growth rate; also, the $\Delta' \delta_{in} \ll 1$ limit is
used to estimate the tearing mode growth rates as $\Delta' \delta_{in}
\rightarrow 1$. Such rates derived in asymptotic limits, however, may not
remain quantitatively accurate as we extend beyond their formal regime of
validity.  We adopt the viewpoint that the growth rates calculated in these
limits are at least  reasonable lowest-order estimates of the rates even
beyond their formally limited range.  Furthermore, the growth rates for
the collisionless tearing mode are determined under smooth conditions,
whereas to determine whether reconnection arises in the current sheets
that naturally arise in plasma turbulence, one needs to determine the
collisionless tearing growth rates in the presence of turbulence; once
again, the estimated tearing mode growth rates used here are taken
simply as the lowest-order estimation.  Computing the tearing mode
growth rates in a turbulent plasma and determining the competition of
the tearing mode with the nonlinear cascade of energy due to
turbulence are both necessary for more accurate calculations, which
likely will require detailed kinetic numerical simulations.  Finally,
alternative proposed mechanisms for the development of current sheets
in plasma turbulence---such as the nonlinear interaction of
counterpropagating \Alfven wave collisions \citep{Howes:2016b}, as
opposed to dynamic alignment \citep{Boldyrev:2006}---may yield a
different scaling for the width and thickness of current sheets; the
resulting predictions for the onset of magnetic reconnection in a
turbulent plasma would thereby also likely change.

In summary, we have followed previous investigations
\citep{Loureiro:2017b,Mallet:2017b} to estimate how the collisionless
tearing mode may arise in the current sheets that develop
self-consistently in plasma turbulence, potentially disrupting the
nonlinear cascade of turbulent energy to small scales.  We restrict
ourselves to the moderately low electron plasma beta limit $\mu^{-1}
\ll \beta_e \ll 1$, which can be expressed in terms of the plasma and turbulence
parameters in \tabref{tab:reduced_params} as the limit $1 \ll \beta_i
\mu/\tau \ll \mu$.  In this limit, magnetic flux conservation is
broken due to electron inertia at electron inertial length scales
$d_e$ \citep{Zocco:2011,Loureiro:2017b,Mallet:2017b}. We consider
only the case when collisionless magnetic reconnection arises in
current sheets with thicknesses $a\sim 1/ k_\perp$ in the MHD regime
$k_\perp \rho_i \lesssim 1$.  We find that the growth rate of
reconnection relative to the turbulent fluctuation frequencies
$\gamma_{RXN}/\omega$ is a function of the isotropic driving
wavenumber $k_0\rho_i$, ion plasma beta $\beta_i$, and ion-to-electron
temperature ratio $\tau$, as well as the ion-to-electron mass ratio
$\mu$.

In general, the normalized collisionless tearing instability growth
rates $\gamma_{RXN}/\omega$---given by \eqref{eq:dpsmall1} in the
$\Delta' \delta_{in} \ll 1$ limit and by \eqref{eq:dplarge1} in the
$\Delta' \delta_{in} \gtrsim 1$ limit---increase as both $k_0\rho_i$ and
$\beta_i$ decrease.  For specified values of the plasma parameters
$\beta_i$ and $\tau$, we obtain different conditions on how small 
$k_0\rho_i$ must be for  reconnection to arise, given by
\eqref{eq:dpsmall_lim1} and \eqref{eq:dpsmall_lim2} in the $\Delta'
\delta_{in} \ll 1$ limit and by \eqref{eq:dplarge_lim1} and
\eqref{eq:dplarge_lim2} in the $\Delta' \delta_{in} \gtrsim 1$ limit.
These scalings mean that reconnection is more likely to arise in
turbulence with a very large driving scale relative to the ion Larmor
radius scale---meaning a smaller value of  $k_0\rho_i \ll 1$---and for low ion plasma beta conditions
$\beta_i \ll 1$.

It is worthwhile emphasizing that these estimates on the importance of
magnetic reconnection in plasma turbulence are calculated only for
current sheets with a thickness $a \sim 1/k_\perp$ in the MHD regime
$k_\perp \rho_i \ll 1$ so that the MHD scaling predictions in
\secref{sec:mhdtheory} apply.  It is also possible that electron-only
magnetic reconnection \citep{Phan:2018} could occur within the range of
scales in the kinetic dissipation range $k_\perp \rho_i \gtrsim 1$, but to
estimate whether this will occur would require the use of modified
turbulence scalings appropriate for the dissipation range
\citep{Howes:2008b,Howes:2011b}.

Finally, how the incorporation of magnetic reconnection into the
turbulent cascade impacts the damping of the turbulence and the
resulting energization of the plasma remains an open question.
Collisionless magnetic reconnection effectively converts magnetic
energy into other forms, specifically the kinetic energy of the bulk
plasma outflows and possibly internal energy in the particle velocity
distribution functions \citep{Drake:2005, Egedal:2008, Egedal:2012,
  Drake:2012, Loureiro:2013b, Jiansen:2018, Pucci:2018, Munoz:2018,
  Dahlin:2020,McCubbin:2022}. But such conversion between magnetic
energy and the bulk kinetic energy of turbulent plasma motions already
occurs ubiquitously in plasma turbulence: the linear terms in the
equations of evolution for the turbulent plasma mediate the ongoing
oscillation of energy from magnetic energy to the kinetic energy of
the turbulent bulk plasma motions and back \citep{Howes:2015b}.  For
example, the physics of undamped \Alfven waves involve the
transformation of the bulk kinetic energy of the perpendicular wave
motions to magnetic energy as the frozen-in magnetic field is
stretched; magnetic tension serves as the restoring force for the
wave, decelerating the perpendicular wave motions and ultimately
re-accelerating them back towards the state where the magnetic field
is not bent, thereby converting magnetic energy back to kinetic
energy.  Thus, collisionless magnetic reconnection does not
necessarily lead directly to damping of the turbulent motions (unless
a significant fraction of the released magnetic energy is converted
into internal energy of the particle velocity distributions), but
rather is simply a channel, in addition to magnetic tension, for
converting magnetic to kinetic energy.  The triggering of magnetic
reconnection, however, will impact the scaling of the characteristic
three-dimensional wave vector of the turbulent motions that result
from the reconnection flow
\citep{Boldyrev:2017,Loureiro:2017a,Loureiro:2017b,Mallet:2017b,Mallet:2017c,Walker:2018},
and these different scalings may channel energy through different
kinetic physical mechanisms that ultimately transfer turbulent energy
to internal particle energy.  Thus, the partitioning of turbulent
energy between ions and electrons, $Q_i/Q_e$, and between degrees of
freedom for both species, $Q_{\perp,i}/Q_{\parallel,i} $ and
$Q_{\perp,e}/Q_{\parallel,e}$, remains to be ascertained in situations
where collisionless magnetic reconnection is triggered in plasma
turbulence.




\section{Phase Diagrams for Plasmas Turbulence}
The scalings of the different proposed kinetic dissipation mechanisms
listed in \secref{sec:proposed} enable us to map the importance of the
different mechanisms for the damping of turbulence as a function of
the fundamental plasma parameters by creating \emph{phase diagrams for
plasma turbulence}, analogous to a similar effort to create a phase
diagram for magnetic reconnection \citep{Ji:2011}.  Here we present
two such phase diagrams based on the plasma and turbulence parameters
estimated for different space and astrophysical plasmas.

\subsection{Collisionality of the Dissipation Range of Plasma Turbulence}
\begin{figure}
\begin{center}\resizebox{4.0in}{!}{\includegraphics*[0pt,0pt][450pt,450pt]{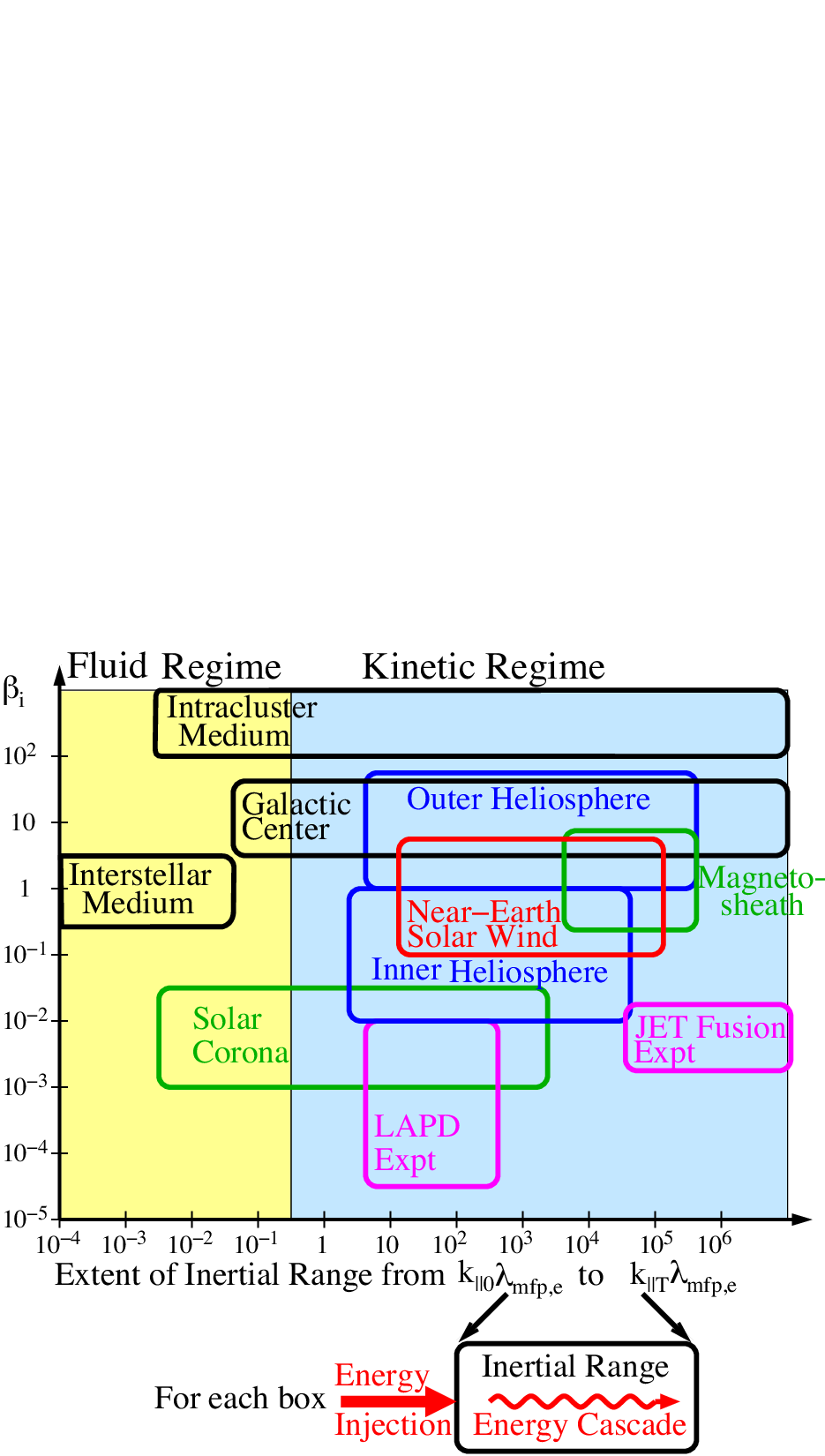}}
\end{center}
\caption{Phase Diagram of the extent of the turbulent inertial range
  from driving scales at $k_{\parallel 0} \lambda_{mfp,e}$ to the
  transition to the dissipation range at $k_{\parallel T}
  \lambda_{mfp,e}$ vs.~the ion plasma beta $\beta_i$. The plot
  indicates that, at the small scale end of the inertial range where
  damping mechanisms can remove energy from the turbulence, for many
  plasma systems of interest the governing dynamics are weakly
  collisional with  $k_{\parallel T} \lambda_{mfp,e} \gg 1$, so
  kinetic plasma theory is necessary to capture the physics of
  turbulent dissipation.
\label{fig:k0rhoi_betai}}
\end{figure}

We first create a phase diagram indicating that, for many plasmas of
interest, even if the turbulence is driven at fluid scales with
$k_{\parallel,0} \lambda_{mfp,e} \ll 1$, the end of the inertial range
generally occurs at scales that are weakly collisional.  Therefore,
the kinetic damping mechanisms analyzed here are responsible for
removing energy from the turbulent cascade, rather than viscosity and
resistivity, which are based on microscopic Coulomb collisions between
particles in the limit of strong collisionality
\citep{Braginskii:1965,Chapman:1970}.  Since the mean free path must
be compared to the parallel length scales of the turbulent
fluctuations characterized by $k_\parallel$, we must first determine
the parallel length scales associated with the small-scale end of the
inertial range at the transition to the dissipation range,
corresponding to a perpendicular scale $k_\perp\rho_i\sim
1$. Specifying the \emph{GS95} scaling with $\alpha=0$ for this
example, the parallel wavenumber at the transition to the dissipation
range $k_{\parallel T}$, determined by \eqref{eq:kpar}, is given by
$k_{\parallel T} \rho_i = (k_0 \rho_i)^{1/3}(k_\perp\rho_i)^{2/3}=(k_0
\rho_i)^{1/3}$, here taking $k_\perp\rho_i\sim 1$ as the end of the
inertial range. For simplicity, we assume here that the turbulence is
driven strongly and isotropically with $\chi_0=1$ and $k_{\parallel
  0}/k_{\perp 0}=1$.  Normalizing this parallel scale in terms of the
electron collisional mean free path $\lambda_{mfp,e}$, we obtain
$k_{\parallel T} \lambda_{mfp,e} = (k_0 \rho_i)^{1/3}
(\lambda_{mfp,e}/\rho_i)$.

The range of plasma and turbulence parameters for a number of plasma
systems of interest are denoted by the colored boxes in
\figref{fig:k0rhoi_betai}.  Plotted on the horizontal axis of
\figref{fig:k0rhoi_betai} is the range of parallel scales bounding
the turbulent inertial range normalized to the electron
collisional mean free path $\lambda_{mfp,e}$: (i) the normalized
parallel driving scale $k_{\parallel 0} \lambda_{mfp,e}$ of the
turbulence is represented by the left end of each box; (ii) the
turbulent energy cascades to smaller scales to the right within each
box (wavy red arrow below plot); and (iii) the parallel scale
corresponding to the small-scale end of the inertial range at
$k_{\parallel T} \lambda_{mfp,e}$ is represented by the right end of
each box.  The vertical extent of the colored boxes indicates the
range of ion plasma beta $\beta_i$ for each of the chosen systems.
The fluid regime (yellow region) indicates conditions under which the
dynamics of the turbulent fluctuations occur on parallel scales that
are strongly collisional with $k_{\parallel 0}\lambda_{mfp,e} \ll 1$; the
kinetic regime (light blue region) indicates conditions under which
the dynamics of the turbulent fluctuations occur on parallel scales
that are weakly collisional with $k_{\parallel 0}\lambda_{mfp,e} \gg
1$.

In \figref{fig:k0rhoi_betai} are shown these parameter ranges for a
variety of plasma systems: the near Earth solar wind (red); the
Earth's magnetosheath and the solar corona (green); the inner
heliosphere at heliocentric radii $R \ll 1$~AU and the outer
heliosphere at $R \gg 1$~AU (blue); the interstellar medium, the
Galactic Center, and the intracluster medium (black),; and terrestrial
laboratory experiments including the Large Plasma Device (LAPD) at UCLA
\citep{Gekelman:1991,Gekelman:2016} and the Joint European Torus (JET)
Fusion experiment \citep{JET:1992}.  The main message of
\figref{fig:k0rhoi_betai} is that, for most turbulent plasma systems
of interest, the parallel length scales associated with the end of the
inertial range correspond to weakly collisional plasma conditions with
$k_{\parallel T} \lambda_{mfp,e} \gg 1$.  Therefore, to explore the
physical damping mechanisms responsible for removing energy from the
turbulent fluctuations and consequently energizing the plasma
particles, plasma kinetic theory is essential.

\subsection{Turbulent Damping Mechanisms on the $(\beta_i,k_0\rho_i)$ Plane}
\begin{figure}
\begin{center}\resizebox{4.5in}{!}{\includegraphics{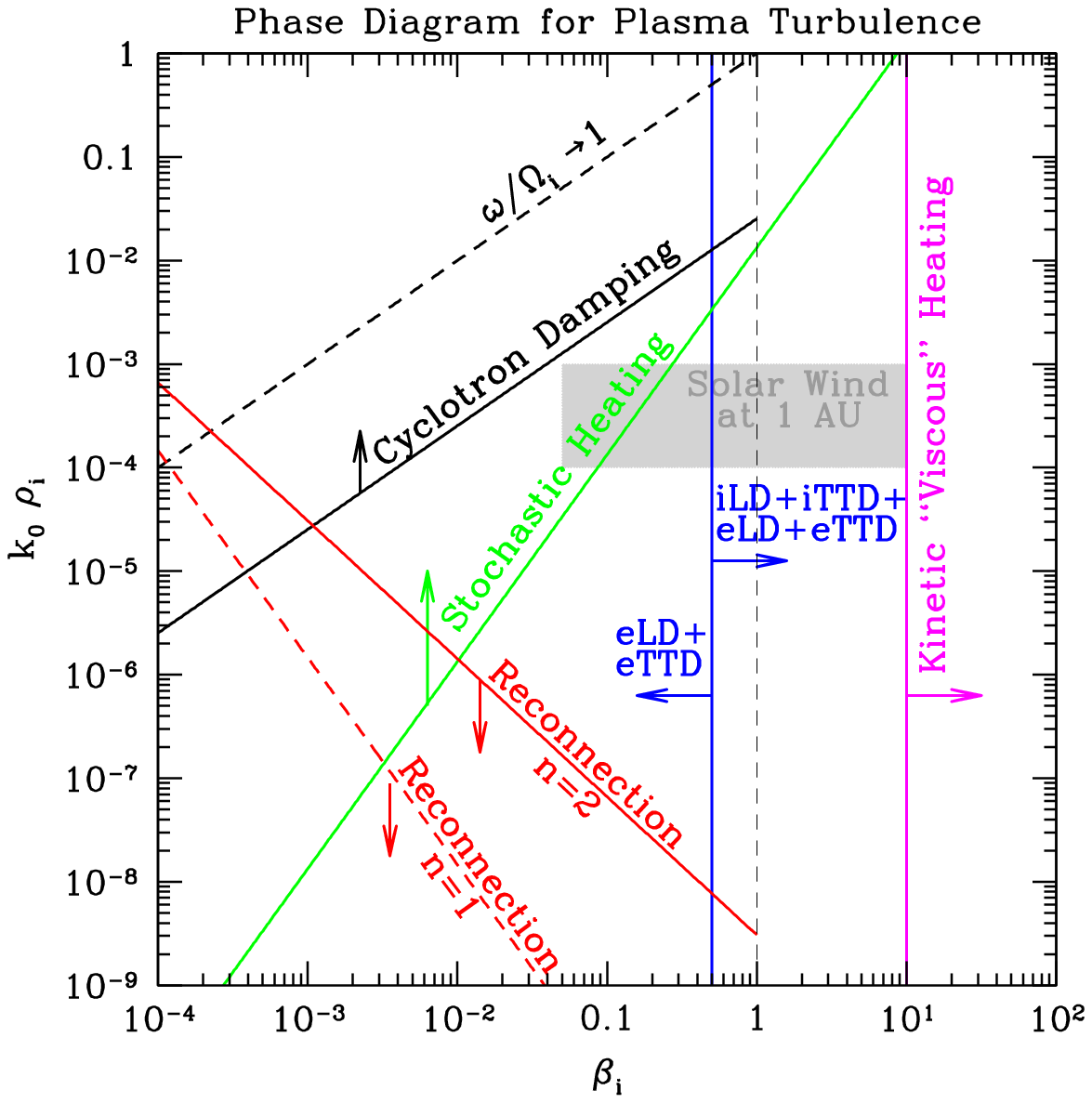}}
\end{center}
\caption{Phase diagram for the kinetic damping mechanisms of weakly collisional plasmas turbulence as a function of  isotropic driving wavenumber $k_0\rho_i$ and ion plasma beta $\beta_i$, showing regions of this parameter space where different mechanisms are likely to contribute to the  heating of the plasma species: ion Landau damping (iLD) and ion transit time damping (iTTD) (blue);  electron  Landau damping (eLD) and electron transit time damping (eTTD) (blue); ion cyclotron damping (iCD) (black); ion stochastic heating (iSH) (green); ion kinetic viscous heating (iVH) (magenta); and collisionless magnetic reconnection (RXN) (red) for intermittent current sheets ($n=1$, red dashed) and sinusoidal current sheets ($n=2$, red solid).  The extent of the turbulent near-Earth solar wind on the  $(\beta_i,k_0\rho_i)$ plane is indicated (gray shading).
\label{fig:phase_k0bi}}
\end{figure}

A single phase diagram that demonstrates the power of using the
dimensionless plasma and turbulence parameters for the isotropic
temperature case presented in \tabref{tab:reduced_params} is a plot of
the predicted regions on the $(\beta_i,k_0\rho_i)$ plane where
different kinetic mechanisms are expected to play a role the damping
of weakly collisional plasma turbulence, presented here in
\figref{fig:phase_k0bi}.  This phase diagram of the turbulent damping
mechanisms is the key result of this study.

For this phase diagram, we specify the ion-to-electron temperature
ratio $\tau=1$ and the ion-to-electron mass ratio $\mu=1836$.  For the
isotropic temperature case, the species temperature anisotropies are
$A_i=A_e=1$, and we take the turbulence to be driven strongly
($\chi_0=0$) and isotropically ($k_{\parallel 0}/k_{\perp 0}=1$) so
that the turbulent driving can be characterized by the isotropic
driving wavenumber $k_0\rho_i$. Furthermore, we assume the turbulence
driving to be balanced ($Z_0^+/Z_0^-=1$) and incompressible
($E_{comp}/E_{inc}=0$).  In addition, the turbulence is assumed to be
weakly collisional with $k_{\parallel 0} \lambda_{mfp,e}\ll 1$,
eliminating magnetic pumping as a potential turbulent damping
mechanism.  Finally, we specifically choose to use the \emph{B06}
scaling for MHD turbulence with $\alpha=1$ to assess the contributions
of the different proposed damping mechanisms.

We use the dependencies discussed in \secref{sec:damping} of the
proposed turbulent damping mechanisms on the dimensionless parameters
in \tabref{tab:reduced_params} to assess the regions of the
two-dimensional $(\beta_i,k_0\rho_i)$ parameter space over which each
mechanism is expected to contribute.

For the Landau-resonant ($n=0$) collisionless damping
mechanisms---Landau damping and transit-time damping---the primary
parameter dependencies summarized in \secref{sec:landaures} are
$\beta_i$ for ion Landau damping (iLD) and ion transit-time damping
(iTTD) and $(\beta_i,\tau,\mu)$ for electron Landau damping (eLD) and
electron transit-time damping (eTTD). Since the sum of LD and TTD for
each species always leads to damping in the case of isotropic
equilibrium temperatures, but separately the net energy transfer by
either mechanism to that species over a full wave period can be
positive or negative---as discussed in \secref{sec:landaures} and
illustrated in \figref{fig:ld_ttd_disp}--- we do not separate the
contributions of LD and TTD for a single species.  Furthermore, since
any turbulent cascade energy not deposited onto the ions by iLD and
iTTD at the ion scales $k_\perp \rho_i \sim 1$ will cascade to
$k_\perp \rho_i \gg 1$ and ultimately be transferred to the electrons,
the partitioning of energy between ions and electrons by these
Landau-resonant mechanisms is simply a function of $\beta_i$.
Solutions of the linear Vlasov-Maxwell dispersion relation for the
normalized ion damping rate by iLD and iTTD,
$(\gamma_{iLD}+\gamma_{iTTD})/\omega$ indicate that ion damping is
non-negligible for $\beta_i\gtrsim 0.5$, so iLD and iTTD will
contribute in this region (blue line with right arrow), independent of
the driving scale $k_0\rho_i$.  For $\beta_i\lesssim 0.5$, eLD and
eTTD will dominate the Landau-resonant damping (blue line with left
arrow), but eLD and eTTD will still contribute to terminate the
remaining turbulent cascade energy not removed by ions at
$\beta_i\gtrsim 0.5$ (blue right arrow).

As discussed in \secref{sec:icd}, ion cyclotron damping (iCD) will
contribute to the damping of turbulence if the turbulent frequencies
approach the ion cyclotron frequency $\omega/\Omega_i \rightarrow 1$,
with the scaling of this ratio given by \eqref{eq:icdscale}.  We can
solve this equation for $k_0\rho_i$ in terms of $\beta_i$ with
$\alpha=1$ for the \emph{B06} scaling, yielding
\begin{equation}
  k_0 \rho_i = \left(  \frac{\omega}{\Omega_i}  \frac{1}{\overline {\omega}}\right)^2
  \beta_i.
  \label{eq:icdthresh}
\end{equation}
Taking the factor $(\omega/\Omega_i)/\overline {\omega}=1$
gives an estimate for when $\omega/\Omega_i=1$ (black dotted line),
but iCD often becomes significant at lower values of the frequency,
 $\omega/\Omega_i \lesssim 1$.  Thus, we estimate the onset of iCD by
taking $(\omega/\Omega_i)/\overline {\omega}=1/2 \pi$ (black solid
line and upward arrow).

The impact of ion stochastic heating (iSH) on turbulent fluctuations
for the \emph{B06} scaling is given by \eqref{eq:gwsh} in
\secref{sec:iSH}.  The exponential dependence in this  equation for
$\gamma_{SH}/\omega$ yields a strong inhibition of ion stochastic heating unless  $c_2  \beta_i^{1/2} (k_0\rho_i )^{-1/4} \lesssim 1$.  Thus, we obtain a condition for $k_0\rho_i$ in terms of $\beta_i$ for the onset of iSH, given by
\begin{equation}
k_0\rho_i \gtrsim c_2^4  \beta_i^{2} .
  \label{eq:ishthresh}
\end{equation}
Note that the dependence of this condition on $c_2^4$ means that an
accurate determination of this constant is critical to assess the role
of iSH on the damping of plasma turbulence.  Taking the value of
$c_2=0.34$ from \citet{Chandran:2010a}, we plot \eqref{eq:ishthresh}
for the onset of ion stochastic heating (green line and upward arrow)
on \figref{fig:phase_k0bi}.

Kinetic viscous heating of ions (iVH) mediated by temperature
anisotropy instabilities\footnote{Recall that this mechanism is
\emph{not} the standard viscous heating mediated by microscopic
Coulomb collisions between charged plasma particles, as emphasized in
\secref{sec:iVH}.}  at high values of ion plasma beta $\beta_i \gg 1$
depends on the $\beta_i$ and the amplitude of the turbulent
fluctuations at the driving scale, as shown by \eqref{eq:gvh} in
\secref{sec:iVH}.  Because this mechanism only contributes when
$\beta_i$ is sufficiently high that the temperature anisotropy can
exceed the threshold values for the proton temperature anisotropy
instabilities given by \tabref{tb:unstableFit} and shown in
\figref{fig:unstableFit}, we take the threshold for this kinetic
damping mechanism to be dependent only on the value of the ion plasma
beta, with an estimated onset at $\beta_i \gtrsim 10$ (magenta line
and right arrow).

Finally, if the isotropic driving wavenumber $k_0\rho_i\ll 1$ and the
ion plasma beta $\beta_i \ll 1$ of the turbulence are sufficiently
low, current sheets with a large width-to-thickness ratio in the plane
perpendicular to the equilibrium magnetic field can be unstable to a
rapid collisionless tearing instability in the large-guide-field
limit, as detailed in \secref{sec:rxn}.  If growth rate of the tearing
instability is faster than the rate of the nonlinear cascade of energy
to small scales in the turbulence $\gamma_{RXN}/\omega_{nl} \gtrsim
1$, collisionless magnetic reconnection can disrupt the turbulent
cascade.  The condition that the normalized collisionless tearing instability
growth rate satisfies $\gamma_{RXN}/\omega \gtrsim 1$ (for the unstable
wavenumber with the maximum growth rate) is given
by either \eqref{eq:dpsmall_lim1} or \eqref{eq:dplarge_lim1} (both
expressions yield the same maximum value at the transitional value of
$\Delta' \delta_{in} = 1$).  In these expressions, $n=1$ corresponds
to a Harris-like current sheet with a hyperbolic tangent profile, and
$n=2$ corresponds to a sinusoidal variation of the reconnecting
magnetic field.  The determination of the tearing growth rate demands
$k_\perp \rho_i \ll 1$ for the validity of this linear tearing growth
rate calculation, so we take $k_\perp \rho_i \rightarrow \mathcal{F}$,
where $\mathcal{F} \ll 1$ is the maximum value of the normalized
perpendicular wavenumber for which the growth rate derivations remain
valid.  Here, we specify a value $\mathcal{F}=0.3$ for our estimation
of the onset of collisionless magnetic reconnection, plotting the threshold
boundaries for reconnection of intermittent current sheets ($n=1$)
(dashed red line and downward arrow) and for reconnection of
sinusoidal current sheets ($n=2$) (solid red line and downward arrow).
Note that the dependence of $k_0\rho_i$ on $\mathcal{F}$ in both
\eqref{eq:dpsmall_lim1} and \eqref{eq:dplarge_lim1} is very strong,
scaling as $\mathcal{F}^9$ for $n=1$ and as $\mathcal{F}^7$ for $n=2$,
so the determination of the maximal value of $k_\perp \rho_i\equiv
\mathcal{F} $ for the tearing instability growth has a strong
influence on whether reconnection will arise.

The phase diagram for the turbulent damping mechanisms on the
$(\beta_i,k_0\rho_i)$ plane in \figref{fig:phase_k0bi} can be used in
the following manner to predict which damping mechanisms will
contribute to the dissipation of turbulence in a specific space or
astrophysical plasma system.  In the near-Earth solar wind, the
observed ion plasma beta values span the range $0.05 \lesssim \beta_i
\lesssim 10$ \citep{Wilson:2018}, taking a 5\% and 95\% limit on the
distribution of $ \beta_i $ values.  Furthermore, the isotropic
driving wavenumber in the near-Earth solar wind spans the typical
values $10^{-4} \lesssim k_0 \rho_i \lesssim 10^{-3}$
\citep{Tu:1990b,Tu:1995,Matthaeus:2005,Howes:2008b,Kiyani:2015}.  This
range of parameters for the solar wind at 1~AU is plotted as the gray
shaded region on \figref{fig:phase_k0bi}. Reading the phase diagram,
the prediction for the damping mechanisms in the near-Earth solar wind
is that Landau-resonant damping for ions (iLD and iTTD) and electrons
(eLD and eTTD) will contribute to the damping over the full range of
solar wind turbulence parameters.  At lower values of $\beta_i
\lesssim 0.3$ and higher values of $k_0 \rho_i $, stochastic heating
is predicted to play a role.  Indeed, in the Earth's magnetosheath
(which has a similar footprint on the $(\beta_i,k_0\rho_i)$ plane as
the near-Earth solar wind), there is direct evidence from
\emph{Magnetospheric MultiScale} spacecraft observations of electron
Landau damping playing a significant role in the damping of the
turbulent fluctuations \citep{Chen:2019,Afshari:2021}. In addition, for
the solar wind in the inner heliosphere, where the values of $\beta_i$
tend to be lower than in the near-Earth solar wind, analysis of
observations from the \emph{Helios} spacecraft \citep{Martinovic:2019}
and \emph{Parker Solar Probe} spacecraft \citep{Martinovic:2020} find
evidence for ion stochastic heating in the turbulent plasma.  Thus,
phase diagrams like \figref{fig:phase_k0bi} can play a valuable role
in predicting the mechanisms that will contribute to the damping of
weakly collisional plasma turbulence in space and astrophysical
plasmas.


\section{Conclusions}
Turbulence is a ubiquitous phenomenon that arises in space and
astrophysical plasmas and mediates the transfer of the energy of
large-scale plasma flows and electromagnetic fields to sufficiently
small scales that dissipation mechanisms can convert that energy into
the microscopic energy of the plasma particles.  A grand challenge
problem in heliophysics and astrophysics is to predict the heating or
acceleration of the different plasma species by the turbulence in
terms of the dimensionless parameters that characterize the plasma and
the nature of the turbulence.  In particular, a long-term goal is to
develop predictive models of the turbulent plasma heating that
determine the partitioning of energy between the ion and electron
species, $Q_i/Q_e$, and between the perpendicular and parallel degrees
of freedom for each species, $Q_{\perp,i}/Q_{\parallel,i} $ and
$Q_{\perp,e}/Q_{\parallel,e}$. An essential step in the development of
such predictive turbulent heating models is to identify the
microphysical kinetic processes that govern the transfer of the
turbulent energy to the particles as a function of the dimensionless
parameters, information that can be summarized in a phase diagram for
the dissipation of plasma turbulence, as shown in
\figref{fig:phase_k0bi}.

The first step in this long term effort is to specify a set of the key
dimensionless parameters upon which the turbulent energy cascade and
its dissipation depend. In this paper, we propose a specific set of
ten dimensionless parameters that characterize the state of the plasma
and the nature of the turbulence,
\begin{equation}
  (\beta_{\parallel,i},\tau_\parallel, A_i, A_e, k_{\parallel 0} \lambda_{mfp,e}; k_{\perp 0} \rho_i,k_{\parallel
  0}/k_{\perp0},\chi_0,Z_0^+/Z_0^-,E_{comp}/E_{inc}),
  \nonumber 
  \end{equation}
summarized in \tabref{tab:params}. Although this set of ten parameters
is sufficient to characterize a wide variety of turbulent space and
astrophysical plasmas, the development of completely general turbulent
heating models on a ten-dimensional parameter space is unlikely to be
successful. Fortunately, we can capture many of the dominant
dependencies of the turbulence and its damping mechanisms by adopting
of reduced set of parameters in the limits that (i) the equilibrium
velocity distribution for each species is an isotropic Maxwellian and
(ii) the turbulence is driven at sufficiently large scale or
sufficiently strongly that the turbulent fluctuations at the
small-scale end of the inertial range $k_\perp \rho_i \sim 1$ satisfy
a state of strong turbulence, with a nonlinearity parameter at those
small scales of $\chi \sim 1$.  With these idealizations we arrive at
the \emph{isotropic temperature case}, where  just three
dimensionless parameters remain
\begin{equation}
(\beta_i,\tau; k_{0} \rho_i), 
  \nonumber 
  \end{equation}
summarized in \tabref{tab:reduced_params}.

A number of kinetic mechanisms have been proposed to govern the
damping of weakly collisional plasma turbulence, as enumerated in
\secref{sec:proposed}.  The critical advance presented in this paper
is to express the dependence of each of these mechanisms on the
\emph{same set of fundamental plasma and turbulence parameters}.  The
result of this analysis of the dependencies for each of the turbulent
damping mechanisms is summarized in \tabref{tab:mechanisms}.

Finally, the scalings of each of the turbulent damping mechanisms with
the isotropic driving wavenumber $k_0 \rho_i$ and the ion plasma beta
$\beta_i$ can be synthesized to generate the first phase diagram for
turbulent damping mechanisms as a function of $(\beta_i,k_0\rho_i)$,
presented in \figref{fig:phase_k0bi}. For a chosen space or
astrophysical plasma system, such as the near-Earth solar wind, the
range of dimensionless plasma and turbulence parameters can be plotted
on the phase diagram, as illustrated by the gray shaded region in the
figure.  One may then read off which mechanisms are predicted to
contribute to the damping of the turbulence in that system.  Kinetic
numerical simulations and spacecraft observations of weakly
collisional turbulence in space or astrophysical plasmas can be used
to update and refine this phase diagram, or to extend its
applicability as additional dimensionless parameters are varied, such
as the ion-to-electron temperature ratio $\tau=T_{i}/T_{e}$.  It is
worthwhile emphasizing if kinetic simulations are used to identify the
turbulent damping mechanisms and to quantify the associated energy
density transfer rates to each plasma species, it is essential that
those simulations be performed in the full three spatial
dimensions. Utilizing reduced dimensionality for such simulations may
unphysically restrict the available channels of energy transfer from
the turbulent fluctuations to the particles, potentially changing
which physical turbulent damping mechanisms are dominant and thereby
artificially altering the resulting partitioning of energy among particle
species and degrees of freedom.

Ultimately, the goal is to produce more accurate and complete
predictive turbulent heating models than those reviewed in
\secref{sec:review}, facilitating improved global modeling of
important turbulent space and astrophysical plasmas, such as the
mesoscale and macroscale energy transport through the heliosphere or
the interpretation of Event Horizon Telescope observations of
accretion disks around supermassive black holes.


\section*{Acknowledgments}
More than fifteen years of turbulent modeling efforts are synthesized
here, having benefited from interactions over many years with Steve
Cowley, Bill Dorland, Eliot Quataert, Alex Schekochihin, Greg Hammett,
Tomo Tatsuno, Ryusuke Numata, Ben Chandran, Stuart Bale, Kevin
Nielson, Jason TenBarge, Jan Drake, Fred Skiff, Craig Kletzing, Steve
Vincena, Troy Carter, Chadi Salem, Kristopher Klein, Chris Chen, Jim
Schroeder, Daniel Told, Frank Jenko, Foaud Sahraoui, Tak Chu Li,
Suranga Ruhunusiri, Jasper Halekas, Sofiane Bourouaine, Jaye Verniero,
Matt Kunz, Andrew McCubbin, Francesco Valentini, Robert Wicks, Daniel
Verscharen, Lloyd Woodham, Sarah Conley, Arya Afshari, Peter Montag,
Daniel McGinnis, Collin Brown, Rui Huang, and Alberto Felix.

\section*{Funding}
Significant funding supporting this long-term project has been
provided by NSF grants PHY-10033446, CAREER AGS-1054061 and
AGS-1842561, DOE grant DE-SC0014599, and NASA grants NNX10AC91G,
80NSSC18K0643, 80NSSC18K1217, and 80NSSC18K1371.


\bibliographystyle{jpp}

\end{document}